\documentclass[english,a4paper]{article}

\expandafter\let\csname equation*\endcsname\relax

\expandafter\let\csname endequation*\endcsname\relax

\usepackage{bm}
\usepackage{graphicx}
\usepackage{amsmath}
\usepackage{amsfonts}
\usepackage{amssymb}
\usepackage{amsthm}
\usepackage{amstext}
\usepackage{amsbsy}
\usepackage{amsopn}
\usepackage{amscd}
\usepackage{amsxtra}
\usepackage[colorlinks=true,allcolors=blue]{hyperref}
\usepackage{color}
\usepackage{soul}
\usepackage[dvipsnames]{xcolor}
\usepackage{listings}

\usepackage{cite}

\usepackage{booktabs} 
\usepackage{enumitem}
\usepackage{makeidx}
\usepackage[most]{tcolorbox}
	\tcbuselibrary{theorems}
	\tcbuselibrary{breakable}
\usepackage[left=1cm,right=1cm,top=2cm,bottom=2cm]{geometry}

\newcommand{\ket}[1]{| #1 \rangle}
\newcommand{\bra}[1]{\langle #1 |}
\newcommand{\braket}[2]{\langle #1 | #2 \rangle}

\newcommand{\sgn}{\mathrm{sgn}}

\newcommand{\Mc}[1]{\mathcal{#1}}

\newcommand{\setR}{\mathbb{R}}

\newcommand{\Id}{\mathbb{I}}

\newcommand{\ii}{\imath}

\newcommand{\sigz}{\hat{\sigma}_z}
\newcommand{\sigx}{\hat{\sigma}_x}
\newcommand{\sigy}{\hat{\sigma}_y}

\begin{document}

\newtcbtheorem[auto counter]{definition}%
  {Definition}{breakable,theorem style=plain,fonttitle=\bfseries\upshape, fontupper=\itshape ,arc=0mm, boxrule=0.5mm,coltitle=black, colback=gray!10!white,colframe=gray!25!black, left=1mm,right=1mm,top=1mm,bottom=1mm}{def}
\newtcbtheorem[auto counter]{example}%
  {Example}{breakable,theorem style=plain,fonttitle=\bfseries\upshape ,
  arc=0mm, boxrule=0.5mm,coltitle=black, colback=gray!10!white,colframe=white, left=1mm,right=1mm,top=1mm,bottom=1mm}{example}
%
%
  \newtcbtheorem[auto counter]{property}%
  {Property}{breakable,theorem style=plain,fonttitle=\bfseries\upshape, fontupper=\itshape ,arc=0mm, boxrule=0.5mm,coltitle=black, colback=gray!10!white,colframe=gray!25!white, left=1mm,right=1mm,top=1mm,bottom=1mm}{property}
  \newtcbtheorem[auto counter]{proposition}%
  {Proposition}{breakable,theorem style=plain,fonttitle=\bfseries\upshape, fontupper=\itshape ,arc=0mm, boxrule=0.5mm,coltitle=black, colback=gray!10!white,colframe=gray!50!white, left=1mm,right=1mm,top=1mm,bottom=1mm}{prop}
\newtcbtheorem[auto counter]{theorem}%
  {Theorem}{breakable,theorem style=plain,fonttitle=\bfseries\upshape, fontupper=\itshape ,arc=0mm, boxrule=0.5mm,coltitle=black, colback=gray!10!white,colframe=gray!75!black, left=1mm,right=1mm,top=1mm,bottom=1mm}{thm}

\tcolorboxenvironment{proof}{
blanker,breakable,left=5mm,
before skip=10pt,after skip=10pt,
borderline west={1mm}{0pt}{gray}}

\lstset{language=Mathematica,numbers=left, numberstyle=\tiny, stepnumber=1, mathescape=true,frame=single}

\title{Introduction to Theoretical and Experimental aspects of Quantum Optimal Control}

\author{Q. Ansel\footnote{Institut UTINAM,  CNRS UMR 6213, Universit\'{e} de Franche-Comt\'{e}, Observatoire des Sciences de l’Univers THETA, 41 bis avenue de l’Observatoire, F-25010, Besançon, France}, E. Dionis, F. Arrouas, B. Peaudecerf\footnote{Laboratoire Collisions Agr\'egats R\'eactivit\'e, UMR 5589, FeRMI, UT3, Universit\'e de Toulouse, CNRS, 118 Route de Narbonne, 31062 Toulouse CEDEX 09, France}, S. Gu\'erin\footnote{Laboratoire Interdisciplinaire Carnot de Bourgogne, CNRS UMR 6303, Universit\'{e} de Bourgogne, BP 47870, F-21078 Dijon, France}, D. Gu\'ery-Odelin\footnote{Laboratoire Collisions Agr\'egats R\'eactivit\'e, UMR 5589, FeRMI, UT3, Universit\'e de Toulouse, CNRS, 118 Route de Narbonne, 31062 Toulouse CEDEX 09, France}, D. Sugny\footnote{Laboratoire Interdisciplinaire Carnot de Bourgogne, CNRS UMR 6303, Universit\'{e} de Bourgogne, BP 47870, F-21078 Dijon, France, dominique.sugny@u-bourgogne.fr}}

\maketitle

\begin{abstract}
Quantum optimal control is a set of methods for designing time-varying electromagnetic fields to perform operations in quantum technologies. This tutorial paper introduces the basic elements of this theory based on the Pontryagin maximum principle, in a physicist-friendly way. An analogy with classical Lagrangian and Hamiltonian mechanics is proposed to present the main results used in this field. Emphasis is placed on the different numerical algorithms to solve a quantum optimal control problem. Several examples ranging from the control of two-level quantum systems to that of Bose-Einstein Condensates (BEC) in a one-dimensional optical lattice are studied in detail, using both analytical and numerical methods. Codes based on shooting method and gradient-based algorithms are provided. The connection between optimal processes and the quantum speed limit is also discussed in two-level quantum systems. In the case of BEC, the experimental implementation of optimal control protocols is described, both for two-level and many-level cases, with the current constraints and limitations of such platforms. This presentation is illustrated by the corresponding experimental results.
\end{abstract}



\section{Introduction to Quantum Control}

\begin{figure}[htbp]
    \centering
    \includegraphics[width=15cm]{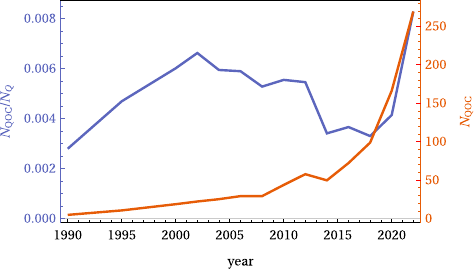}
    \caption{Growth in the number of scientific articles $N_{QOC}$ (red curve) with the keywords ``quantum" and ``optimal control" published each year on the internet, during the period 1990-2022, and ratio between $N_{QOC}$ and the number of scientific articles $N_Q$ with the keywords ``quantum" (blue curve). Despite an increasing number of annual publications, the ratio decreased during the  period 2000-2015 (the number of publications in quantum science grew very rapidly during these years), but since 2018 the trend has reversed. At the beginning of 2023 (not shown in the graph), the trend is confirmed with almost 1\% of publications concerning QOC. Data were collected using Google Scholar.}
    \label{fig:number_of_publications}
\end{figure}

The design and development of quantum technologies requires the use of many advanced techniques in order to tackle the fragility of quantum information, and the difficulty of isolating and manipulating quantum entities~\cite{raimond2006exploring,kurizki2015quantum,acin2018quantum,becher20232023,shore-book,ricebook,RMPsugny}. These challenges have led in particular to the development of Quantum Optimal Control (QOC)~\cite{guerin2003,werschnik2007,brif2010,bonnard_optimal_2012,altafini2012,cat,
kochroadmap,rembold2020introduction,PRXQuantumsugny,kuprov2023spin,dalessandro-book,stefanatos2021}, a branch of optimal control theory whose aim is to adapt and apply the tools of optimal control to quantum systems. As its name suggests, optimal control~\cite{pontryaginbook,leemarkusbook,bryson1975applied,kirk2004optimal,trelat2012optimal,agrachev-book,bressan-piccoli,schaettler-book,liberzon-book,boscain-book,jurdjevic-book} is a mathematical theory that refers to the design of time-varying controls to manipulate dynamical systems in order to ideally achieve specific goals (encapsulated in a figure of merit). QOC was initiated in the eighties, and has since then seen an exponential development. It nowadays comprises a very powerful toolbox of analytical and numerical methods for designing control protocols under various experimental constraints and limitations. QOC is therefore not only a mature theoretical field, but also a very efficient approach from an experimental point of view on many quantum platforms. A key aspect of QOC is that it is an open-loop procedure that provides the optimal control process without any feedback from the experiment. Successful applications therefore require a very accurate modeling of the dynamical system and careful consideration of the controls that are experimentally available. Such QOC techniques have recently been successfully applied in many different areas of quantum technologies, ranging from quantum computing and simulation to sensing~\cite{kochroadmap,rembold2020introduction}, making QOC a key tool of growing importance in the development of quantum technologies. This can be seen by its growing presence in the literature, as illustrated in Fig.~\ref{fig:number_of_publications}. QOC is not the only control method that is available to manipulate quantum systems in an optimized fashion. Among others, we can mention adiabatic passage techniques~\cite{vitanov2001,RMPstirap,guerin2011}, shortcut to adiabaticity approaches~\cite{RMPSTA,STA,campo2013,whitty2020,torosov2021} or composite pulses~\cite{vitanovCP,dridi2020,Ivanov2022,torosov2011}. While quantum optimal control in most cases deals
with time-dependent pulses, it can be applied to other issues such as the control of Hamiltonian structure~\cite{rabitz2001,rabitz2011}. Optimal control can also be performed in the frequency domain as, e.g., shown in~\cite{rabitz2016}. Furthermore, the framework of the quantum speed limit (QSL)~\cite{deffner2017quantum,frey2016,bukov2019,oconnor2021,poggi2019,campo2013b}, which seeks to establish lower bounds on the minimum time required to steer a system from a given initial state to a target state, is closely linked to QOC~\cite{deffner2017quantum,calarco2009,diaz2020,hegerfeldt2013driving}. The corresponding time is expressed as a ratio between the distance to the target state and the dynamical speed of evolution. This approach, like QOC, has been the subject of intense development in recent years.

Several textbooks and reviews have introduced and described the basic elements and provided a fairly complete picture of the field of QOC~\cite{bonnard_optimal_2012,rembold2020introduction,
PRXQuantumsugny,dalessandro-book,wilhelmreview}. Others have focused on its practical numerical implementation or on the description of various optimization algorithms~\cite{werschnik2007,bonnans2006numerical,koch2012,borzi-book}. However, either their starting points are at a high mathematical level, or they only consider a specific aspect of the QOC toolbox, so that it can be difficult for a newcomer to get a complete overview of the field and to apply such techniques to their own system. This tutorial paper aims to fill that gap. This introduction covers a wide range of aspects, from the description of the Pontryagin Maximum Principle (PMP) which can be seen as the central mathematical result of the theory~\cite{pontryaginbook,leemarkusbook,liberzon-book}, to the analytical or numerical computations of optimal control and their experimental implementation. Mathematical issues are treated with a minimum of rigor, but with several reminders of basic notions. The various numerical algorithms available in the literature are described. Special emphasis is given to two numerical methods, namely the shooting and the gradient-based algorithms, which allow solving low- and high-dimensional quantum optimal problems respectively. Numerical codes are provided in the supplementary material. We point out that the efficiency of such algorithms is due to the absence of traps in most of quantum control landscapes, as originally proposed in~\cite{rabitz2004,rabitz2006} and then discussed and extended in~\cite{pechen2011there,moore2012exploring,chakrabarti2007quantum,pechen2012quantum,larocca2020}. Due to the large number and diversity of optimization methods, some interesting topics are not covered in this tutorial. Among others, we can mention machine learning techniques~\cite{HushScience2017,bukovPRX2018,dayPRL2019,mehtaPhysRep2019,carleoRMP2019,giannelli2022,khalid2023}, the Hamilton-Jacobi-Bellman approach~\cite{bertsekasbook}, second-order optimization algorithms~\cite{grape2,tannor2011, sherson2020,goodwin2023}, closed-loop control~\cite{naturerouchon,egger2014,porotti2023}, controllability and accessibility of quantum systems~\cite{infinity-1,sachkov_controllability_2000,schirmer_complete_2001,schirmer2003controllability,Albertini_notions_of_controllability_2003}, quantum control landscapes~\cite{pechen2011there,moore2012exploring,chakrabarti2007quantum,pechen2012quantum,larocca2020}, robust optimal control~\cite{kobzar2004exploring,k1,k2,kobzar2012exploring,daems:2013,chen2014sampling,Van_Damme_robust_2017,zeng2018,wu2019,dridi2020b,dong2021,Ansel_2021,propson2022,guerin2022,guerin2022b,harutyunyan2022robust,nelson2023,schirmer2023,carolan2023}, optimal control of linear systems~\cite{liberzon-book,vardan2020,vardan2019,li2017,li2011,evangelakos2023b},
optimal control and quantum sensing~\cite{Degen_review_2017,poggiali2018,wittler2021,Lin_optimal_2021,ansel2023optimal,Liu_optimal_QM_review_2022} and the control of open quantum systems~\cite{sugny07,dirr2009,bonnard2009,addis2016,Koch2016,fux2021,pechen2021,pechen2023}.

The success of QOC has been demonstrated theoretically in a large number of quantum systems~\cite{kochroadmap}. Although such control protocols are interesting from a theoretical point of view, e.g. to know the physical limits of a dynamical system in terms of control time or fidelity, they are generally not the final answer to a control problem: the ultimate goal is to implement the calculated control process experimentally. Several difficulties have to be overcome in order to transfer the theoretical result to the experiment, which can be divided into two categories. The first problem comes from the model system which must be sufficiently precise in an open-loop framework to describe the physical system, but relatively simple to apply the optimization algorithms numerically. A second obstacle is related to the family of control pulses that can be implemented experimentally. Depending on the experimental setup, different constraints may arise, ranging from the control time and the maximum intensity of the control available, to its Fourier bandwidth or its time discretization. It is now possible to take these constraints into account in optimization algorithms. This development makes QOC more useful in
terms of experimental applications and helps to bridge the gap between control theory and control
experiments. Such a project has been successfully carried out on various platforms such as superconducting circuits, NV centers, magnetic resonance and trapped atoms, ions and molecules~\cite{kochroadmap}. The experimental implementation of optimal control protocols is highly dependent on the system under study, and we cannot cover all possible situations here. In this paper, we focus on a specific example for which we provide numerical solutions, namely the control of a Bose-Einstein Condensate (BEC) in a one-dimensional optical lattice~\cite{RMP1999,RMP2008,gross2017}. This system has attracted a lot of interest in recent years from a control point of view~\cite{Hohenester2007,Jager2014,Sorensen2018,bason2012,zhou2018,Weidner2018,BEC2021,BEC2023,BEC2024,frank2016}, especially for quantum simulation applications~\cite{altman,noriRMP,cirac2012}, for which QOC can be  used to efficiently prepare the initial state of the system~\cite{dupont2023}. It allows us to illustrate the implementation of optimal control on an infinite-size system, as well as to emulate a two-level system. We discuss the different steps of the experimental implementation from the modeling of the system dynamics, to the experimental constraints on the control design and the measurement of the final state of the system. The impact of various experimental limitations is highlighted and discussed. Experimental results based on optimal control protocols are presented.

This tutorial  paper is organized as follows. In Sec.~\ref{chap:optimal control theory}, we introduce the optimal control theory based on the PMP. The analogy with classical Lagrangian and Hamiltonian systems is used to describe the basic concepts of this theory. In Sec.~\ref{sec:QOC}, we show how such results can be adapted to quantum systems. The time-optimal control of a two-level quantum system is used as an illustrative example. The connection between QOC and QSL in this system is discussed. In Sec.~\ref{sec:numerical_methods}, a detailed introduction to numerical methods is given with particular emphasis on two different  approaches, namely the shooting method and gradient-based optimization algorithms which are also illustrated in a two-level quantum system. Pseudo-codes describing the structure of the algorithms are provided in the main text, and Python codes for specific control processes can be found in the supplementary material. These numerical approaches are used in Sec.~\ref{sectheoexp} to manipulate a BEC in an optical lattice. A complete description of an experimental implementation is given as well as the constraints and limitations of the experimental apparatus used. A number of examples in classical and quantum physics are given throughout the text. The simplest ones are placed in a gray box and can be ignored by a reader already experienced in optimal control. A conclusion is given and prospective views are suggested in Sec.~\ref{sec:conclusion}.~\ref{secappA} gives a list of mathematical symbols and acronyms used in the article. \ref{app_lagrange} and~\ref{appendixPMP} contain mathematical results used in the main text. A description of the numerical codes of the supplementary material is also provided in~\ref{app_code}.


\section{Optimal Control Theory and Pontryagin Maximum Principle}
\label{chap:optimal control theory}

\subsection{Introduction}

Optimal Control Theory (OCT) has its roots in the calculus of variations which is over 300 years old~\cite{Courant_Hilber_vol_1,Courant_Hilber_vol_2,gelfand2000calculus}.
Interest in this branch of mathematics grew rapidly with the advent of computer science in the early 1960s. A rigorous mathematical framework for OCT was given by L. Pontryagin and his co-workers in 1960 with the introduction of the Pontryagin Maximum Principle (PMP)~\cite{pontryaginbook,leemarkusbook,bryson1975applied,kirk2004optimal,liberzon-book}, which then led to a variety of applications~\cite{bryson1996optimal}. In particular, OCT was at the origin of optimal trajectory prediction in aeronautics~\cite{bonnard_optimal_2012,trelat2012optimal}. Today, OCT is used in a wide range of fields ranging from economics, to physics and electronics, to name but a few. The PMP transforms the optimal control problem into a
generalized Hamiltonian system subject to a maximization condition and some boundary conditions. In this framework,
the goal is to find the Hamiltonian trajectory that reaches the target state, while minimizing the cost functional which
defines the optimization procedure. A key advantage of the PMP is that it reduces the initial infinite-dimensional control
landscape to a finite low-dimensional space. This brief description also shows that optimal control is closely related  to classical mechanics and its Lagrangian or Hamiltonian formalism.

Before we present the PMP at the end of this section, let us take a step back to basics. As the above brief description of optimal control shows, a first observation is that optimal control problems are very similar to those in classical mechanics~\cite{goldsteinbook,arnoldbook}, where the goal is to find the trajectory of a classical system that minimizes a certain quantity, the action.
The same direction is followed in OCT, but instead of the usual action, other quantities relevant to the control procedure are minimized. Another important modification is the presence of a time-dependent parameter in the dynamical equations which can be shaped to some extent by an external operator. In the case of an aeroplane, for example, the control parameters include all possible actions that can be performed on the aircraft, such as modifying the engine thrust or changing the orientation of elevators and ailerons. In quantum physics, the control agent is generally a shaped electromagnetic field. The formal concepts are illustrated in this section using a simple example, namely a point particle controlled by a time-dependent force. We begin by recalling some key elements of the calculus of variations applied to a point particle.
\begin{example}{}{ex1}
 A basic idea of the calculus of variation and of the principle of least action is to derive the equation of motion of a physical system from a single quantity, the action~\cite{goldsteinbook}. Mathematically, the latter takes the form of a functional~\cite{gelfand2000calculus} i.e. a function of functions. In the case of a free one-dimensional particle, the action $S$ can be expressed as
\begin{equation}
S[x]= \int_0^{t_f} \frac{1}{2} m \dot x^2(t)~dt,
\end{equation}
where $x(t)\in \setR$ is the position of the particle at time $t$ with $t\in [0,t_f]$, $t_f$ the duration of the dynamics, $m$ its mass, and the dot symbolizes the time derivative. The physical motion is associated with the least action $S$. From the condition $\delta S=0$ where $\delta S$ is the functional derivative of $S$ taken for two trajectories close to each other~\cite{goldsteinbook}, it can be shown that the physical trajectory satisfies the Euler-Lagrange equation
\begin{equation}
\frac{\partial \Mc L}{\partial x} -\frac{d}{dt} \frac{\partial \Mc L}{\partial \dot x} = 0,
\label{eq:Euler_Lagrange_point_particle}
\end{equation}
where $\Mc L = m\dot x^2/2$ is the Lagrangian, here equal to the kinetic energy of the particle. An explicit calculation leads to the expected result $m \Ddot x(t) = 0$.

The model system can be extended by introducing a control $f(t)\in\setR$, which takes the form of a  force applied to the particle. With this modification, the action becomes
\begin{equation}
S[x]= \int_0^{t_f} \left( \frac{1}{2} m \dot x^2 (t) + f(t) x(t) \right)~dt,
\end{equation}
where the extra term $-f(t)x(t)$ is a potential energy. The Lagrangian is then defined as the difference between the kinetic energy and the potential energy. The equation of motion, calculated using Eq.~\eqref{eq:Euler_Lagrange_point_particle} with the new Lagrangian is
$m \Ddot x(t) = f(t)$. The latter can be determined from Newton's law, but also from the Hamiltonian formalism, in which the Lagrangian is replaced by the Hamiltonian $H$. The Hamiltonian is a function on the phase space i.e. position and momentum ($\Gamma = \setR^2$ for a point particle), rather than a function of position and velocity for the Lagrangian. The phase space can be constructed by defining a conjugate variable, the momentum, as $p = \tfrac{\partial \Mc L}{\partial \dot x}$, and $H$ can be expressed as $H= p \dot x - \Mc L$~\cite{goldsteinbook}. This change of variables is called a Legendre transformation. The equations of motion are given by Hamilton's first-order differential equations
\begin{equation}
\dot x=\frac{\partial H}{\partial p},~\dot p=-\frac{\partial H}{\partial x},
\label{eq:Hamilton_equation_point_particle}
\end{equation}
which allow us to recover the system dynamics. In the case of a point particle, we have $p = m \dot x$ and the Hamiltonian is $H = \tfrac{p^2}{2m} - f x$, which leads from Eq.~\eqref{eq:Hamilton_equation_point_particle} to $\dot p = f$ and $m\dot x = p$, and finally to $m\Ddot x = f$ as expected.

Now suppose that a particular time-dependent force can be generated. It is clear that the dynamics of the system are modified, the corresponding trajectory being solution of the equation $m\Ddot x(t)=f(t)$. The particle is then steered from the point $x(0)$ to $x(t_f)$. This example can be reformulated in terms of optimal control. The idea is to reverse the procedure by fixing the initial and final states of the system a priori and by designing the corresponding control. Since this problem can have a very large number of solutions, a specific control is selected based on an additional criterion as described below.
\end{example}

An optimal control problem is usually defined as follows. The first step is to introduce the system to be controlled, whose state $X(t) = \{ X_a(t)\}_{a=1,\cdots,n}$ is a real vector, $X(t)\in \setR^n$ with coordinates $X_a$. We assume that the system dynamics are described by a real first-order differential equation of the form
\begin{equation}\label{eqcontrol}
\dot X(t) = F (X(t),u(t),t),
\end{equation}
as provided e.g. by Eq.~\eqref{eq:Hamilton_equation_point_particle}, where $u(t) \in U \subset \setR^m$ is the control with $m$ real components and $F = \{ F_a\}_{a=1,\cdots,n}$ is a vector function that should be smooth enough. The set $U$ corresponds to the admissible values of the control and is determined by the operator. This choice is usually dictated by the experimental limitations of the device or by the hypotheses used to derive the model system. Note that $U=\mathbb{R}^m$ if there is no specific constraint. A standard example for a one-dimensional control parameter is $U=[u_{\min},u_{\max}]$ where $u_{\min}$ and $u_{\max}$ are respectively the minimum and maximum of allowed control values.
A solution to Eq.~\eqref{eqcontrol} is well-defined from a mathematical point of view if the control $u$ belongs to a particular set of functions called the \textbf{\textit{admissible}} controls. In many cases, continuous or piecewise continuous functions are sufficient to guarantee the existence of a solution, but not of the optimal control, as discussed in~\cite{PRXQuantumsugny}. We also emphasize that there are problems for which the optimal control is not a piecewise continuous function but presents for instance a chattering process, characterized by an infinite number of switchings between two extreme values in a finite time interval~\cite{schaettler-book,fuller}. Note that this phenomenon can also be observed in QOC as recently shown in~\cite{Robin2022}.


The optimal control protocol can be found by introducing a cost functional $\mathcal{C}$ that should be minimized. Note that a maximization can also be considered by minimizing $-\mathcal{C}$. The functional $\mathcal{C}$ can be expressed as the sum of a terminal cost $G$ and a running cost $F_0$ depending respectively on the final state and the trajectory followed by the system:
\begin{equation}
\mathcal{C} = G(X(t_f),t_f) + \int_0^{t_f} F_0 (X(t),u(t),t)dt ,
\label{eq:def_cost_fun}
\end{equation}
where $t_f$ is the control time, which can be fixed or free. In standard applications, the terminal cost $G$ can describe the distance to the target state, while the second term can penalize either the control time or the energy of the control. Such running costs correspond to $F_0=1$ and $F_0=\frac{1}{2}u^2(t)$, respectively, up to a constant factor.
\begin{example}{}{ex2}
We consider the problem of steering a one-dimensional point particle with minimum energy consumption between the points $x=0$ and $x=1$ in a given time $t_f$, such that the initial and final velocities are zero. The equations of motion in the phase space $\mathbb{R}^2$ are
\begin{equation}
\frac{d}{dt} \left(\begin{array}{c}
x \\
p
\end{array} \right) = \left( \begin{array}{c}
p/m \\
f
\end{array} \right),
\end{equation}
with initial $(0,0)$ and final $(1,0)$ states respectively. The cost functional associated with this optimal control problem is chosen to be
\begin{equation}
\mathcal{C} = (x(t_f)-1)^2+p(t_f)^2+ \int_0^{t_f} \frac{1}{2}\alpha f^2(t) ~dt,
\end{equation}
where $\alpha$ is a constant factor such that the second term has an energy dimension. Note that $\alpha$ can also be used to weight the relative importance of the two terms in the cost functional. The minimum of $\mathcal{C}$ corresponds to a compromise between the distance of the final state to the target and the energy consumed along the trajectory to reach the final state.
\end{example}


In summary, the  task in an optimal control problem is to find the control $u^*$ that minimizes the cost functional $\mathcal{C}$ under the constraint that $\dot X=F(X,u,t)$. It can be mathematically described as an infinite dimensional constrained optimization problem since all amplitudes $u(t)$ in a continuous time interval are optimized. As in a finite dimensional constrained optimization problem, the main difficulty comes from the condition~\eqref{eqcontrol} to be satisfied at any time $t$. The functional $\mathcal{C}$ cannot be minimized directly due to this additional constraint. The idea is then to increase the dimension of the state of the system to obtain an unconstrained optimization problem~\cite{bryson1975applied,ito2008lagrange,contreras_dynamic_2017}. An extended space is defined by doubling the number of variables of the system state. The adjoint state $\Lambda (t)\in\setR^n$ is introduced and the new state can be written as $(X,\Lambda,u)\in\setR^{2n+m}$. The state $\Lambda(t)$ plays the same role as Lagrange multipliers in a finite dimensional problem, except that the static constraints and a finite number of Lagrange multipliers are respectively replaced by a dynamical constraint~\eqref{eqcontrol} and a time-dependent function. We refer the reader to~\ref{app_lagrange} for details on this approach.

We then consider the action associated with the optimal control problem:
\begin{definition}{Action of the optimal control problem}{Action of the optimal control problem}
Let $X(t) \in \setR^n$ be the state of a physical system at time $t$, $\Lambda(t)\in \setR^n$ its adjoint state, and $u(t) \in U \subset \setR^m$ a control. Let $G$ be a terminal cost function, $F_0$ a running cost, and $F$ a vector function describing the system dynamics. The optimal control action is defined as
\begin{equation}\label{eq_def_action}
S = G(X(t_f),t_f) + \int_0^{t_f} dt \underbrace{\left[ F_0 (X(t),u(t),t) +\Lambda(t)\cdot \left(\dot X(t)-F(X(t),u(t),t)\right)\right]}_{\Mc L (X,\dot{X},u,t,\Lambda)}.
\end{equation}
The function in the integral is the Lagrangian $\Mc L$ of the optimal control problem.
\end{definition}
From this definition, it is straightforward to deduce that $\Lambda$ is the conjugate coordinate of $X$ using
$$
\frac{\partial \Mc L}{\partial \dot X} = \Lambda.
$$

\begin{example}{}{ex3}
A direct application of Def.~\ref{def:Action of the optimal control problem} to the system defined in Example~\ref{example:ex2} leads us to the following Lagrangian:
\begin{equation}
\Mc L = \frac{1}{2}\alpha f^2 + \Lambda_x( \dot x - p/m) + \Lambda_p(\dot p -f),
\end{equation}
where $\Lambda_x$ and $\Lambda_p$ are respectively the adjoint states of $x$ and $p$, $\Lambda=(\Lambda_x,\Lambda_p)$.
\end{example}

\subsection{First-order variation and Pontryagin Maximum Principle}
\label{sec:first order variation}

\subsubsection{Lagrangian formulation.}
\label{sec:Lagrangian approach}

The action introduced in Def.~\ref{def:Action of the optimal control problem} is very similar to the action of a usual classical system and it contains all the information needed to find the solutions of the optimal control problem. We use the same approach as in classical mechanics by calculating the conditions that the extremals of $S$ must fulfill when considering a small variation of the control~\cite{Courant_Hilber_vol_1,Courant_Hilber_vol_2,gelfand2000calculus}. The analysis is more demanding than in classical mechanics because the set $U$ can be closed as in the case $U=[u_{\min},u_{\max}]$. One must be careful at the boundary of $U$, because it leads to extremals that are not specified by functional derivatives of $S$. This is similar to the case of a function $f$ defined on a closed interval which has extrema on the boundary of the interval that are not given by a zero of its derivative.

Here, we examine a system with unconstrained controls. The case of a closed set is discussed below with the Hamiltonian formalism. Extremals are characterized by the condition $\delta S=0$ for small variations $\delta X$, $\delta \Lambda$, and $\delta u$ which are assumed to be independent. These variations are chosen to be as general as possible, and they are not a priori restricted to satisty $\dot X = F$.

The functional derivative of $S$ with respect to the three variables, $X$, $\Lambda$ and $u$ is
$$
\delta S=\frac{\partial G}{\partial X(t_f)}\delta X(t_f)+\int_0^{t_f}\left[\frac{\partial F_0}{\partial X}\delta X+\frac{\partial F_0}{\partial u}\delta u
+\delta \Lambda\cdot (\dot{X}-F)+\Lambda\cdot\left(\delta \dot{X}-\frac{\partial F}{\partial X}\delta X-\frac{\partial F}{\partial u}\delta u\right)\right].
$$
Integrating by part the term $\Lambda\cdot\delta \dot{X}$, we obtain
\begin{align}
&\delta S=\left(\frac{\partial G}{\partial X(t_f)}+\Lambda(t_f)\right)\delta X(t_f)+\Lambda(0)\delta X(0)\nonumber \\
& +\int_0^{t_f}dt\left(\left[\frac{\partial F_0}{\partial X}-\dot{\Lambda}-\Lambda\frac{\partial F}{\partial X}\right]\delta X+\left[\dot{X}-F\right]\delta \Lambda+
\left[\frac{\partial F_0}{\partial u}-\Lambda\frac{\partial F}{\partial u}\right]\delta u \right).\nonumber
\end{align}
Next, we deduce that the extremals of $S$ fulfill the following conditions
\begin{align}
& \dot{\Lambda}=\frac{\partial F_0}{\partial X}-\Lambda\frac{\partial F}{\partial X},~\Lambda(t_f)=-\frac{\partial G}{\partial X(t_f)},\nonumber \\
& \dot{X}=F,~X(0)=X_0,\nonumber \\
& \Lambda\frac{\partial F}{\partial u}-\frac{\partial F_0}{\partial u}=0.\nonumber
\end{align}
The initial condition on $X$ comes from the fact that $\delta X(0)=0$, i.e. we consider trajectories with the same initial state.


\noindent The results are summarized in the following theorem.
\begin{theorem}{Extremals of the action}{Extremals of the action}
\label{th:extremal}
Extremals of the action of the optimal control problem  at a fixed final time with the initial state $X(0)=X_0$ satisfy the following conditions:
\begin{align}
\label{eq_S_p}
& \dot X_a (t)  = F_a (t) ~\textrm{(Euler-Lagrange equation for $\Lambda$)},\\
\label{eq_S_x}
& \dot \Lambda_a (t)  = \frac{\partial F_0}{\partial X_a(t)} - \Lambda(t)\cdot \frac{\partial F}{\partial X_a(t)}~\textrm{(Euler-Lagrange equation for $X$)},\\
& \Lambda_a (t_f) = -\frac{\partial G}{\partial X_a(t_f)}~\textrm{(Boundary condition for $X$)},\\
\label{eq_S_u}
& \frac{\partial F_0}{\partial u_a}  = \Lambda(t)\cdot \frac{\partial F}{\partial u_a(t)}~\textrm{(Euler-Lagrange equation for $u$)}.
\end{align}
\end{theorem}

Several comments on these results can be made. Equation~\eqref{eq_S_p} is the equation of motion, and thus, any solution of $\delta S =0$ must correspond to a physical trajectory of the system. Equation~\eqref{eq_S_x} is derived from the Euler-Lagrange equation $\frac{d}{dt}\frac{\partial \mathcal{L}}{\partial \dot{X}}-\frac{\partial \mathcal{L}}{\partial X}=0$ for $X$. It returns a differential equation for the dynamics of $\Lambda(t)$, whose value at final time is given by the gradient of the terminal cost $G$ over the final state $X(t_f)$. Equation~\eqref{eq_S_u} leads to a strong condition for the control $u$. As can be seen in the examples, Eq.~\eqref{eq_S_u} can often be used to express the control as a function of $X$ and $\Lambda$, i.e. $u(t) = u(X(t),\Lambda(t))$. The control is then completely determined by the dynamics of the state and its adjoint state. Using the $n$ initial conditions of $X$ and the $n$ final conditions of $\Lambda$ parameterizing $X(t)$ and $\Lambda(t)$, we obtain that the optimal control is a function of a finite number of parameters, transforming thus an infinite-dimensional optimization problem into a finite one.

We emphasize that identities of Th.~\ref{thm:Extremals of the action} are satisfied by all the extremals of the action. It is therefore only a necessary optimality condition that selects trajectory candidates to be optimal.
In practice, it is straightforward to establish equations of Th.~\ref{thm:Extremals of the action} which can be expressed as a set of non-linear coupled differential equations with two-side boundary conditions. The difficult task is to find the solutions of such equations. This can be done analytically in the simplest cases, otherwise numerically. Some examples of applications to quantum systems are given in Sec.~\ref{sec:QOC}.

\begin{example}{}{ex4}

Theorem~\ref{thm:Extremals of the action} can be applied directly  to the optimal control Lagrangian $\Mc L =\frac{1}{2}\alpha f^2 + \Lambda_x( \dot x - p/m) + \Lambda_p(\dot p -f)$ of Example~\ref{example:ex3}. The derivatives of $\Mc L$ with respect to the states $(x,p)$ and the adjoint states $(\Lambda_x,\Lambda_p)$ give the equations of motion
\begin{equation}
\begin{split}
& \dot x = p/m, \\
& \dot  p = f, \\
& \dot \Lambda_x = 0, \\
& \dot \Lambda_p = -\frac{\Lambda_x}{m},
\end{split}
\end{equation}
while control can be obtained from the derivative of $\Mc L$ with respect to the control $f$
$$
f=\frac{\Lambda_p}{\alpha}.
$$
Boundary conditions can also be deduced using the terminal cost $G=(x(t_f)-1)^2+p(t_f)^2$, introduced in Example~\ref{example:ex2}. We obtain:
\begin{align*}
    \Lambda_x(t_f)& = - 2(x(t_f)-1),\\
    \Lambda_p(t_f) &= - 2p(t_f).
\end{align*}
For an optimal trajectory reaching exactly the target, we have $x(t_f)=1$ and $p(t_f)=0$, and thus $\Lambda_x(t_f) = \Lambda_p(t_f)=0$.
\end{example}

\subsubsection{Hamiltonian formulation.}

\begin{figure}[htp]
\begin{center}
\includegraphics[width=7.5cm]{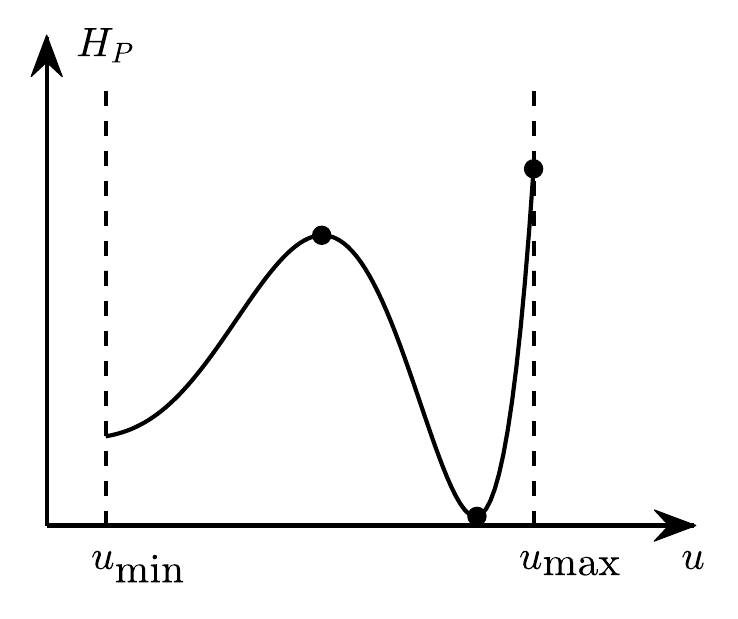}
\end{center}
\caption{Schematic plot of the Pontryagin Hamiltonian $H_P$ as a function of the control $u$ in the interval $U=[u_{\textrm{min}},u_{\textrm{max}}]$ represented by the vertical dashed lines. The dots indicate the position of the extremal values of $H_P$. In this example, the global maximum of $H_P$ lies on the edge of $U$, while a local maximum lies inside $U$.}
\label{fignew}
\end{figure}

As in classical mechanics~\cite{goldsteinbook}, we can adopt either a Lagrangian or a Hamiltonian formalism to derive the equations of motion. The Hamiltonian structure can be derived from the optimal control action introduced in Def.~\ref{def:Action of the optimal control problem}. Since $\Lambda$ is the conjugate momentum of $X$, the Hamiltonian $H_P$, called the Pontryagin Hamiltonian, can be defined as $H_P = \Lambda\cdot \dot X - \Mc L = \Lambda\cdot  F - F_0$. When $X$ and $\Lambda$ satisfy the extremal equations, the functional derivative of $S$ given by
$$
\delta S=\int_0^{t_f}\left(\frac{\partial F_0}{\partial u}-\Lambda\cdot \frac{\partial F}{\partial u}\right)\cdot \delta u(t) dt
$$
can be then written as
\begin{equation}\label{eqSPont}
\delta S=-\int_0^{t_f}\frac{\partial H_P}{\partial u}\cdot \delta u(t) dt .
\end{equation}
Equation~\eqref{eqSPont} allows us, for an open set $U$, to find the extremal condition for the control, $\frac{\partial H_P}{\partial u}=0$. Here only the \textbf{\textit{normal}} extremals are obtained. Under certain conditions, control processes that do not depend on the running cost $F_0$ can also be solutions of the optimal control problem. Such extremals are called \textbf{\textit{abnormal}}. To take them into account, a negative constant $\Lambda_0\leq 0$ is added to the definition of the Hamiltonian $H_P$ which is given in its final form as follows~\cite{bonnard_optimal_2012,PRXQuantumsugny,kirk2004optimal}.
\begin{definition}{Pontryagin Hamiltonian}{Pontryagin Hamiltonian}
The Pontryagin Hamiltonian $H_P$ is given by:
\begin{equation}
H_P = \Lambda\cdot F+ \Lambda_0 F_0.
\label{eq:def_Hamiltonian_OC}
\end{equation}
\end{definition}
For $\Lambda_0<0$, we find the previous definition of $H_P$ by noting that $\Lambda_0$ can be normalized to -1 without loss of generality. The Pontraygin Hamiltonian does not depend on $F_0$ when $\Lambda_0=0$, which leads to the family of abnormal solutions. We stress that abnormal solutions also occur in finite-dimensional optimization problems as a special case of Lagrange multipliers. This point is discussed in~\ref{app_lagrange} which shows that abnormal extremals are not restricted to optimal control problems. An example of abnormal control is described in Example~\ref{example:ex8}.
\begin{example}{}{ex5}
The Pontryagin Hamiltonian for the optimal control Lagrangian $\Mc L =\frac{1}{2}\alpha f^2 + \Lambda_x( \dot x - p/m) + \Lambda_p(\dot p -f)$ can be written as
\begin{equation}
H_P = \Lambda_x\frac{p}{m} + \Lambda_p f +\frac{\Lambda_0}{2}\alpha f^2,
\end{equation}
where $\Lambda_0$ is a negative constant. In the abnormal case, the Pontryagin Hamiltonian is given by
$$
H_P=\Lambda_x\frac{p}{m} + \Lambda_p f.
$$
\end{example}
Optimal trajectories are given by Hamilton's equations, once again highlighting the link between classical mechanics and optimal control. They can be written as:
\begin{equation}
\begin{split}
& \dot \Lambda_a = -\frac{\partial H_P}{\partial X_a},  \\
& \dot X_a = \frac{\partial H_P}{\partial \Lambda_a},  \\
& \frac{\partial H_P}{\partial u_a} =0,
\end{split}
\label{eq:weak_PMP}
\end{equation}
A straightforward calculation reveals that these equations are equivalent to dynamical equations of Th.~\ref{thm:Extremals of the action}. Equations~\eqref{eq:weak_PMP} that can be used with unconstrained control ($U$ is an open set) correspond to \textit{"the weak Pontryagin principle"}. The theory can be extended to consider the general case. This leads to the \textit{Pontryagin Maximum Principle} (PMP) which can be stated as follows~\cite{bonnard_optimal_2012,PRXQuantumsugny,pontryaginbook,bryson1975applied,
kirk2004optimal,trelat2012optimal,liberzon-book}.
\begin{theorem}{Pontryagin Maximum Principle}{Pontryagin Maximum Principle}
We consider the dynamical system defined by
$$
\dot X(t) = F(X(t),u(t)) ,
$$
where $F$ is a smooth vector function and  $u:[0,t_f]\rightarrow U \subset \setR^m$ the control. The goal of the control protocol is to steer the system from $X_0$ to $X_f$ at time $t_f$ which is fixed or free. The optimal control problem $u^\star$ is defined from the cost functional $\mathcal{C}$ to minimize
$$
\mathcal{C}= G(X(t_f)) + \int_0^{t_f} F_0 (X(t'),u(t'))dt',
$$
where $F_0$ and $G$ are two smooth functions. The Pontryagin Hamiltonian $H_P$ is defined as
$$
H_P(X,\Lambda,\Lambda_0,u) = \Lambda\cdot F(X,u)+ \Lambda_0F_0(X,u),
$$
where $\Lambda(t) \in \setR^n$ is the adjoint state, and the abnormal multiplier $\Lambda_0\leq 0$ a constant. The pair $(X,u^\star)$ is optimal if there exists a non zero continuous pair $(\Lambda,\Lambda_0)$ such that the trajectories of the extended system are given by Hamilton's equations $\dot \Lambda = -\partial_{X} H_P$ and $\dot X = \partial_{\Lambda}H_P$, and the maximization condition can be written almost everywhere on $[0,t_f]$  as:
\begin{equation}\label{eqmaxPMP}
H_P (X,\Lambda,\Lambda_0,u^\star) = \max_{u \in U} H_P (X,\Lambda,\Lambda_0,u).
\end{equation}
The state and adjoint state satisfy respectively the initial and final conditions
$$
X(0)=X_0,~\Lambda(t_f)=\Lambda_0\frac{\partial G(X(t_f)}{\partial X(t_f)}.
$$
\end{theorem}
Some mathematical details and a geometric interpretation of the PMP can be found in~\ref{appendixPMP}. We observe that the first-order condition $\frac{\partial H_P}{\partial u}=0$ is replaced by a stronger maximization condition~\eqref{eqmaxPMP} of the Pontryagin Hamiltonian along the optimal trajectory. This modification is crucial to treat the case of a closed set $U$ such as $U=[u_{\textrm{min}},u_{\textrm{max}}]$ since the maximum of $H_P$ can be defined on an open or a closed set. A schematic description of this maximization is given in Fig.~\ref{fignew}.

Some additional comments and extensions can be made on Theorem~\ref{thm:Pontryagin Maximum Principle}. The Hamiltonian $H_P$ is generally called a \textbf{\textit{pseudo-Hamiltonian}} because it depends on the control $u$. It can be shown that the Hamiltonian $H_P$ is constant in time and it is constantly zero if $t_f$ is free. A brief description of this condition is given in~\ref{appendixPMP}. A solution $X$ of this optimal control problem is called an extremal trajectory and is candidate to be optimal. This means that the PMP is only a necessary condition for optimality and several extremal trajectories may exist. Additional work is required to select the optimal solution among such trajectories. As in a finite-dimensional optimization problem (see~\ref{app_lagrange} for details), two different sets of trajectories can be extremals, namely the \textbf{\textit{normal}} and the \textbf{\textit{abnormal}}, respectively for $\Lambda_0< 0$ and $\Lambda_0=0$. The pair $(\Lambda,\Lambda_0)$ is defined up to a constant factor, but cannot simultaneously be equal to 0. This degree of freedom comes from the fact that $H_P$ and the adjoint states are abstract quantities which have no physical meaning. Multiplying $(\Lambda,\Lambda_0)$ by a constant factor amounts to multiplying $H_P$ by the same factor without changing the optimal control problem. This allows in the normal case to normalize the constant $\Lambda_0$ to a specific value such as $-1/2$ or $-1$. The target state here is only a point, but a set of points or a subset of $\mathbb{R}^n$ can be chosen as target. The final condition for the adjoint state must then be adapted. It is also possible to consider the trajectories that reach exactly the state $X_f$ at time $t_f$. In this case, the final conditions are satisfied by the state as $X(t_f)=X_f$ and not by the final adjoint state $\Lambda(t_f)$. In a general situation, the optimal solutions of the Hamiltonian's equations of the PMP are defined from $2n$ conditions given at initial or final times on the state or the adjoint state.

Based on Th.~\ref{thm:Pontryagin Maximum Principle}, a systematic way to solve the optimal equations given by the PMP can be formulated. The different steps must be followed for both normal and abnormal extremals.
The first objective is to use the maximization condition to express the control in terms of the state and adjoint state as $u=v(X,\Lambda)$. If this is possible, the control is said to be \textbf{\textit{regular}}, otherwise it is \textbf{\textit{singular}}. Note that the two situations can be mixed for a given trajectory in the sense that the control can be regular at some times and singular at others. Different quantum optimal control problems in which singular extremals are optimal have been found in the literature~\cite{PRXQuantumsugny,lapert2010singular,Lapert_exploring_2012}. In the regular situation, the second step is to insert the expression of the control into the Hamiltonian's equation. If the function $v$ is smooth, we get a well-defined Hamiltonian system as in classical mechanics except that the desired trajectory is defined by two-side boundary conditions on $X(0)$ and $\Lambda(t_f)$. A straightforward way to solve this problem is to use a shooting technique which consists of finding the initial value $\Lambda(0)$ such that the final condition on the adjoint state at time $t_f$ is satisfied. This approach faces two main difficulties due to the possible complexity of the system dynamics. The first is related to the non uniqueness of the solution and the second to the potentially strong sensitivity to initial conditions in the case of chaotic dynamics. The latter is found especially in high-dimensional systems. This observation justifies the use of shooting techniques only for simple control problems of low dimension. Other numerical optimization methods have been developed to solve more complicated issues. Such optimization algorithms are described in Sec.~\ref{sec:numerical_methods}.

The following examples illustrate different situations that can be encountered with the PMP.

\begin{figure}[htbp]
\begin{center}
\includegraphics[width=7.5cm]{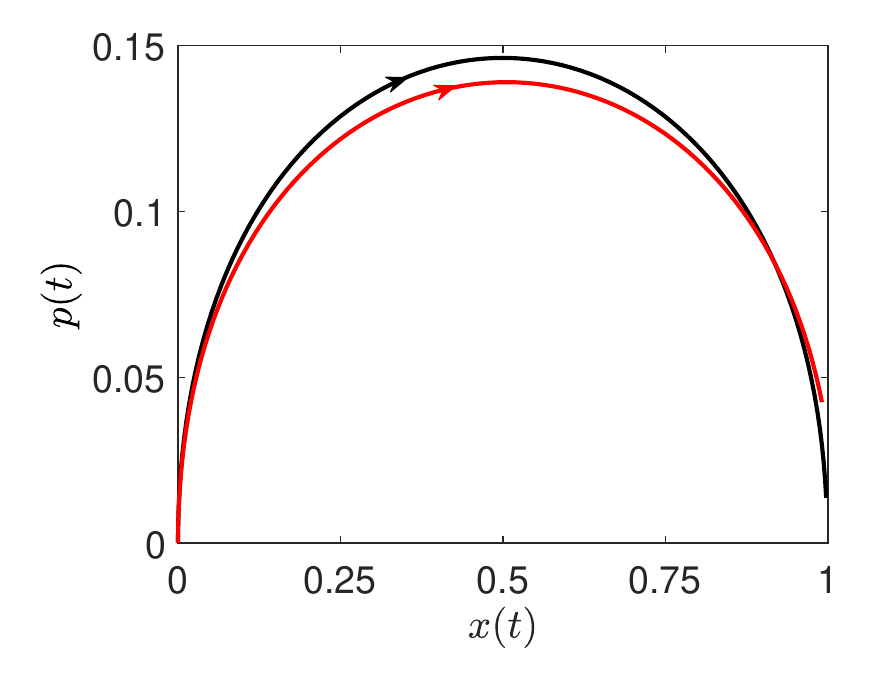}
\includegraphics[width=7.5cm]{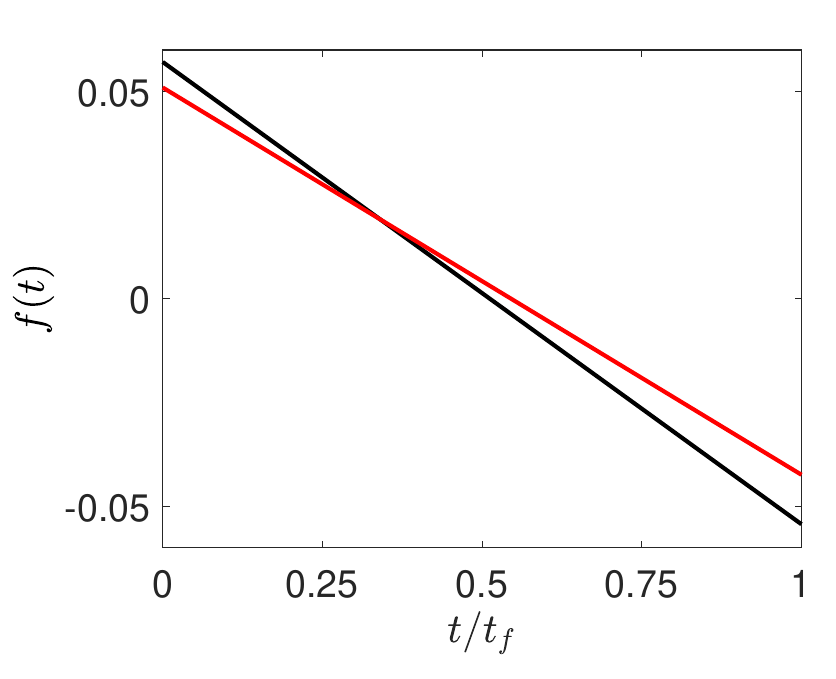}
\end{center}
\caption{Plot of the optimal trajectories in the space $(x,p)$ and of the corresponding control $f(t)$ for $\alpha=0.25$ (black) and $\alpha=1$ (red). Numerical parameters are set to $t_f=10$ and $m=1$.}
\label{fig3}
\end{figure}

\begin{example}{}{ex6}

We consider the control of a point particle in the energy minimum case from state $(0,0)$ to $(1,0)$ in a fixed time $t_f$. The cost functional to minimize is given by
$$
\mathcal{C}=\frac{1}{2}(x(t_f)-1)^2+\frac{1}{2}p(t_f)^2+\frac{\alpha}{2}\int_0^{t_f}f^2(t)dt.
$$
The Pontryagin Hamiltonian can be expressed as
$$
H_P=\Lambda_x\frac{p}{m}+\Lambda_p f-\frac{\alpha f^2}{2},
$$
where $\Lambda_0$ has been set to -1 (the abnormal case plays no role in this problem). Hamilton's equations can then be written as $\dot x = p/m$, $\dot p= f$, $\dot \Lambda_x =0$, $\dot \Lambda_p = - \Lambda_x/m$, and the maximization condition reads $f=\Lambda_p/\alpha$. The extremal trajectories are therefore regular. Note that we find the same expressions as in Example~\ref{example:ex4}.

Plugging the expression of $f$ into $H_P$, a true Hamiltonian $H$ is obtained
$$
H=\Lambda_x\frac{p}{m}+\frac{\Lambda_p^2}{2\alpha},
$$
and the optimal trajectories are given by the Hamiltonian's equations derived from $H$. Since $\Lambda_x$ is a constant of motion, it is straightforward to show that
$$
f(t)=\frac{\Lambda_p(t)}{\alpha}=-\frac{\Lambda_x}{\alpha m}t+\frac{\Lambda_p(0)}{\alpha},
$$
and the system dynamics read
\begin{equation}
\begin{split}
p(t)&=-\frac{\Lambda_x}{2\alpha m}t^2 +  \frac{\Lambda_p(0)}{\alpha} t, \\
x(t)&=-\frac{\Lambda_x}{6\alpha m^2}t^3+\frac{\Lambda_p(0)}{2\alpha m}t^2,
\end{split}
\end{equation}
where we use the initial state $(x(0),p(0))=(0,0)$. The optimal trajectory is derived from the final condition $\Lambda_x(t_f)=1-x(t_f)$ and $\Lambda_p(t_f)=-p(t_f)$. A linear system of equations then allows to find $\Lambda_x$ and $\Lambda_p(0)$
\begin{equation}
\begin{split}
& \left(1-\frac{t_f^3}{6\alpha m^2}\right)\Lambda_x+\frac{t_f^2}{2\alpha m}\Lambda_p(0)=1, \\
& \left(-\frac{t_f^2}{2\alpha m}-\frac{t_f}{m}\right)\Lambda_x+\left(\frac{t_f}{\alpha}+1\right)\Lambda_p(0)=0.
\end{split}
\end{equation}
Optimal trajectories are represented in Fig.~\ref{fig3} for different values of the parameters.
\end{example}

\begin{figure}[htbp]
\begin{center}
\includegraphics[width=7.5cm]{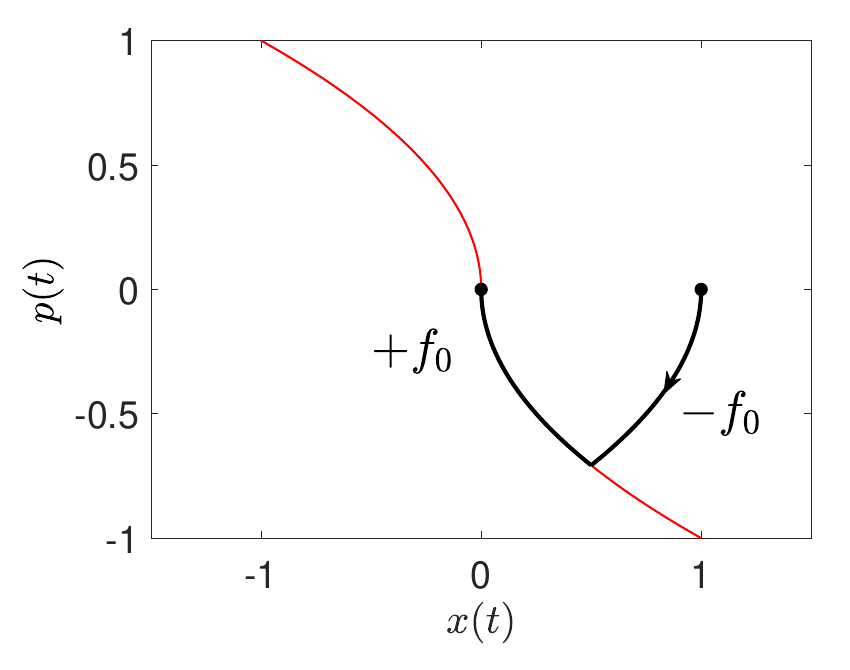}
\end{center}
\caption{
Plot of the time-optimal trajectory to go from the state $(1,0)$ to $(0,0)$ represented as black dots. The optimal control is bang-bang with first a control equal to $-f_0$ and then to $+f_0$. The red curve depicts the set $\mathcal{P}$.
}\label{fig4}
\end{figure}

\begin{example}{}{ex7}
We consider the same control problem but in minimum time. To simplify the description of the optimal solution, we exchange the roles of the initial and target states which in this case are respectively $(1,0)$ and $(0,0)$. There is no constraint on the control energy, only on its amplitude, $f(t)\in [-f_0,f_0]$ where $f_0$ is the maximum force allowed. For time-optimal problems, it is often preferable to reach the target exactly, the cost functional is defined from a running cost as $\mathcal{C}=\int_0^{t_f}dt=t_f$. Again, the abnormal extremals are not relevant and the parameter $\Lambda_0$ is normalized to -1. The Pontryagin Hamiltonian can be written as
$$
H_P=\Lambda_x\frac{p}{m}+\Lambda_pf-1,
$$
and Hamilton's equations are the same as in the preceding example. Since the final time is free, the Pontryagin Hamiltonian is zero at any time $t$. We deduce that $\Lambda_x\frac{p}{m}+\Lambda_pf=1$ is a constant along the optimal trajectory. In the open interval, the maximization condition $\frac{\partial H_P}{\partial f}=0$ leads to $\Lambda_p(t)=0$ on a non-zero time interval and thus to $\dot{\Lambda}_p(t)=0=-\Lambda_x/m$. We deduce that the condition $\Lambda_x\frac{p}{m}+\Lambda_pf=1$ cannot be satisfied and that the optimal control takes values on the boundary of the interval, i.e. $f(t)=\pm f_0$. The sign of $f$ has to be chosen to maximize $H_P$ at any time $t$. The only Hamiltonian term depending on $f$ being $\Lambda_p f$, we deduce that the optimal control can be expressed as $f(t)=\textrm{sign}[\Lambda_p]f_0$. The latter is a square wave  with switchings at times for which $\Lambda_p(t)=0$. Such a solution is referred to as \emph{bang-bang} in the control literature. The Hamiltonian's equations can be directly integrated which leads for a trajectory in the interval $[t_i,t]$ to
\begin{equation}
\begin{split}
x(t)&=f\frac{(t-t_i)^2}{2m}+\frac{p(t_i)}{m}(t-t_i)+x(t_i), \\
p(t)&=f(t-t_i)+p(t_i),\nonumber
\end{split}
\end{equation}
where $f$ is a constant equal to $+f_0$ or $-f_0$. Note that $p$ is an increasing (resp. decreasing) function of time when $f=+f_0$ (resp. $f=-f_0$).
We also deduce that the optimal trajectories in the space $(x,p)$ are parabolas. For the adjoint state, we arrive at
\begin{equation}
\begin{split}
\Lambda_x(t)&=\Lambda_x, \\
\Lambda_p(t)&=-\frac{\Lambda_x}{m}(t-t_i)+\Lambda_p(t_i),\nonumber
\end{split}
\end{equation}
where $\Lambda_x$ is a constant. $\Lambda_p$ is a linear function of time with at most one zero. The optimal control therefore has at most one switching and the candidates to optimality are of the form:\\
- $f(t)=+f_0$ for $t\in[0,t_f]$,\\
- $f(t)=-f_0$ for $t\in[0,t_f]$,\\
- $f(t)=+f_0$ for $t\in[0,t_s[$ and $f(t)=-f_0$  for $t\in ]t_s,t_f]$,\\
- $f(t)=-f_0$ for $t\in[0,t_s[$ and $f(t)=+f_0$  for $t\in ]t_s,t_f]$,\\
where $t_s$ is the switching time to be determined. The two bang solutions, i.e. the trajectories without switching, correspond to the arcs of parabolas passing through the target state $(0,0)$ as shown in Fig.~\ref{fig4}. We denote this set of points by $\mathcal{P}$. If the initial state belongs to $\mathcal{P}$ then the time-optimal solution is a constant control equal to $+f_0$ when $x(0)>0$ and $-f_0$ otherwise. In the general case, two arcs must be concatenated. This is the situation of the example for which $(x(0),p(0))=(1,0)$. A first bang with $f=-f_0$ is used to reach $\mathcal{P}$ and then $f=+f_0$ to attain the target state, as depicted in Fig.~\ref{fig4}.
\end{example}

\begin{example}{}{ex8}
We now give an example of abnormal control. We follow the example presented in~\cite{Arutyunov}. The control problem is almost the same as in Example~7 where the minimum time $t^\star$ to reach the origin is derived. The unique optimal control switches at time $t_s$ from $-f_0$ to $+f_0$. The time $t_f$ is set to $t^\star$ and we consider a running cost $\mathcal{C}$ to minimize
$$
\mathcal{C}=\int_0^{t^\star}f(t)\sqrt{|t-t_s|}dt.
$$
Since the control time is equal to the minimum time, the constraints on the control being the same, it is clear that the only solution to the optimal control problem is the one previously derived and that this process does not depend on the chosen cost functional.

Using the PMP in this case, we arrive at the following Pontryagin Hamiltonian
$$
H_P=\Lambda_x\frac{p}{m}+\Lambda_pf(t)+\Lambda_0f(t)\sqrt{|t-t_s|}.
$$
The Hamiltonian equations are the same as before, but the maximization condition is modified due to the new cost functional. Note that $\Lambda_p(t)$ is a linear function of time. The control $f$ is chosen to maximize the expression $f(t)(\Lambda_p+\Lambda_0\sqrt{|t-t_s|})$. Since the optimal solution has a switching at $t=t_s$, we deduce that $\Lambda_p(t_s)=0$. If $\Lambda_0\neq 0$ then there is a small interval around $t=t_s$ such that $\Lambda_p(t)+\Lambda_0\sqrt{|t-t_s|})\neq 0$ because a square root function grows faster than a linear function at the origin. There is a contradiction because $t_s$ is a switching time. We conclude that $\Lambda_0=0$ and that the optimal solution is an abnormal extremal of the control problem.
\end{example}

\section{Quantum Optimal Control}
\label{sec:QOC}

\subsection{From the Schr\"odinger equation to the Pontryagin Hamiltonian}
\label{sec:from_QM_to_PMP}

We are now interested in applying OCT to a quantum system. Let $\Mc H$ be the Hilbert space of the system, and assume that the quantum Hamiltonian operator of the system can be written as
\begin{equation}
\hat H(t) = \hat H_0 + \sum_{k = 1}^m u_k(t) \hat H_k ,
\end{equation}
with real control parameters $u_k$. The quantum operators are denoted by a hat in the rest of the paper. We restrict the discussion to bilinear systems whose dynamics are linear with respect to the state and the control. This kind of model can be applied to many experimental situations with a good accuracy. However, optimal control can also be applied to other configurations, when the control enters non-linearly in the Hamiltonian~\cite{lapert2008,ohtsuki2008,rabitz2011b} or in non-linear systems~\cite{zhang2011time,chen2016,dorier2017,guerin2020,zhu2023}. The state of the system at time $t$ is described by the vector $|\psi(t)\rangle\in\mathcal{H}$ whose trajectory is the solution of the time-dependent Schr\"odinger equation (in units where $\hbar=1$)
\begin{equation}
\frac{d \ket{\psi}}{dt} = -\ii \hat H(t) \ket{\psi},
\label{eq:schrodinger_eq}
\end{equation}
with a fixed initial state $|\psi_0\rangle$. The norm of $|\psi\rangle$ is a constant of motion equal to one at any time $t$, $\langle\psi|\psi\rangle=1$. Equation~\eqref{eq:schrodinger_eq} is a complex-valued first-order differential equation, and a method must be found to define real-valued optimal control quantities. An idea is to define the adjoint state as an element of the Hilbert space, $\ket{\chi}\in \Mc H$. Then, the Lagrangian can be defined as:
\begin{equation}
\Mc L = F_0(\ket{\psi},u,t) + \Re \left( \braket{\chi}{\dot \psi} +\ii \bra{\chi} \hat H(t) \ket{\psi}\right),
\end{equation}
where $\Re(\cdot)$ and $\Im(\cdot)$ denote respectively the real and imaginary parts of a complex number. The conjugate variables $\Re(\langle\chi|)$ and $\Im(\langle\chi|)$ of respectively $\Re(|\psi\rangle)$ and $\Im(|\psi\rangle)$ satisfy the Euler-Lagrange equations
$$
\Re(\langle\chi|)=\frac{\partial \mathcal{L}}{\partial \Re(|\dot{\psi}\rangle)},~
\Im(\langle\chi|)=\frac{\partial \mathcal{L}}{\partial \Im(|\dot{\psi}\rangle)},
$$
in which we use the convention that the derivative with respect to a ket gives a bra, and vice versa. We emphasize here that $\ket{\chi}$ is not necessary normalized to one. Using the Euler Lagrange equations for the variables $\Re( \langle\chi|)$ and $\Im( \langle\chi|)$, i.e. here
$$
\frac{\partial \mathcal{L}}{\partial \Re(\langle\chi |)}=0=\frac{\partial \mathcal{L}}{\partial \Im(\langle\chi |)},
$$
we arrive at the following equations of motion
\begin{equation}
\begin{split}
\frac{d \Re (\ket{\psi})}{dt} &= \Re \left( - \ii \hat H(t) \ket{\psi} \right), \\
\frac{d \Im (\ket{\psi})}{dt} &= \Im \left( - \ii \hat H(t) \ket{\psi} \right),\nonumber
\end{split}
\end{equation}
which correspond to the Schr\"odinger equation~\eqref{eq:schrodinger_eq}.
An optimal control problem for a quantum system can then be defined using the action:
\begin{equation}
S = G(\ket{\psi(t_f)}) + \int_0^{t_f}dt \left( F_0 (\ket{\psi},u,t) + \Re \left( \braket{\chi}{\dot \psi} +\ii \bra{\chi} \hat H(t) \ket{\psi}\right)\right).
\label{eq:action_QOC_schrodinger}
\end{equation}
The Pontryagin Hamiltonian can be expressed as
$$
H_P=\Re\left(\langle\chi|\dot\psi\rangle\right)+\chi_0 F_0(\ket{\psi},u,t),
$$
which can be transformed into
\begin{equation}
H_P = \Im \left(\bra{\chi} \hat H(t) \ket{\psi}\right) + \chi_0 F_0(\ket{\psi},u,t),
\label{eq:Hamilotnian_QOC_schrodinger}
\end{equation}
where $\chi_0$ is the abnormal multiplier, $\chi_0\leq 0$. The Hamilton equations can be written as
\begin{equation}\label{eqcomplexadj}
\begin{split}
|\dot{\psi}\rangle &=2\frac{\partial H_P}{\partial \langle \chi|}=-\ii \hat{H}|\psi\rangle ,\\
\langle\dot{\chi}| &=-2\frac{\partial H_P}{\partial |\psi\rangle}=\ii \langle\chi|\hat{H}-2 \chi_0\frac{\partial F_0}{\partial |\psi\rangle},\\
\end{split}
\end{equation}
where the factor 2 comes from the definition of the derivative with respect to $|\psi\rangle$ as $\partial/\partial |\psi\rangle=\frac{1}{2}(\partial/\partial \Re(|\psi\rangle)-\ii \partial/\partial \Im(|\psi\rangle)$. We observe that the dynamics of $\ket{\chi}$ are governed by the Schr\"odinger equation plus an additional term depending on the running cost. It reduces to the Schr\"odinger equation if $F_0$ does not depend on $|\psi\rangle$. In this case, since the pair $(|\chi\rangle,\chi_0)$ is defined up to a multiplicative constant, it is always possible to assume that $\langle\chi|\chi\rangle(t)=1$ and to interpret $|\chi\rangle$ as an abstract wave function. In addition, the adjoint state satisfies a boundary condition at final time
\begin{equation}\label{eqcomplexfinaladj}
\langle\chi(t_f)| =2\chi_0 \frac{\partial G}{\partial |\psi(t_f)\rangle}.
\end{equation}
When there is no constraint on the control, i.e. $U=\mathbb{R}^m$, the condition $\frac{\partial H_P}{\partial u_k}=0$ of the PMP yields for $k=1,\cdots,m$:
\begin{equation}\label{eqcomplexmax}
\frac{\partial H_P}{\partial u_k}=\Im \left(\bra{\chi} \hat H_k(t) \ket{\psi}\right) + \chi_0 \frac{\partial F_0}{\partial u_k}(\ket{\psi},u,t)
=0.
\end{equation}
\begin{example}{}{ex9}
\label{ex9}
We consider several $G$ and $F_0$ functions that can be used in quantum control. The role of $G$ is often to measure the distance of the final state $|\psi(t_f)\rangle$ to the target state $|\psi_f\rangle$. Standard terminal costs are
$$
G_1(|\psi(t_f)\rangle)=1-|\langle\psi_f|\psi(t_f)\rangle|^2,~G_2(|\psi(t_f)\rangle=1-\Re(\langle\psi_f|\psi(t_f)\rangle).
$$
The function $G_2$ is related to the minimization of the square modulus of $|\psi_f\rangle-|\psi(t_f)\rangle$ since
$$
|| |\psi(t_f)\rangle-|\psi_f\rangle ||^2=2(1-\Re(\langle\psi(t_f)|\psi_f\rangle).
$$
In the case of the cost $G_1$, the target state is reached up to a global phase, whereas this phase is fixed to 0 in the second case. Different choices of the running cost $F_0$ are possible to penalize either the control parameter $u$ or the trajectory followed by the system. An example is given by
$$
F_0(|\psi(t)\rangle,u(t))=|\langle\psi_1|\psi(t)\rangle|^2+\frac{u^2}{2},
$$
where the goal is to minimize both the energy of the pulse and the projection onto a forbidden state $|\psi_1\rangle$. Using Eq.~\eqref{eqcomplexadj}, we deduce that the adjoint state is the solution of the differential equation
$$
|\dot{\chi}\rangle =-\ii \hat{H}|\chi\rangle-2\chi_0 \langle \psi_1|\psi(t)\rangle |\psi_1\rangle,
$$
with the final condition given respectively by $|\chi(t_f)\rangle = -2\chi_0 \langle\psi_f|\psi(t_f)\rangle |\psi_f\rangle$ and $|\chi(t_f)\rangle = -\chi_0|\psi_f\rangle$ for $G_1$ and $G_2$. The maximization condition~\eqref{eqcomplexmax} can be written as
$$
\frac{\partial H_P}{\partial u_k}=\Im (\bra{\chi} \hat H_k(t) \ket{\psi}) + \chi_0 u_k .
$$
\end{example}

This formulation of the optimal control problem is of practical interest because it remains close to usual quantum mechanical equations. Nonetheless other procedures exist to transform complex-valued functions into real-valued ones. Alternatives are preferred when the system is described in terms of a density matrix or an evolution operator. As an illustrative example, we discuss here the case of a two-level quantum system whose state is given by the Bloch vector.
For a pure state, the density matrix $\hat\rho$ of a two-level quantum system can be written as
\begin{equation}
\hat \rho = \ket{\psi}\bra{\psi} = \frac{1}{2}\left( \hat \Id_2 + x \sigx + y \sigy + z \sigz \right)=\frac{1}{2}\left( \begin{array}{cc}
1+z & x- \ii y \\
x+ \ii y & 1-z
\end{array} \right),
\label{eq:spin_density_matrix}
\end{equation}
where $x,y,z$ are real numbers such that $x^2 + y^2 + z^2 = 1$ and $\sigx, \sigy,\sigz$ are Pauli matrices. In the case of a mixed state, a similar result can be established, but with $x^2 + y^2 + z^2 \leq 1$. The vector $q=(x,y,z)$ is called the Bloch vector. The interesting point is that $x= \langle \sigx \rangle = \textrm{Tr} (\sigx \hat \rho)$, and similarly for $y$ and $z$. It is then straightforward to derive the equations of motion for the three real variables $x$, $y$, $z$ by computing $\frac{d}{dt}\langle\sigx \rangle$, $\tfrac{d}{dt}\langle \sigy \rangle$, and $\tfrac{d}{dt}\langle\sigz \rangle$. Assume, for instance, that the dynamics of $|\psi\rangle$ are governed by the following Schr\"odinger equation
$$
i|\dot{\psi}\rangle = \left[\Delta\frac{\sigz}{2}+u_x\frac{\sigx}{2}+u_y\frac{\sigy}{2}\right]|\psi\rangle,
$$
where $\Delta$ is a fixed parameter. It can be shown that
\begin{equation}\label{eqbloch}
\begin{split}
\dot{x}&=-\Delta y+u_y z, \\
\dot{y}&=\Delta x-u_x z, \\
\dot{z}&=u_x y-u_y x.
\end{split}
\end{equation}
The PMP can then be applied to this real differential system. The Pontryagin Hamiltonian can be written as
$$
H_P=\Delta(p_y x-p_x y)+u_x(p_z y-p_y z)+u_y(p_xz-p_zx),
$$
where $(p_x,p_y,p_z)$ are the real coordinates of the adjoint state. This approach is particularly well adapted to study the optimal control of open quantum systems~\cite{bonnard2009,lapert2010singular,Lapert_exploring_2012}, whose dynamics are governed by, e.g., the Lindblad-Kossakovski equation~\cite{breuer2002theory,gardiner2004quantum}.

\subsection{Time-optimal control of a two-level quantum system}
\label{sec:OCT_spin}

In this section, we are interested in the design of time-optimal controls for state-to-state transfer of a two-level quantum system which is one of the most relevant and simple systems in quantum technologies~\cite{raimond2006exploring,
levitt2013spin,cheng2023noisy}. We also discuss the results from the point of view of the QSL~\cite{deffner2017quantum,frey2016,hegerfeldt2013driving}. In order to keep the discussion as simple as possible, we consider situations in which any interaction with the environment can be neglected so that the system is described by a pure state. We can leave aside the density matrix formalism.

We consider the time-optimal control of a two-level quantum system with respectively one and two independent control parameters. This issue has been considered in detail in references~\cite{PRXQuantumsugny,d2001optimal,boscain2002,PhysRevA.63.032308,boscain2006time,PhysRevA.85.012317,sugny08,garon2013time,dionis2023time,evangelakos2023}. When the approximation of a closed quantum system is not satisfied, the optimal control problem has to be extended to account for interactions with the environment~\cite{bonnard_optimal_2012,lapert2010singular,zhang2011time,rebentrost2009optimal,roloff2009optimal,lapert2011towards,PhysRevA.82.063418,
floether2012robust,mukherjee2013,lapert2013,riaz2019optimal,basilewitsch2019reservoir,fischer2019time,Ansel_2022}.

\subsubsection{The case of two controls.}
\label{sec:Time optimal control with two inputs and without drift}

As a first example, we consider the time-optimal control of a two-level quantum system whose Hamiltonian can be written as:
\begin{equation}
\label{eq:Ham_2_inputs_no_drift}
\hat H(t) = \frac{ u_z (t)}{2}\sigz + \frac{ u_x(t)}{2} \sigx.
\end{equation}
The control $u(t) =(u_x(t),u_z(t))$ is assimilated to a time-dependent electromagnetic field, with components in two different directions $x$ and $z$. In experiments, the strength of the field is limited such that $\Vert u(t) \Vert  \in [0,u_0]$, where $u_0$ is the maximum available amplitude. The Hilbert space is spanned by the basis $\{\ket{\uparrow},\ket{\downarrow}\}$ with $\ket{\uparrow}=(1,0)$ and $\ket{\downarrow}=(0,1)$.

We focus on the control process that transforms the state $\ket{\uparrow}$ into $\ket{\psi_{f}}=e^{\ii \theta} \ket{\downarrow}$, where $\theta$ is a phase factor which is not relevant for this quantum state. The cost functional to minimize is given by
\begin{equation}
\mathcal{C} = \int_0^{t_f}dt.
\end{equation}
We are interested in control protocols reaching exactly the set of target states, so there is no terminal cost in this case. Moreover, control time is kept free in a time-optimal process. The Pontryagin's Hamiltonian reads
\begin{equation}
\begin{split}
H_P & = \Im \left(\bra{\chi} \hat H(t) \ket{\psi}\right)+\chi_0 \\
& = \frac{ u_z (t)}{2} \Im \left(\bra{\chi} \sigz \ket{\psi}\right) + \frac{ u_x(t)}{2}  \Im \left(\bra{\chi} \sigx \ket{\psi}\right)+\chi_0 \\
& =  u_z (t) H_z(t) +   u_x(t) H_x(t)+\chi_0,
\end{split}
\end{equation}
with
\begin{equation}
\label{eq:def_Hx}
H_{x,y,z}(t) = \frac{1}{2}\Im\left(\bra{\chi(t)} \hat{\sigma}_{x,y,z} \ket{\psi(t)}\right).
\end{equation}
The next step is to compute Hamilton's equations which can be expressed as
\begin{align}
\label{eq:EOM_psi}
& \frac{d \ket{\psi(t)}}{dt} = -\ii \hat H(t) \ket{\psi(t)}, \\
\label{eq:EOM_chi}
& \frac{d \ket{\chi(t)}}{dt} = -\ii \hat H(t)\ket{\chi(t)}.
\end{align}
The time derivatives of the quantities $H_{x,y,z}$  satisfy the following differential system
\begin{equation}\label{eq:dH}
\begin{split}
\dot{H}_x &= - u_z(t) H_y(t), \\
\dot{H}_y &= u_z(t) H_x(t)-  u_x(t) H_z(t), \\
\dot{H}_z &= u_x(t) H_y (t).
\end{split}
\end{equation}
from which it can be shown that $H_x^2+H_y^2+H_z^2$ is a constant of motion.

The maximization condition on $H_P$ leads to both regular and singular trajectories. In the interior of $U$, we have $\partial H_P/\partial u_x=\partial H_P/\partial u_z=0$, which leads to $H_x(t)=H_z(t)=0$ on a non-zero time interval. This corresponds to a singular trajectory. Plugging these conditions into Eq.~\eqref{eq:dH}, we obtain that $H_y(t)$ is constant. The case $H_y(t)=0$ is not relevant since it gives $|\chi(t)\rangle=0$ using \eqref{eq:def_Hx}. When $H_y(t)\neq 0$, we have $u_x(t) = u_z(t) =0$, which is obviously not optimal. Consequently, the optimal control is regular.
The regular trajectory corresponds to the boundary of $U$ for which $u_x^2+u_z^2=u_0^2$. Introducing the angle $\theta$ such that $u_x=u_0\cos\theta$ and $u_z=u_0\sin\theta$, we get $H_P=u_0(\cos\theta H_x+\sin\theta H_z)+\chi_0$. The maximization condition implies that $\frac{\partial H_P}{\partial \theta}=0$, i.e. $\tan\theta =\frac{H_z}{H_x}$ and finally
$$
u_x=u_0\frac{H_x}{\sqrt{H_x^2+H_z^2}},~u_z=u_0\frac{H_z}{\sqrt{H_x^2+H_z^2}}.
$$
The maximum value of $H_P$ is given by $H_P=u_0\sqrt{H_x^2+H_z^2}+\chi_0$ which is a constant of motion. The abnormal multiplier can be set to -1 which leads to
$$
H_x^2+H_z^2=\frac{1}{u_0^2},
$$
using $H_P=0$ for this time-optimal control. The optimal solution can then be described as follows. We emphasize that the control $u_z$ only produces a modification of the phase associated with each state $\ket{\uparrow}$ and $\ket{\downarrow}$ and cannot influence the population transfer which is only due to $u_x$.  We deduce that the control $u_x(t) = \pm u_0$, $u_z(t)=0$ (for all $t$) allows us to reach the state with the maximum speed (if $u_z \neq 0$, we have $u_x \neq u_0$, and thus the velocity of rotation around the direction $x$ is not maximum). When $u_x(t)=+u_0$, it is straightforward to integrate the Schr\"odinger equation as $\ket{\psi(t)} = \cos(u_0 t /2) \ket{\uparrow} + \ii \sin(u_0 t/2)\ket{\downarrow}$. Simultaneously, the adjoint state can be determined to be $\ket{\chi(t)}=-(2\imath/u_0)\hat{\sigma}_x \ket{\psi(t)}$ from the constraints on $H_{x,y,z}$ (notice that $\ket{\chi}$ is not necessarily normalized here). For a given control time $t_f$, we deduce that $|\langle\psi(t_f)|\downarrow\rangle|^2=\sin(u_0 t_f/2)^2$.  The minimum time $t^\star$ to exactly reach the target state is therefore given by
\begin{equation}
t^\star= \frac{\pi}{u_0},
\end{equation}
the optimal trajectory corresponding to a $\pi$- pulse, i.e. a pulse with a time-integrated area equal to $\pi$ for any maximum pulse amplitude $u_0$.
\subsubsection{The case of one control with a drift.}
\label{sec:OCT_spin_time_1_input_and_offset}

As a second example of application, we consider the same control problem as in Sec.~\ref{sec:Time optimal control with two inputs and without drift}, but with the following Hamiltonian:
\begin{equation}\label{eqhamdrift}
\hat H = \frac{\Delta}{2}\sigz + \frac{u(t)}{2} \sigx,
\end{equation}
where $\Delta \in \setR$ is a constant (a drift, also called frequency offset), and $u(t) \in [-u_0,u_0]$ is a one-dimensional control parameter. The analysis is only slightly modified as follows. The Pontryagin Hamiltonian can be written as
\begin{equation}
H_P = \Delta H_z(t) + u(t) H_x(t),
\end{equation}
and the Hamiltonian's equations are given by:
\begin{align}
\label{eq:EOM_psi_v2}
& \frac{d \ket{\psi(t)}}{dt} = -\ii \hat H(t) \ket{\psi(t)}, \nonumber\\
& \frac{d \ket{\chi(t)}}{dt} = -\ii \hat H(t)\ket{\chi(t)}.\nonumber
\end{align}
The maximization condition of the PMP consists in maximizing the only term of $H_P$ depending on $u(t)$, i.e. $u(t)H_x$ with the constraint $-u_0\leq u(t)\leq u_0$. When $H_x(t)\neq 0$, it is straightforward to show that the control satisfies $u(t)=u_0~\sgn [H_x(t)]$, while we cannot conclude if $H_x(t)=0$. We deduce that we have a regular trajectory when $u(t)=\pm u_0$ and singular arcs when $H_x(t) =0$ on a non-zero time interval. We start the study with the description of regular controls.

\paragraph{Regular controls.}

From $u(t) = u_0~\sgn[H_x(t)]$, we observe that the control is a piecewise constant function and a change of sign  occurs when $H_x (t)= 0$ in an isolated point. In control theory, the function $H_x$ is called a switching function.  We obtain a bang-bang control that suddenly jumps from a control of maximum (or minimum) amplitude to its opposite. We can go one step further by calculating the time derivatives of $H_{x,y,z}$, which yields
\begin{equation}\label{eqdH2}
\begin{split}
\dot{H}_x &= -\Delta H_y(t), \\
\dot{H}_y &=\Delta H_x(t)- u(t) H_z(t), \\
\dot{H}_z &= u(t) H_y .
\end{split}
\end{equation}
We observe that Eq.~\eqref{eqdH2} gives a closed-form system of differential equations, similarly to Eq.~\eqref{eq:dH}, that can be integrated on a bang (between two switchings). By denoting $t_i$ the switching time $i$, we have
\begin{equation}
\left( \begin{array}{c}
H_x(t) \\
H_y(t) \\
H_z(t)
\end{array} \right) = \exp \left[ (t-t_i)\left( \begin{array}{ccc}
0 & -\Delta & 0 \\
\Delta & 0 & - u \\
0 & u & 0
\end{array} \right) \right] . \left( \begin{array}{c}
0 \\
H_y(t_i) \\
H_z(t_i)
\end{array} \right),
\end{equation}
with $u=u_0~\sgn[H_x(t)]$ a constant, $t \in [t_i,t_{i+1}]$ and we use the fact that $H_x(t_i)=0$. The explicit calculation of the matrix exponential leads us to
\begin{equation}\label{eqHx}
H_x(t) = -\frac{\Delta}{\Omega^2} \left[\Omega H_y(t_i) \sin(\Omega (t-t_i) )- u H_z(t_i) (1-\cos(\Omega (t-t_i)))  \right],
\end{equation}
with $\Omega = \sqrt{u_0^2 + \Delta^2}$. The duration of the bang, given by $t_{i+1}-t_i$ can be determined from Eq.~\eqref{eqHx}. We arrive at
\begin{equation}
t_{i+1}-t_i =\left\lbrace \begin{array}{ll}
\frac{\pi n}{\Omega} & \text{ if } \frac{u H_z(t_i)}{\Omega H_y(t_i)}=0, \\
\frac{2}{\Omega}\left[ \pi n + \arctan\left( \frac{\Omega H_y(t_i)}{u H_z(t_i)}\right) \right] & \text{ if } \frac{u H_z(t_i)}{\Omega H_y(t_i)} \neq 0,
\end{array} \right.
\label{eq:switching_times_bang_bang}
\end{equation}
where $n$ is an integer chosen such that $t_{i+1} -t_i$ is the smallest positive non zero solution. We also observe that the function $H_{x}$ is $2\pi/\Omega$ periodic. We deduce that the maximum duration between two switchings is $2\pi/\Omega$.
Similar calculations as for $H_x(t)$ show that $H_z(t_{i+1}) = H_z(t_i)$ and $H_y(t_{i+1})=H_y(t_i)$. This means that at switching $i+1$, the system is in the same configuration as at switching $i$. All bangs have the same duration. For an arbitrary regular trajectory, we cannot start or end at a switching. Therefore, such a trajectory has the following structure: a first bang with control amplitude $u = \pm u_0$ and duration $t_1$, followed by a series of bangs of amplitudes $\pm u_0$ and duration $t_b$ and a final bang of duration $t_2$. Note that $t_1$ and $t_2$ are both less than or equal to $t_b$.

\paragraph{Singular controls.}

The control is singular when $H_x(t) = 0$ on a non-zero time interval, which gives $\tfrac{d}{dt}H_x = \tfrac{d^2}{dt^2}H_x = 0$. From Eq.~\eqref{eqdH2}, we deduce that $H_y(t) = 0$ and $-u(t) H_z(t) = 0$. Therefore, the control is always equal to $0$, or the vector $(H_x,H_y,H_z)$ is zero. The latter condition is only satisfied if the adjoint state is always zero, which is a trivial solution.

\paragraph{Calculation of the optimal control.}
The final step is to determine the time-optimal trajectory. First, we argue that singular control (or concatenation of regular and singular controls) cannot be optimal. For a singular control equal to zero, the Hamiltonian operator can be simplified to $\hat H = \tfrac{\omega}{2}\sigz$, which cannot produce a population transfer between the initial and target states, and thus is not optimal. The optimal trajectory is therefore regular. To find the switching times of the optimal procedure, we analytically  compute the evolution operator for an increasing number of switchings, stopping when the corresponding control can steer the system to the target. If there are several solutions with the same number of switchings, we keep the one(s) with the shortest duration.

\begin{figure}[htbp]
\centering
\includegraphics[width=0.5\textwidth]{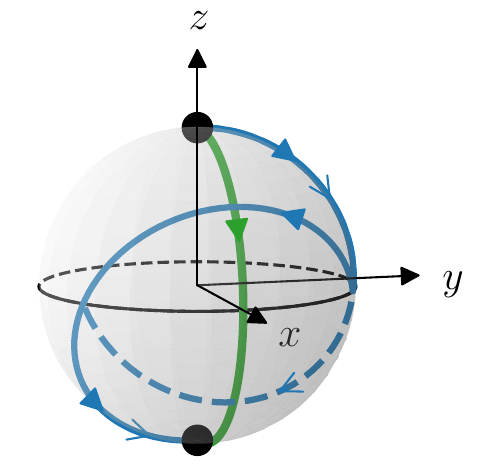}
\caption{Plot of the time-optimal trajectories on the Bloch sphere in the case of a two-level quantum system with one control and a drift. The two optimal trajectories are represented in blue (solid and dashed lines) while the green curve depicts a geodesic going from the north to the south pole. The black solid line represents the equator of the sphere.}
\label{fig:blochsphere}
\end{figure}

We integrate the Schr\"odinger equation on a single bang with a control amplitude $u=\pm u_0$. The evolution operator associated with this bang is given by
\begin{equation}
\hat U_{\text{bang}} = e^{-\ii (\Delta \sigz + u \sigx )t/2}= \cos\left(\frac{\Omega t}{2}\right)\hat \Id - \ii\left( \frac{\Delta}{\Omega}\sigz+  \frac{u}{\Omega}\sigx\right)\sin\left(\frac{\Omega t}{2}\right).
\label{eq:U_bang}
\end{equation}
Since $\Omega = \sqrt{\Delta^2 + u_0^2}$, we observe that the maximum for a population transfer with a single bang is $u^2/\Omega^2<1$ unless $\Delta =0$, hence, there is at least one switching. With two bangs we have:%
\begin{equation}
\hat U_{\text{2-bangs}} = e^{-\ii (\Delta \sigz - u \sigx )t_2/2} e^{-\ii (\Delta \sigz + u \sigx )t_1/2},
\end{equation}
where $t_1$ and $t_2$ are the respective duration of the two bangs. The final distance to the target state can be expressed as
\begin{equation}
\left\vert\bra{\downarrow}\hat U_{\text{2-bangs}} \ket{\uparrow}\right\vert^2 = \frac{u^2}{\Omega^4} \left[ 4 \Delta^2 \sin^2\left(\frac{\Omega t_1}{2}\right) \sin^2\left(\frac{\Omega t_2}{2}\right) + \Omega^2 \sin^2\left(\frac{\Omega (t_1-t_2)}{2}\right)\right].
\end{equation}
We have to maximize this function such that $t_1+t_2$ is minimum. The solution to this problem is given by
\begin{align}
t_1 &= \frac{1}{\Omega}(\pi - \arccos(\Delta^2/u_0^2)),\\
t_2 &= \frac{1}{\Omega}(\pi + \arccos(\Delta^2/u_0^2)),
\end{align}
when $|\Delta|<u_0$, with a total control time given by
\begin{equation}
t_f=t_1+t_2= \frac{2\pi}{\Omega}.
\end{equation}
Note that $t_1$ and $t_2$ can be permuted and the initial sign of the control plays no role. The optimal trajectories are represented in Fig.~\ref{fig:blochsphere} using the Bloch vector representation. There are two equivalent solutions with a switching on the equator of the sphere. To show that this solution is the global optimal solution, it must also be compared with controls with a larger number of switchings. This is the case as described in~\cite{PRXQuantumsugny,boscain2006time}. The number of bangs of the optimal solution is strictly larger than 2 when $|\Delta|>u_0$~\cite{boscain2006time,assemat2010,evangelakos2023}.

\subsubsection{Quantum speed limit.}
\label{sec:QSL}

In this section, we present the results obtained from the point of view of QSL~\cite{deffner2017quantum,hegerfeldt2013driving,PhysRevLett.130.010402,PhysRevA.92.062106}. First, we review the main idea on which QSL is based, and then we discuss its relation to the optimal control formalism.

QSL can be described as a way to generate, as quickly as possible, a state orthogonal to the initial one. Since this concept does not take into account dynamical constraints, it provides only a lower bound on the minimum time to reach the orthogonal state. This statement can be formalized rigorously as follows. The speed in the Hilbert space is naturally given by the Schr\"odinger equation: $\ket{\dot \psi} = -\ii \hat H \ket{\psi}$. Similarly to classical dynamical systems, the velocity vector can be decomposed into parallel and transverse contributions
\begin{align}
\ket{\dot \psi_\parallel} &=\ket\psi\bra\psi \dot{\psi}\rangle= -\ii \bra{\psi}\hat H \ket{\psi}~ \ket{\psi}, \\
\ket{\dot \psi_\perp} &=\ket{\dot \psi} - \ket{\dot \psi_\parallel}.
\end{align}
By construction, we have $\braket{\dot \psi_\perp}{\psi}=0$. The transfer is maximized when the the norm of $\ket{\dot \psi_\parallel}$ is minimized and the one of $\ket{\dot \psi_\perp}$ maximized. The latter can be expressed as
\begin{equation}
|\dot \psi_\perp|^2 =  |\dot \psi|^2 - |\dot \psi_\parallel|^2 = \langle \hat H ^2 \rangle_{\psi} - \langle \hat H \rangle^2_\psi,
\end{equation}
which is the variance of $\hat H$, denoted below $\Delta H^2$. In the case of Hamiltonian~\eqref{eq:Ham_2_inputs_no_drift} for two independent controls, the variance reads
\begin{equation}
\Delta H^2 = \frac{u_0^2}{4} - \frac{1}{4}\left( u_z z + u_x x\right)^2,
\label{eq:Delta E2}
\end{equation}
in which the Bloch coordinates $(x,y,z)$ are used. The maximum of this variance, $u_0^2/4$, is obtained when $u_z z + u_x x =0$. The maximum value of $\Delta H$ gives the QSL of Mandelstam-Tamm~\cite{deffner2017quantum}. Note that other estimates have been proposed in the literature~\cite{MARGOLUS1998188,PhysRevLett.130.010402} with similar properties as the Mandelstam-Tamm bound. The lower bound on the minimum time given by the QSL is
$$
t_{\textrm{QSL}}=\frac{\pi/2}{\sqrt{\Delta H^2}}=\frac{\pi}{u_0}.
$$
We observe that $t^\star=t_{\textrm{QSL}}$. This point can also be verified with the optimal solution derived from the PMP.
For the optimal control $u_x=u_0, u_z=0$, we find from the state evolution that the quantity $x=0$ at all times, therefore the condition $u_z z + u_x x =0$ is satisfied.
The corresponding trajectory on the Bloch sphere is the half of the great circle from the north to the south pole, in the $zOy$ plane. It is also a geodesic connecting the initial and the target states for which the speed of travel is maximum.

However, the relation $t^\star=t_{\textrm{QSL}}$ is not generic, as shown in the case with one control of Sec.~\ref{sec:OCT_spin_time_1_input_and_offset}. For the Hamiltonian~\eqref{eqhamdrift}, the variance is given by
$$
\Delta H^2=\frac{\Omega^2}{4}-\frac{1}{4}(\Delta z+u x)^2.
$$
The maximum of $\Delta H^2$ is $\Omega^2/4$ which leads to $t_{\textrm{QSL}}=\pi/\Omega$, whereas the minimum time is $t^\star=2\pi/\Omega$. In this example, the lower bound cannot be saturated because the geodesic does not correspond to a dynamical trajectory of the system. In other words, the condition $\Delta z+u x=0$ to reach the maximum speed cannot be satisfied by the control $u$. The geodesic corresponding to the trajectory used in the QSL approach is represented for this example in Fig.~\ref{fig:blochsphere}.
\section{Numerical methods}
\label{sec:numerical_methods}

In this section, we present the state-of-the-art numerical methods that can be used to solve quantum optimal control problems. The literature in this field is important~\cite{bonnans2006numerical,qin2008differential,ab2015comprehensive,
feng2021monarch,maros2012computational,davis1987genetic,goldberg2007genetic,pham2012intelligent} and we do not claim to be exhaustive. Our aim is to introduce the main ideas and technical details that can be found in the references mentioned. Numerical examples are described below with pseudo-codes, while Python codes are provided in the supplementary material.

We begin with a general presentation of optimization algorithms found in scientific computing software. These algorithms can be used as black boxes and are generally effective when the number of parameters to be optimized is relatively small. A list of software to solve quantum optimal control problems is given in a second step. Special emphasis is then placed on the description of two algorithms, namely the shooting techniques and the gradient-based algorithms which are two direct applications of the PMP.

\subsection{A general overview of standard optimization algorithms}
\label{sec:std_opt_algo}

\begin{figure}
\begin{center}
\includegraphics[width=0.85\textwidth]{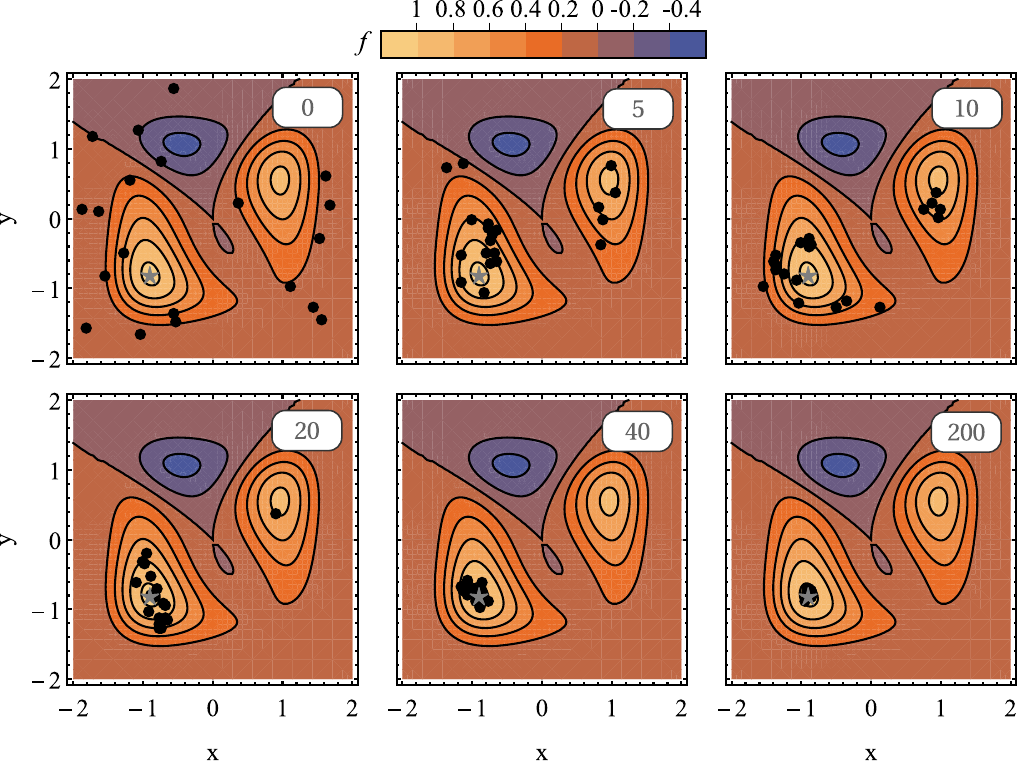}
\end{center}
\caption{Example of particle search of the global minimum of the function $f:(x,y)\mapsto -(x^2 + x y - y^3/2 )e^{-(x^4 + y^4)/2}$ in the domain $x \in [-2,2]$, $y\in[-2,2]$ using the JAYA algorithm~\cite{venkata_rao_jaya:_2016}. The position of the particles is given by black points while the global minimum is located by a gray star. The boxed numbers at the top right of each panel correspond to the number of iterations of the algorithm. Initially, the positions of all particles are generated randomly. They gradually move to regions of the control landscape where the function is likely to be a minimum. In the end, all the particles are located around the global minimum of the function.
}
\label{fig:JAYA}
\end{figure}

Optimization algorithms are generally implemented in most scientific computing softwares~\cite{2020SciPy-NMeth,eaton2013gnu,MATLAB,Mathematica}. An optimization algorithm can be described as a systematic iterative method to find the minimum or  maximum of a function. The function must have a finite number of entries, and thus, continuous controls must be discretized in some way using e.g. piecewise constant functions, Taylor or Fourier expansions, etc. The optimization algorithms are usually presented in the following form
\begin{equation}
\texttt{OptAlgo}(\texttt{Cost function},\texttt{ Constraints},\texttt{ Initial guess}, \texttt{ Options})
\end{equation}
where $\texttt{Cost function}$ is the cost function or figure of merit $\mathcal{C}$ to minimize or maximize, defined in Eq.~\eqref{eq:def_cost_fun}, $\texttt{ Constraints}$ are constraints that limit the domain of definition of the control, the set $U$, $\texttt{Initial guess}$ is an initial guess of the control, used by the algorithm to find a better solution. $\texttt{Options}$ are additional parameters that depend on the algorithm. They correspond either to a choice of a numerical method or to parameters that influence the convergence of the algorithm. We can roughly divide the algorithms into two different families, namely the gradient-free and the gradient-based algorithms~\cite{kochroadmap}. As their name suggests, they are based either on the calculation of the gradient of the figure of merit or on some other procedure that allows the control protocol to be improved at each step of the iterative process. Both approaches have clear advantages depending on the system dimension and on the precision of the control process.
\begin{itemize}
\item[$-$]~\textbf{Gradient-free algorithm.}
This set of methods includes different algorithms. As the number of variants is very large, only a few examples are  mentioned. In the \textit{Simplex method}~\cite{maros2012computational}, the space of control parameters $\setR^n$ is explored with a $n$-simplex whose size decreases recursively around a minimum of the cost function. This approach has been applied in QOC with the algorithm CRAB~\cite{calarco2011}, which can be used to control many body systems~\cite{Calarco2022}. Another optimization procedure is based on \textit{Evolutionary algorithms}~\cite{davis1987genetic,goldberg2007genetic,pham2012intelligent,
venkata_rao_jaya:_2016} (differential evolution, genetic algorithms, simulated annealing, JAYA...). The basic idea of this approach is to iteratively explore the parameter space with $N$ particles, also called "walkers". The method used to update the position of the particles is specific to each method, but usually the particles are moved randomly with a probability law, designed  to favour a direction towards the global minimum of the cost function. An example of a particle path towards the minimum of a function using the JAYA algorithm~\cite{venkata_rao_jaya:_2016} is shown in Fig.~\ref{fig:JAYA}. Evolutionary algorithms are well-known in quantum control and are used experimentally to design a control protocol in a model-free approach~\cite{rabitz1992}.

\item[$-$]~\textbf{Gradient-based algorithm.} The gradient of the function with respect to the control parameters is calculated, and it allows to know in which direction the cost takes a lower value~\cite{bonnans2006numerical,kelley1999iterative}. Based on the gradient, a new control is designed, and the minimum of the cost function is found iteratively. For the optimization of piecewise constant controls, two different procedures have been developed in QOC~\cite{DYNAMO}, namely the Krotov-type methods~\cite{koch2012,krotov1983iterative,palao2002,palao2003,goerz2020} and the GRAPE-type procedures~\cite{khaneja_optimal_2005}. While GRAPE updates the control in all time slices concurrently, the Krotov procedure updates the control sequentially from one time slice to the next. A full comparison of the two methods is given in~\cite{Jager2014,DYNAMO}. Such approaches can be improved by using the second order derivative of the cost function with respect to the control parameters~\cite{bryson1975applied,grape2,tannor2011,sherson2020}. These improvements are not discussed in the main text but are included in some Python codes that can be found in the supplementary material. The GRAPE algorithm and its connection to the PMP is presented in Sec.~\ref{sec:GRAPE_algo}.
\end{itemize}

We conclude this section by some general comments about the application of such algorithms to quantum control. First, we point out that machine learning techniques such as Reinforcement Learning can also be applied to find optimal control protocol~\cite{carleoRMP2019,giannelli2022}. The coefficients of a neural network are optimized by trial and error in a learning step using in general gradient procedures. This network is then used to map each state of the system to the best action (or control) to take to reach the target state. From a more general point of view, Reinforcement Learning can be described as a kind of dynamic programming approach that can be mathematically justified in the continuous limit by the Hamilton-Jacobi-Bellman formulation of optimal control~\cite{bertsekasbook}.

For quantum control problems, simplex methods are generally not the most effective, but they are able to design control procedures in very complex dynamics or when strong control constraints are considered. The accuracy achieved with evolutionary or gradient-based algorithms is generally better. Evolutionary algorithms are interesting when the number of parameters is small ($\lesssim$ 20) or when the cost function cannot be differentiated. Among the different methods, we particularly recommend the simulated annealing method, which is a good compromise between the computation time and its ability to find the global minimum. Another interesting choice is the JAYA algorithm~\cite{venkata_rao_jaya:_2016} which is very simple. Its main advantage lies in the absence of external parameters that influence the convergence (in most algorithms, one or more parameters have to be adjusted manually to obtain a good and fast convergence). Gradient-based algorithms are interesting when the number of parameters to be optimized is very large ($\gtrsim 100$). They can be complemented by second-order formulations to improve convergence to a local extremum~\cite{bryson1975applied,grape2}. Gradient-based algorithms are particularly efficient when the gradient can be computed from an analytic expression. Otherwise, it can be estimated using a finite difference approach, at the expense of computational time. Many factors can influence the choice of algorithm. The way the library is coded and the different options available are also important factors. It is generally advised to test a few different algorithms on a given QOC problem, as a preliminary step. This can also provide a good insight into the control problem since different algorithms may return different solutions (which can be only local extremums of the cost function). Many optimization algorithms are provided in scientific computing softwares. MATLAB~\cite{MATLAB} proposes a global optimization toolbox with e.g. the function \texttt{simulannealbnd} (simulated annealing with constraints) or the function \texttt{fmincon} (gradient algorithm which takes into account the Hessian). In Python/ScyPy~\cite{2020SciPy-NMeth}, different approaches can be found in the routine \texttt{scipy.optimize} (mostly simplex and gradient-based methods). In Mathematica~\cite{Mathematica}, the function \texttt{NMinimize} can be used. It has built-in methods that cover all the families of algorithms listed here above. An advantage of this function is that it is not necessary to provide an initial guess control, the algorithm works simultaneously on several optimization processes in which the initial controls are chosen automatically.

\paragraph{Packages and Software for Quantum Optimal Control.} We provide below a list of freely available packages and softwares designed for quantum optimal control applications. The list is given in alphabetical order.
\begin{itemize}
\item[$-$]~Bocop~\cite{Bocop}: Bocop is an open-source toolbox for solving optimal control problems. It is a general optimization software that can be used for any optimization problem. Its key features are global optimization for both deterministic and stochastic systems, computation of switching and stopping times, parallel execution with OpenMP, and Matlab/Python interfaces.

\item[$-$]~DYNAMO: Dynamic Framework for Quantum Optimal Control~\cite{DYNAMO}. DYNAMO is a Matlab package with integrated GRAPE and Krotov methods.
\item[$-$]~HamPath~\cite{caillau2012differential,HamPath}: HamPath is a general optimization package written in Fortran, and compatible with Matlab, Python, and GnuPlot. It is specifically designed to solve the equations of the PMP (shooting algorithm). Advanced higher-order optimal control methods are implemented, such as the computation of conjugated points~\cite{bonnard_optimal_2012}.

\item[$-$]~Krotov~\cite{goerz2020}: An open-source Python package based on the Krotov algorithm to solve problems of QOC extending from state-to-state transfer to quantum gate implementation.

\item[$-$]~OCTBEC: A Matlab toolbox for optimal quantum control of Bose–Einstein condensates~\cite{HOHENESTER2014194}.

\item[$-$]~QEngine: A C++ library for quantum optimal control of ultracold atoms~\cite{sorensen2019qengine}. QEngine is a C++ library for performing optimal control of Bose–Einstein condensates, Bose–Hubbard type models, and two interacting particles.

\item[$-$]~QOPT: An Experiment-Oriented Software Package for Qubit Simulation and Quantum Optimal Control~\cite{teske2022qopt}. QOPT is a Python library with a scope similar to the optimization package of QuTiP, but it supports computations with stochastic noises.

\item[$-$]~Quandary: An open-source C++ package for high-performance optimal control of open quantum systems~\cite{gunther2021quandary}. Quandary is fully compiled and supports parallel computation of system dynamics. It is therefore possible to optimize quantum controls for a system of large dimension. The optimization algorithm is a homemade gradient algorithm that has some similarities with GRAPE.

\item[$-$]~QuOCS: The quantum optimal control suite~\cite{rossignolo2023quocs}. QuOCS is a Python library that combines several gradient-based optimization algorithms, such as GRAPE and AD-GRAPE. The QuOCS is initially developed and mainly used as part of the gradient-free dCRAB method. The software performs both open- and closed-loop optimizations, and it can be connected to real-time quantum experiments.

\item[$-$]~Qocttols: A Python code that performs optimization calculations on quantum systems using  gradient-based algorithms~\cite{castro}. It is an open and free software that works on closed and open systems. Floquet formalism with periodic perturbations can also be used.

\item[$-$]~QuTiP: Quantum Toolbox in Python~\cite{johansson2012qutip}. QuTiP is an open-source Python package for simulating the dynamics of open quantum systems. It contains an optimization toolbox with integrated GRAPE and CRAB algorithms.

\item[$-$]~Spinach: Spinach is an open-source Matlab package~\cite{spinach} for computations in Nuclear Magnetic Resonance (NMR), Electron Paramagnetic Resonance (EPR), Magnetic Resonance Imaging (MRI), Dynamic Nuclear Polarization (DNP) and Magic Angle Spinning (MAS). Optimal Control, and other topics of Magnetic Resonance spectroscopy are available.

\item[$-$]~SpinDrops~\cite{Spindrops}: SpinDrops is an interactive quantum spin simulator. The software includes several state-of-the-art pulse sequences, but controls can also be created with an editor or with an optimizer based on GRAPE. This is a standalone software with android, iOs, Linux, macOS, and windows supports. An online version is also available.
\item[$-$]~Travolta: An open-source software for parallelized quantum optimal control computations in photo-excited systems using  gradient-based algorithms~\cite{travolta}.

\item[$-$]~WavePacket: A Matlab package for numerical quantum dynamics~\cite{schmidt2018wavepacket}. WavePacket is designed to solve a wide range of quantum dynamical and optimization problems. The optimization is based on the gradient-based algorithm described in~\cite{Zhu1998,Zhu1998b,Ohtsuki2004}.

\end{itemize}

The majority of the packages incorporate GRAPE and sometimes, one or two other algorithms are also included. In this list, QuTiP and Spinach are the libraries with the largest number of functionalities. Their scope goes beyond control  optimization. 

\subsection{Shooting techniques}
\label{sec:shooting_algo}

Behind the name "shooting techniques" hides a large class of more or less sophisticated algorithms~\cite{trelat2012optimal,bryson1996optimal,Bocop,bock1984multiple,
trelat2005controle,giftthaler2018family}. They are also called indirect methods in the literature~\cite{trelat2012optimal}. The general idea is inspired by a shooting problem, where the goal is to place a bullet or an arrow on a target. The only parameters that can be modified are the orientation of the gun and the initial velocity of the bullet, i.e. the dynamics of the system can only be influenced by a precise choice of initial conditions. Random shots can be taken to get an estimate of these initial conditions, but more sophisticated strategies can be used for this purpose.

Using Hamiltonian's equations of the PMP, the optimal control problem can be cast into this form. The parameters to be optimized correspond to the initial value of the adjoint state (see Sec.~\ref{sec:Lagrangian approach}). The estimation of this initial adjoint state may be improved by optimization algorithms such as those described in Sec.~\ref{sec:std_opt_algo}.
The pseudo-code of the algorithm can be written as follows:

\begin{lstlisting}
Function $(X(t),u(t))$=Xtraj($\Lambda_0$,$t_f$)
(* Integrate from 0 to $t_f$ the equations of motion $\dot \Lambda = -\partial _{X} H_P$ and
 $\dot X = \partial_{\Lambda} H_P$, where the control is expressed using the maximization condition of
 the PMP in the form $u(t)=u(X(t),\Lambda(t))$.
 The initial conditions for $\Lambda$ are given by $\Lambda_0$, and
 the ones for $X$ are fixed by the definition of the control problem. *)

Function $\mathcal{C}$=Cost($X(t)$,$u(t)$,$t_f$)
(* Return the value of the cost functional*)
$\Lambda_{\textrm{opt}}$ = Minimize(Cost(Xtraj($\Lambda_0$,$t_f$),$t_f$),$\Lambda_0 \in \Mc D$)
$u_{\textrm{opt}}(t)$ = Xtraj($\Lambda_{\textrm{opt}}$,$t_f$)$[2]$
\end{lstlisting}

\begin{figure}[htbp]
    \begin{center}
    \includegraphics[width=0.75\textwidth]{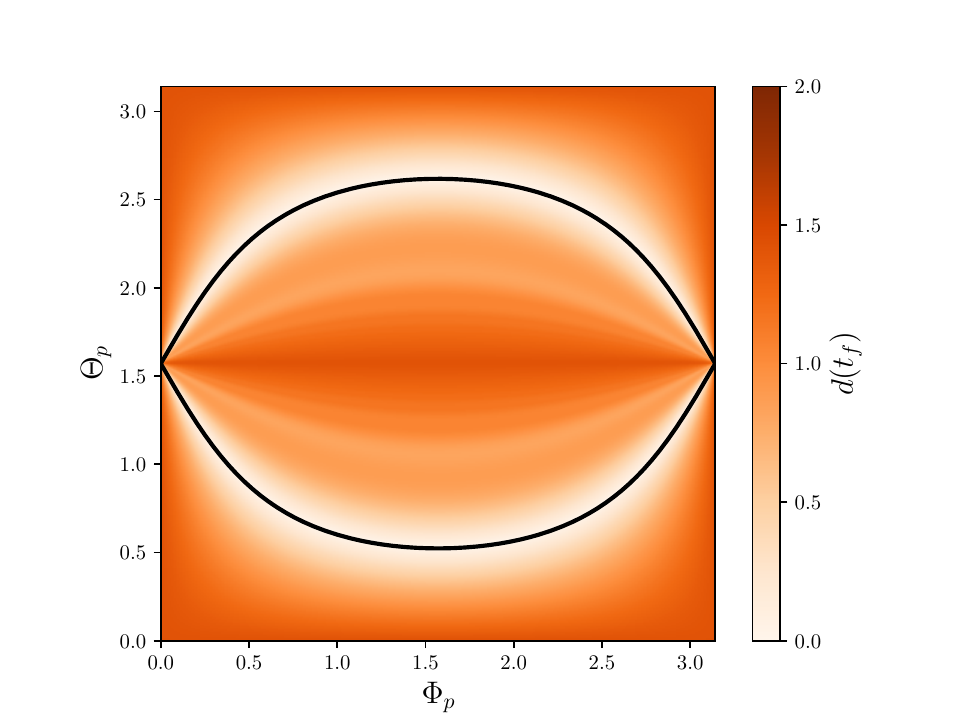}
    \end{center}
  \caption{Euclidean distance $d$ to the target state $(0,1,0)$ at time $t_f=t^\star$ for the trajectories with different initial adjoint states parameterized by the angles $\Theta_p$ and $\Phi_p$. The minimum time solutions corresponding to the minimum of $d$ are represented by solid lines.}
  \label{figcontour}
\end{figure}

Here above, \texttt{Minimize} is an arbitrary optimization algorithm (see e.g. Sec.~\ref{sec:std_opt_algo}), and $\Mc D$ is the domain of definition of $\Lambda_0$. The shooting technique is efficient when the solutions of the PMP are regular and without switching. The singular case can be difficult to solve numerically and switching times may not be detected by the algorithm~\cite{bonnard2003singular,martinon2007using}. Efficient numerical codes have been developed to tackle these kinds of computation, such as Bocop~\cite{Bocop} or Hampath~\cite{caillau2012differential,HamPath}. 
\begin{example}{}{ex10}
We illustrate the application of the shooting algorithm on the state-to-state transfer of a two-level quantum system with two controls and no drift. The Hamiltonian of the system is given by
$$
\hat H = \frac{u_x}{2} \sigx+\frac{u_y}{2} \sigy.
$$
We describe the state of the system in terms of Bloch coordinates $(x,y,z)$. Using Eq.~\eqref{eqbloch}, we get:
\begin{equation}
\begin{split}
\dot{x}&=u_y z, \\
\dot{y}&=-u_x z, \\
\dot{z}&=-u_yx+u_xy.\nonumber
\end{split}
\end{equation}
The goal of the control problem is to bring the system from $(1,0,0)$ to $(0,1,0)$ in minimum time with the constraint $u_x^2+u_y^2\leq 1$. The cost functional is thus $\mathcal{C}=t_f$. It can be shown that the optimal protocol consists of saturating the bound at any time, that is $u_x^2+u_y^2=1$~\cite{PRXQuantumsugny}. The idea is that the fastest control protocol requires generally maximum control intensity. The Pontryagin Hamiltonian can be expressed as
$$
H_P=u_x(p_zy-p_yz)+u_y(p_xz-p_zx)+p_0,
$$
where $(p_x,p_y,p_z)$ is the adjoint state which also satisfies the Bloch equation, and $p_0$ the abnormal multiplier. Since the final time is free, we have $H_P=0$. The maximization condition of $H_P$ leads in the regular case to the following optimal controls
\begin{equation}
\begin{split}
u_x &= (p_zy-p_yz)/H,\\
u_y &= (p_xz-p_zx)/H, \nonumber\\
\end{split}
\end{equation}
where $H=\sqrt{(p_zy-p_yz)^2+(p_xz-p_zx)^2}$.

As described in~\cite{PRXQuantumsugny}, the control problem can be integrated exactly. The minimum time is $t_f=t^\star=\frac{\pi\sqrt{3}}{2}$ and the corresponding initial adjoint state at time $t=0$ has the coordinates $(p_x(0),\frac{1}{\sqrt{3}},\pm 1)$. Note that the first component is not fixed but the same control process is obtained for any value of $p_x(0)$. The two possible values of $p_z(0)=\pm 1$ correspond to the two symmetric trajectories with respect to the equator on the Bloch sphere. A global description of the control landscape can be made by representing the initial adjoint state on a sphere as $p_x=R_p\sin\Theta_p\cos\Theta_p$, $p_y=R_p\sin\Theta_p\sin\Phi_p$ and $p_z=R_p\cos\Theta_p$. We choose to normalize $R_p$ to one, i.e. the adjoint state $(p_x,p_y,p_z)$ belongs to the sphere of unit radius, while the abnormal multiplier is not known. For each couple of initial values $(\Theta_p(0),\Phi_p(0))$, we numerically compute the corresponding optimal trajectory and the euclidean distance $d$ to the target state at time $t^\star$. By definition, the optimal solutions verify $d=0$. The corresponding control landscape is plotted in Fig.~\ref{figcontour}.

The same problem can also be solved numerically by using an iterative shooting technique. In this case, the goal is not to calculate all the trajectories to obtain the control landscape, but to iteratively find the right solution that reaches the target state. We choose to normalize the abnormal multiplier to -1 which leads to $H_P=1$, but the modulus of $(p_x,p_y,p_z)$ is not fixed. A very good convergence of the algorithm is observed. We find the optimal solution and the corresponding initial values of the adjoint state with high numerical precision. The corresponding Python code \emph{shooting.py} is provided in the supplementary material.
\end{example}
\subsection{Gradient-based algorithms designed for quantum control}
\label{sec:GRAPE_algo}

In this paragraph, we focus on a gradient-based algorithm called \textbf{GRAPE}, for \textbf{Gr}adient \textbf{A}scent \textbf{P}ulse \textbf{E}ngineering~\cite{rebentrost2009optimal,PhysRevA.82.063418,khaneja_optimal_2005,auckenthaler2010matrix,goodwin2016modified,PhysRevA.102.042612,ding2019robust}. This algorithm was used to solve a large amount of quantum control problems. The references~\cite{kobzar2004exploring,kobzar2012exploring,Van_Damme_robust_2017,Ansel_2021,BEC2021,BEC2023,Lapert_exploring_2012,Ansel_2022,haidong2017,
ansel_2017,VANREETH201739,
Van_Damme_time_optimal_2018,Ansel2023} is a very short list of studies in which GRAPE plays a key role. This list is by no means  exhaustive~\cite{cat,kochroadmap}. GRAPE and its variants are one of the most famous gradient-based algorithms in the quantum control community.

\paragraph{A direct gradient algorithm.} We first present a direct derivation of a gradient-based algorithm. The explicit relation to the PMP is discussed in a second step. The basic assumption which may be experimentally relevant is that the control is a piecewise constant function with a finite number $N$ of time steps of duration $\Delta t=t_f/N$. We denote by $u_n$ the amplitude of the control in the time interval $[(n-1)\Delta t,n\Delta t)$ with $n\in \{1,\cdots, N\}$. The control $u$ is now described by a set of $N$ real values $(u_1,u_2,\cdots,u_N)$. The core of the algorithm is based on a standard gradient descent algorithm for finite-dimensional optimization~\cite{bonnans2006numerical,kelley1999iterative}, which greatly simplifies its derivation.
The basic principle comes from the observation that if the cost function $\mathcal{C}(u)$ is differentiable then $\mathcal{C} ( u ) $ has the fastest decrease in the direction $-\partial_{u_n} \mathcal{C}(u)$ given by the gradient of the cost function with respect to the control parameters. An iterative algorithm to estimate the optimal control can then be designed. Starting from a given control $u$, a new control $u'$ such that $\mathcal{C}(u')<\mathcal{C}(u)$ can be computed from the relation
\begin{equation}\label{eqnewcontrol}
u'_n = u_n - \epsilon \frac{\partial \mathcal{C}(u) }{\partial u_n},
\end{equation}
where $\epsilon>0$ is a parameter chosen small enough to guarantee the convergence of the algorithm, but large enough  to limit the number of iterations. The cost $\mathcal{C}(u')$ can be estimated at first order in $\epsilon$ as follows
\begin{equation}
\begin{split}
\mathcal{C}(u')&=\mathcal{C}\left(u-\epsilon\frac{\partial \mathcal{C}(u)}{\partial u}\right) \\
&= \mathcal{C}(u)-\epsilon \left(\frac{\partial \mathcal{C}(u)}{\partial u}\right)^2 + O(\epsilon^2) \\
&\lesssim \mathcal{C}(u).
\end{split}
\end{equation}
The strict inequality is achieved for $\epsilon$ small enough. The parameter $\epsilon$ is a free parameter that must be set by hand. Its automatic adjustment can be performed using a line-search algorithm, which aims to find the optimal value of $\epsilon$ at each iteration step. A line-search improves convergence but at the cost of increased computation time. Another way to speed up the convergence of the algorithm is to replace $\epsilon$ by gradients of previous iterations as in conjugate gradient approaches or by the inverse of the Hessian in second-order methods. The latter is the core of Newton's type algorithms. The Hessian can be difficult to compute in practice, but it can be estimated simply by knowing  the gradient and the cost function. Such algorithms are called quasi-Newton~\cite{gill1972quasi}. We will not discuss these questions about the choice of $\epsilon$, and we assume in the following that this parameter is known. General ideas can be adapted to more advanced algorithms.

To summarize, the pseudo-code for a standard gradient algorithm is the following:

\begin{lstlisting}
Choose an initial guess control $u_0$.
Function $(\mathcal{C}_1,u)$ = GradientMinimize($u_0,\epsilon,k_{max}$)
	$u=u_0$
	$\mathcal{C}_1 = \mathcal{C}(u)$
	For [$1\leq k\leq k_{max}$,
		$u' = u - \epsilon \frac{\partial \mathcal{C}(u) }{\partial u}$
		$\mathcal{C}_2 = \mathcal{C}(u')$
		If[$\mathcal{C}_2>\mathcal{C}_1$,
			Break,
			$\mathcal{C}_1 = \mathcal{C}_2$
			$u = u'$]]

\end{lstlisting}
The constant $k_{max}$ which corresponds to the maximum number of iterations is set large enough such that the convergence to a minimum of $\mathcal{C}$ is reached.

\paragraph{Calculation of the gradient.} We have not yet clarified how $\tfrac{\partial \mathcal{C}(u) }{\partial u_n}$ is calculated, and this is the difficult part of the algorithm. GRAPE provides a clever calculation trick to save computational time and derive analytical formulas. For simplicity, we consider that the Hamiltonian of the quantum system is of the form:
\begin{equation}
\hat H(t) = \hat H_0 + u(t) \hat H_1,
\end{equation}
where $\hat H_0$ and $\hat H_1$ are two time independent Hamiltonian operators which do not commute and $u$ is a one-dimensional piecewise constant unbounded function (i.e. $u(t) \in \setR$, and on the step interval $n$, $u(t) = u_n$). The generalization to several control parameters is straightforward. The evolution operator during a time step can be expressed as
\begin{equation}
\hat U_n=\hat U(n \Delta t,(n-1)\Delta t) = e^{-\ii  \Delta t (\hat H_0 + u_n \hat H_1)},
\label{eq:U_evol_piecewise_constant}
\end{equation}
and $|\psi(t_f)\rangle = \hat U_N \hat U_{N-1}\cdots \hat U_1 |\psi_0\rangle$. We denote by $|\psi_n\rangle =\hat U_n\cdots \hat U_1|\psi_0\rangle$ the state at time $t=n\Delta t$.

To simplify the description, we consider only the case of a terminal cost of the form $\mathcal{C}=G(|\psi(t_f)\rangle) =G(\hat U_N \hat U_{N-1}\cdots \hat U_1 |\psi_0\rangle)$. Since $\hat{U}_n$ is the only term depending on $u_n$, we deduce that the derivative of $G$ with respect to $u_n$ can be expressed as
$$
\frac{\partial{G}}{\partial u_n}=\frac{\partial{G}}{\partial |\psi(t_f)\rangle}\frac{\partial |\psi(t_f)\rangle}{\partial u_n}+\frac{\partial \langle\psi(t_f)|}{\partial u_n}\frac{\partial{G}}{\partial \langle\psi(t_f)|},
$$
which leads to
\begin{equation}
\begin{split}
\frac{\partial{G}}{\partial u_n}&=\frac{\partial{G}}{\partial |\psi(t_f)\rangle}
\hat U_N \cdots \frac{\partial \hat U_n}{\partial u_n}\cdots \hat U_1 |\psi_0\rangle \\
&+ \langle\psi_0|
\hat U_1^\dagger \cdots \frac{\partial \hat U_n^\dagger}{\partial u_n}\cdots \hat U_N^\dagger \frac{\partial{G}}{\partial \langle\psi(t_f)|}.
\end{split}
\end{equation}
It remains to compute $\frac{\partial \hat U_n}{\partial u_n}$ which is not trivial because $[\hat H_0,\hat H_1]\neq 0$. A formal expression can be obtained from the Wilcox formula~\cite{wilcox1967exponential,wilcox} which gives the derivative of the exponential of a matrix $A(\theta)$ with respect to a parameter $\theta$
$$
\frac{\partial e^{tA}}{\partial \theta}=e^{tA}\int_0^t e^{-t' A}\frac{\partial A}{\partial \theta}e^{t'A}dt'.
$$
We deduce that
$$
\frac{\partial \hat U_n}{\partial u_n}=-\ii\Delta t \hat U_n\overline{\hat H}_1,
$$
where $\overline{\hat H}_1=\frac{1}{\Delta t}\int_0^{\Delta t}e^{\ii t'(\hat H_0+u_n\hat H_1)}dt'\hat H_1 e^{-\ii t'(\hat H_0+u_n\hat H_1)}dt'$ is the average of $\hat H_1$ in the Heisenberg representation over the duration $\Delta t$. For a sufficiently small time step $\Delta t$, $\overline{\hat H}_1$ can be identified to $\hat H_1$ and we arrive at
$$
\frac{\partial \hat U_n}{\partial u_n}\simeq -\ii \Delta t \hat H_1\hat U_n.
$$
Introducing an adjoint state $|\chi\rangle$ defined by a backward propagation in time from the target state such that $\langle\chi (t_f)|=\langle\chi_N|=\frac{\partial G}{\partial |\psi(t_f)\rangle}$ and $\langle\chi_n|=\langle \chi_N|\hat U_N\cdots \hat U_n$, the gradient can be expressed as
$$
\frac{\partial G}{\partial u_n}=-\ii\Delta t(\langle \chi_n|\hat H_1|\psi_n\rangle-\langle\psi_n|\hat H_1|\chi_n\rangle),
$$
which can be simplified into
\begin{equation}\label{eqgradientfinal}
\frac{\partial G}{\partial u_n}=2\Delta t \Im(\langle\chi_n|\hat H_1|\psi_n\rangle).
\end{equation}
We finally arrive at
\begin{equation}\label{eqfinalgrape}
\delta u_n=u_n'-u_n=-\epsilon \Im(\langle\chi_n|\hat{H}_1|\psi_n\rangle ).
\end{equation}

\begin{example}{}{ex11}\label{ex11}
We consider the same terminal costs $G_1$ and $G_2$ as in Example~\ref{example:ex9}. We recall that $G_1=1-|\langle\psi_f|\psi(t_f)\rangle|^2$ and $G_2=1-\Re\left(\langle\psi_f|\psi(t_f)\rangle\right)$. We deduce that the final condition of the adjoint state is given respectively by $|\chi_N\rangle=-\langle\psi_f|\psi(t_f)\rangle |\psi_f\rangle$ and $|\chi_N\rangle=-\frac{|\psi_f\rangle}{2}$ for $G_1$ and $G_2$.
\end{example}
Using Eq.~\eqref{eqgradientfinal}, we observe that the gradient can be computed in a very efficient way with a forward and a backward propagation in time of the dynamics from the initial and the target states, respectively. The pseudo-code for this computation can be written as follows.
\begin{lstlisting}
Function $\ket{\psi(t)}$ = PropforwardSchrodingerEq($u$,$\ket{\psi_{0}}$)
(*Propagate forward the Schrodinger equation using the control $u$ and the
initial state $\ket{\psi_{0}}$. Compute and store the quantum state at each time step, *)

Function $\ket{\chi(t)}$ = PropbackwardSchrodingerEq($u$,$\ket{\chi(t_f)}$)
(*Propagate backward the Schrodinger equation using the control $u$ and the
final state $\ket{\chi(t_f)}=2\chi_0\frac{\partial G}{\partial \langle\psi(t_f)|}$. Compute and store $\ket{\chi(t)}$ at each time step.*)

Function $\delta u$=GradientGRAPE($u,\epsilon$)
	$\ket{\psi(t)}$ = PropforwardSchrodingerEq($u(t)$,$\ket{\psi_{0}}$)
	$\ket{\chi(t)}$ = PropbackwardSchrodingerEq($u(t)$,$\ket{\chi(t_f)}$)
	For[$1\leq n\leq N$, $\delta u_n= -\epsilon\Im (\langle\chi_n|  \hat H_1  |\psi_n\rangle)$]
	

\end{lstlisting}
Using this method, the time evolution of a quantum state is computed twice at each evaluation of the gradient, while $N + 1 \gg 2$ evaluations of the dynamics are necessary when the gradient is estimated from a finite difference (i.e. with a formula $\partial_{u_n}G \approx (G(u_n + \Delta u_n) - G(u_n))/\Delta(u_n)$). When the number of time steps is large (hundreds or thousands of time steps), this approach can significantly reduce the computational time.

\paragraph{GRAPE versus PMP} The relation between GRAPE and the PMP can be established in the limit when the time step becomes infinitesimally small. In this case, the standard derivative is replaced by a functional one. Starting from Eq.~\eqref{eqSPont}, we see that the choice $\delta u=\epsilon \frac{\partial H_P}{\partial u}$ leads to
$$
\delta S=-\epsilon \int_0^{t_f} \left(\frac{\partial H_P}{\partial u}\right)^2dt,
$$
which is negative to first order in $\epsilon$. Using
$H_P=\Im(\langle \chi|\hat H|\psi\rangle)$, we deduce that
$$
\delta u=-\epsilon \Im(\langle \chi|\hat{H}_1|\psi\rangle),
$$
which is the continuous version of the time-discretized gradient derived in Eq.~\eqref{eqfinalgrape}. In other words, a gradient-based algorithm can be viewed as a time digitalization of an equivalent algorithm derived from the PMP. Note that this equivalence is only valid when $U$ is open, i.e. in the weak PMP.

Over the years, several versions of GRAPE have been developed. Some possible modifications are presented below.


\paragraph{Parameterized continuous functions.}
An interesting modification of the algorithm is to replace a piecewise constant control by parameterized functions~\cite{skinner_optimal_2010}. The control $u$ is then expanded over a set of functions $\{f_k(t)\}_{k=1,\cdots,k_{\textrm{max}}}$ such as $u=v(\{a_kf_k(t)\})$ where $v$ is a smooth function. The goal is to find the real coefficients $(a_k)$ to minimize the cost functional. In the case of  Fourier series, this would be:
\begin{equation}
u(t) = a_0+\sum_{k=1}^{k_{max}} a_k \cos(k\omega   t)+b_k \sin(k\omega   t),~\omega = 2\pi/t_f.
\label{eq:param_control_fourier}
\end{equation}
At each step of the iterative algorithm, the values of the parameters are corrected from the gradients $\frac{\partial G}{\partial a_k}$, $\frac{\partial G}{\partial b_k}$. Using the chain rule derivation, the gradient $\frac{\partial G}{\partial a_k}$ for example can be expressed as
$$
\frac{\partial G}{\partial a_k}=\sum_{n=1}^N\frac{\partial G}{\partial u_n}\frac{\partial u_n}{\partial a_k}=\sum_{n=1}^N \frac{\partial G}{\partial u_n}\frac{\partial v}{\partial a_k}f_k(t_n),
$$
where the functions $f_k$ are approximated by piecewise constant functions with the same time step as $u$. The following pseudo-code can be used to implement this variant of GRAPE.

\begin{lstlisting}
Function $u$ = Control($\{a_k,b_k\}$)
(* Compute $u_k = u(t_k)$ for each $t_k$ using the parameters $\{a_k,b_k\}_{k=0,\cdots,k_{\textrm{max}}}$  *)

Function $\delta a, \delta b$ = GradientGRAPE2($\{a_k,b_k\}$)
	$u$ = Control($\{a_k,b_k\}$)
	$\ket{\psi(t)}$ = PropforwardSchrodingerEq($u(t)$,$\ket{\psi_{0}}$)
	$\ket{\chi(t)}$ = PropbackwardSchrodingerEq($u(t)$,$\ket{\chi(t_f)}$)
	For[$1\leq n\leq N$,
		$\delta u_n= -\epsilon\Im \bra{\chi(t_{n})}  \hat H_1  \ket{\psi(t_n)}$]
	For[$1\leq k\leq k_{\textrm{max}}$,
		$\delta a_k=\sum_{n=1}^N \delta u_n\frac{\partial u}{\partial a_k}(t_n)$
                $\delta b_k=\sum_{n=1}^N \delta u_n\frac{\partial u}{\partial b_k}(t_n) $]
		
\end{lstlisting}

\paragraph{Exact gradient.}
A slight modification of the system can be used to compute the gradient exactly. The idea is to consider a larger system given by the Hamiltonian $\tilde{H}$
\begin{equation}
\tilde{H}=\begin{pmatrix}
\hat{H} & 0 \\
\partial_u\hat{H} & \hat{H}
\end{pmatrix} .
\end{equation}
In the standard case, we have $\partial_u\hat{H}=\hat{H}_1$. The matrix exponential of $\tilde{H}$ has interesting properties such as
$$
e^{\alpha\tilde{H}}=
\begin{pmatrix}
e^{\alpha \hat H} & 0\\
\partial_u e^{\alpha \hat{H}} & e^{\alpha\hat{H}}
\end{pmatrix},
$$
which can be shown by using a Taylor series expansion of the exponential function. The parameter $\alpha$ is a complex constant. We deduce that the evolution operator $\tilde{U}$ corresponding to the Hamiltonian $\tilde{H}$ can be expressed as
$$
\tilde{U}=
\begin{pmatrix}
\hat{U} & 0\\
\partial_u \hat{U} & \hat{U}
\end{pmatrix}.
$$
This result suggests that the propagator $\hat{U}$ and its gradient can be computed simultaneously by solving the Schr\"odinger equation associated to the Hamiltonian $\tilde{H}$ with the initial condition $(\hat I, 0)$. The gradient can then be used directly in GRAPE to correct the control at each step of the algorithm. This variant is known in the literature as the auxiliary matrix approach of GRAPE.

\begin{example}{}{ex12}
To illustrate the method, we study the control of a state-to-state transfer in a two-level quantum system with minimal energy. The goal is to steer the system from the initial state $\ket{\uparrow}$ to the target $\ket{\downarrow}$ at time $t_f$ fixed  with the Hamiltonian $\hat H(t) = \frac{ \Delta}{2}\sigz + \frac{ u}{2} \sigx$. The cost functional to minimize can be expressed as
$$
\mathcal{C}=1-|\langle \uparrow|\psi(t_f)\rangle |^2+\frac{p_0}{2}\int_0^{t_f}u(t)^2dt.
$$
where $p_0$ is a factor allowing to adjust the relative weight of the two terms in the cost. The control time $t_f$ is set to $2\pi/\sqrt{1+\Delta^2}$. The Hamiltonian's equations of the PMP show that the dynamics of the state $|\psi(t)\rangle$ and adjoint state $|\chi(t)\rangle$ are governed by the Schr\"odinger equation with respectively the initial condition $|\psi(0)\rangle =|\uparrow\rangle$ and the final condition $|\chi(t_f)\rangle = \langle \uparrow |\psi(t_f)\rangle |\uparrow\rangle$ with $\chi_0=-1/2$ (see Example~\ref{example:ex9}). In the GRAPE algorithm, the correction $\delta u$ at each step can be written as
$$
\delta u =-\epsilon\left(\Im(\langle\chi(t)|\frac{\sigx}{2}|\psi(t)\rangle +\frac{p_0}{2}u(t)\right).
$$
The convergence of the algorithm is very good in this example. An example is plotted in Fig.~\ref{figgrape}. The corresponding Python code \emph{GRAPE.py} is provided in the supplementary material. The Python code \emph{GRAPE2.py} solves the same optimal control problem but with a polynomial parameterization of the control pulse.
\end{example}
\begin{figure}[htbp]
    \centering
    \includegraphics[width=8cm]{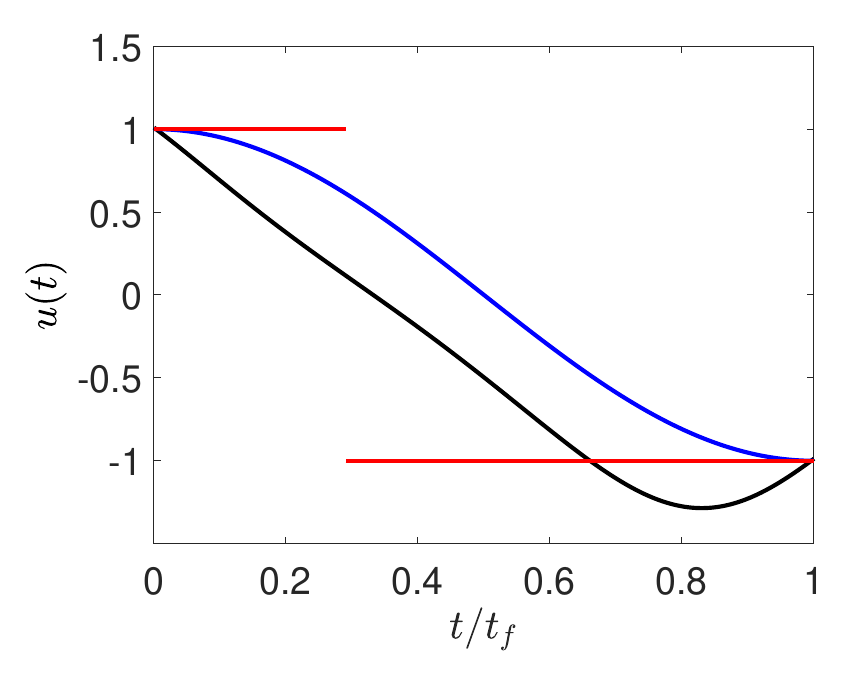}
  \caption{Optimal control $u$ (black line) designed by GRAPE for Example~\ref{example:ex12}. Numerical parameters are set to $u_0=1$ and $\Delta=\frac{u_0}{2}$. The red line represents the time-optimal solution with the constraint $|u(t)|\leq u_0$. The blue line depicts the guess control used in the algorithm. The parameter $p_0$ is set to $0.1/t_f$.}\label{figgrape}
\end{figure}

\section{From theory to experiment: Optimal control of a Bose-Einstein condensate in an optical lattice}\label{sectheoexp}
We propose to demonstrate the different steps of the application of QOC to a specific experimental system, namely the control of a Bose-Einstein Condensate (BEC) in an optical lattice. We refer the reader to the references~\cite{BEC2021,BEC2023,BEC2024} for additional theoretical and experimental details on these results.


\subsection{The model system}
\label{BEC_model}
Dilute Bose-Einstein condensates are routinely produced in cold atom experiments: a gas of identical bosons, cooled to a temperature close to absolute zero condenses into a single quantum state, and is therefore described by a single wave function~\cite{RMP2008}. On typical experimental timescales (see Sec.~\ref{sec:Experimental_optimal_control}), we can neglect both the interactions between the atoms of the gas and the parabolic magnetic potential used to confine the system. Such conditions allow us to simplify the description of the dynamics of the system which are governed by the  Schr\"odinger equation with a periodic sinusoidal potential~\cite{Eckardt2017}. This trapping potential originates from a one-dimensional optical lattice produced by two interfering laser beams of wavelength $\lambda$ propagating in opposite directions, whose relative phase and amplitude can be modulated in time with good precision. The goal of the control is then to manipulate the motional state of the BEC in this potential.

The wave function $\ket{\psi(t)}$ which belongs to the Hilbert space $\mathcal{H}= \mathcal{L}^2(\mathbb{R}, dx)$, evolves in time according to the Schrödinger equation,
       \begin{align}
  \imath \hbar \frac{d\ket{\psi(t)}}{dt} = \left( \frac{\hat{p}^2}{2m} - \frac{s(t) E_L}{2}\cos\left( k_L \hat{x} + \varphi(t) \right) \right) \ket{\psi(t)},\label{BEC_SE}
\end{align}
where $\hat{p}=-\imath\hbar\frac{\partial}{\partial x}$ and $\hat{x}=x$ are respectively the momentum operator and the position operator in the position representation with $x$ the spatial coordinate along the optical lattice axis. We denote by $m$ the mass of the atom, $k_L=2\pi/(\lambda/2)$ the wave vector and $E_L=\frac{\hbar^2 k_L^2}{2m}=4E_R$ (with $E_R$ the recoil energy) the energy associated with the lattice. 
The control parameters are the dimensionless depth $s(t)$ and the phase $\varphi(t)$. In this section, we focus on using $\varphi(t)$ as a single control parameter, while the dimensionless lattice depth $s$ is kept constant. Varying $\varphi$ as a function of time corresponds to moving or shaking the lattice position along the $x$- axis. We consider the following change of variables to obtain dimensionless coordinates:
    \begin{align}
      t &\rightarrow \frac{E_L}{\hbar}t, \nonumber\\
      x &\rightarrow k_L x.\nonumber
    \end{align}
This yields,
       \begin{align}
  \imath \frac{d\ket{\psi(t)}}{dt} \equiv \hat{H}(t)\ket{\psi(t)}=\left(\hat{p}^2 - \frac{s}{2}\cos\left( \hat{x} + \varphi(t) \right)\right)\ket{\psi(t)},
\end{align}
with $\hat{p}=-\imath\frac{\partial}{\partial x}$ and $\hat{x}=x$  in the position representation.

We denote by $\ket{\phi_{\alpha}}$ the eigenvectors of the momentum operator with eigenvalue $\alpha$ and wavefunction $\phi_{\alpha}(x)=\frac{1}{\sqrt{2\pi}}e^{\imath \alpha x}$. Since the potential is periodic in $x$, the Bloch theorem states that the parameter $\alpha$ can be expressed as $\alpha = n + q$, where $n \in \mathbb{Z}$ and $q\in [-0.5,0.5]$ is the quasimomentum. The quasimomentum can formally take any real value, but due to the periodicity of the potential, two quasimomenta separated by an integer are equivalent. 
Furthermore, this periodicity implies that the quasimomentum $q$ is conserved during the  control process. In the subspace of a given quasimomentum $q$, we can expand a generic state on the plane wave basis as,
       \begin{align}
  \ket{\psi} = \sum_{n\in\mathbb{Z}}c_{q,n} \ket{\phi_{q+n}}.\nonumber
\end{align}
Using this decomposition, the dynamic is given in terms of the coefficients $c_{q,n}$ as
    \begin{align}
   \label{BEC_coef} \imath\Dot{c}_{q,n} &= \left( n+q \right)^2c_{q,n} - \frac{s}{4}\left(e^{\imath\varphi(t)}c_{q,n-1} + e^{-\imath\varphi(t)}c_{q,n+1} \right).\nonumber
\end{align}
The Schr\"odinger equation can be written in matrix form as follows,
       \begin{align}
   \imath\frac{d\ket{\psi(t)}}{dt} = \left( \hat{H}_0 + \cos\left( \varphi(t)\right)\hat{H}_1 + \sin\left( \varphi(t)\right)\hat{H}_2\right) \ket{\psi(t)},\nonumber
\end{align}
where
       \begin{align}
   \ket{\psi(t)} = \begin{pmatrix}
  \vdots \\
  c_{q,n-1} \\
  c_{q,n} \\
  c_{q,n+1} \\
  \vdots
  \end{pmatrix},
\end{align}
       \begin{align}
       \centering
  \label{BEC_vector} \hat{H}_0 =  \begin{pmatrix}
  & \ddots &  &  &  \\
  \dots & 0 & \left((n-1)+q\right)^2 & 0 & 0 & 0 & \dots \\
  \dots & 0 & 0 & \left(n+q\right)^2 & 0 & 0 & \dots \\
  \dots & 0 & 0 & 0 & \left((n+1)+q\right)^2 & 0 & \dots \\
   &  &  &  &  & \ddots &
  \end{pmatrix}, 
\end{align}
and
       \begin{align}
       \centering
  \hat{H}_1 = \begin{pmatrix}
  \ddots &  & \ddots &  &  \\
  \dots & -\frac{s}{4} & 0 & -\frac{s}{4} & 0 & 0 & \dots \\
  \dots & 0 & -\frac{s}{4} & 0 & -\frac{s}{4} & 0 & \dots \\
  \dots & 0 & 0 & -\frac{s}{4} & 0 & -\frac{s}{4} & \dots \\
   &  &  &  & \ddots &  & \ddots
  \end{pmatrix}, \
  \hat{H}_2 = \begin{pmatrix}
  \ddots &  & \ddots &  &  \\
  \dots & -\imath\frac{s}{4} & 0 & \imath\frac{s}{4} & 0 & 0 & \dots \\
  \dots & 0 & -\imath\frac{s}{4} & 0 & \imath\frac{s}{4} & 0 & \dots \\
  \dots & 0 & 0 & -\imath\frac{s}{4} & 0 & \imath\frac{s}{4} & \dots \\
   &  &  &  & \ddots &  & \ddots
  \end{pmatrix}.\nonumber
  \end{align}
\paragraph{Landau-Zener-type Hamiltonian.} Interestingly, a two-level approximation can be derived from the BEC system when $s\ll 1$.

\begin{figure}
         \centering
    \includegraphics[width=0.5\textwidth]{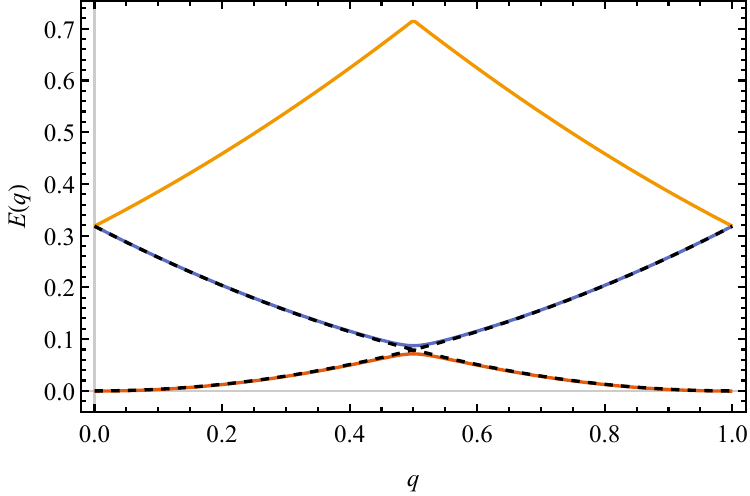}
    \caption{Eigenvalues $E(q)$ of $\hat{H}$ (colored solid lines) and $\hat{p}^2$ (black dashed lines) as a function of $q$.}
   \label{levels_figure}
\end{figure}

Calculating the eigenvalues of $\hat{H}$ as a function of the quasimomentum yields the lattice band structure $E_m(q)$ ($m\in\mathbb{N}$), as shown in Fig.~\ref{levels_figure}. We denote the corresponding Bloch eigenfunctions $\ket{\Psi_m(q)}$. For a small value of $s$ (typically $s<0.5$) and $q$ close to $0.5$, the first two energy bands $E_0(q)$ and $E_1(q)$ are well-separated in energy from the others, and may be considered as an effective two-level system. If the BEC is initially prepared in the subspace formed by these first two bands, it will remain in this subspace.

Experimentally, while the quasimomentum can be tuned from its initial value by the application of a force (inducing a Bloch oscillation), it is most often equal to zero, corresponding to a homogeneous BEC, or to the ground state of the lattice $\ket{\Psi_0(0)}$ (prepared adiabatically). We therefore do not consider the quasimomentum as a control parameter here, and assume $q=0$ for the rest of this section. Using the unitary transformation,
    \begin{align}
      \ket{\psi} &\rightarrow \hat{U}\ket{\psi}, \nonumber \\
      \hat{H} &\rightarrow \hat{U}\hat{H}\hat{U}^{\dagger} + \imath \dot{\hat{U}} \hat{U}^{\dagger},\nonumber
    \end{align}
where $\hat{U}=e^{-\imath \hat{p}\varphi(t)}$, we obtain the following Hamiltonian,
    \begin{align}
      \hat{H} = \left(\hat{p} + \frac{\dot{\varphi}(t)}{2} \right)^2 - \frac{s}{2} \cos \left( \hat{x} \right) ,\nonumber
    \end{align}
where $\frac{\dot{\varphi}(t)}{2}$ plays the role of a controlled quasi momentum (effectively, $\frac{\dot{\varphi}(t)}{2}$ is the quasimomentum of the BEC in the reference frame of the moving lattice). If we set $\frac{\dot{\varphi}(t)}{2} = \frac{1}{2} + \frac{\dot{\tilde{\varphi}}(t)}{2}$, with $\dot{\tilde{\varphi}}(t)$ a control parameter close to $0$, we can isolate the first two energy levels and write the Hamiltonian in the basis $\left( \ket{\phi_{0-1}}, \ket{\phi_{0+0}} \right)$ such that,
       \begin{align}
   \hat{H} = \begin{pmatrix}
  \left( \frac{1}{2} - \frac{\tilde{\dot{\varphi}}(t)}{2} \right)^2 & -\frac{s}{4}\\
   -\frac{s}{4} & \left( \frac{1}{2} + \frac{\tilde{\dot{\varphi}}(t)}{2} \right)^2  \\
  \end{pmatrix}.
\end{align}
Up to a term proportional to the identity, we obtain a two-level quantum system whose Hamiltonian can be expressed as
    \begin{align}
      \hat{H} = \frac{\Delta(t)}{2}\hat{\sigma}_z + \frac{\omega}{2}\hat{\sigma}_x ,
    \end{align}
where $\omega=-\frac{s}{2}$ and $\Delta(t) = - \dot{\tilde{\varphi}}(t)$.
This expression is similar to the Hamiltonian of a two-level quantum system with an offset term and a single control, for which time optimal control strategies are studied in Sec.~\ref{sec:OCT_spin_time_1_input_and_offset}. With the following basis change,
\begin{align}
\ket{\phi_{0-1}}\rightarrow\frac{1}{\sqrt{2}}(\ket{\phi_{0-1}}+\ket{\phi_{0-0}})\nonumber \\
\ket{\phi_{0-0}}\rightarrow\frac{1}{\sqrt{2}}(\ket{\phi_{0-1}}-\ket{\phi_{0-0}})\nonumber
\end{align}
the correspondence is exact and allows to implement optimal solutions in this system as shown in Sec.~\ref{sec:Experimental_optimal_control}.

\subsection{Numerical optimal control}
\label{sec:Numerical-optimal-control}
We consider the application of GRAPE to the control of a BEC in an optical lattice. The BEC system has a Hamiltonian given by
       \begin{align}
\hat{H} = \hat{H}_0 + \cos\left(u(t)\right) \hat{H}_1 + \sin\left(u(t)\right)\hat{H}_2,
\end{align}
where we set $u(t)=\varphi(t)$ to use the same notation as in the general description of GRAPE. The goal is to find a control that minimizes the cost function $\mathcal{C}$ at a fixed final time $t_f$
       \begin{align}
  \mathcal{C} = 1 - \lvert \braket{\psi_{f}}{\psi(t_f)} \rvert^2,
  \label{eq:cost_bec}
\end{align}
where $\ket{\psi(t_f)}$ is the state at final time, and $\ket{\psi_{f}}$, the target state. The application of the PMP yields a Pontryagin Hamiltonian of the form \eqref{eq:Hamilotnian_QOC_schrodinger} with $F_0=0$. The adjoint state $\ket{\chi(t)}$ whose time evolution is also governed by the Schr\"odinger equation, has the final condition
       \begin{align}
  \ket{\chi(t_f)} = -2\chi_0\braket{\psi_{f}}{\psi(t_f)}\ket{\psi_{f}}.
	\end{align}
The abnormal multiplier is set to $\chi_0=-1/2$ in the numerical simulation. Using the maximization condition of the PMP, the control is iteratively updated such that
       \begin{align}
 u'_n = u_n - \epsilon \Im \left(\bra{\chi(t_{n})}{\left(-\sin\left(u_n\right)\hat H_1 + \cos\left(u_n\right)\hat{H}_2\right)}\ket{\psi(t_n)}\right)
\end{align}
The control time is set to a multiple duration characteristic of the dynamical timescale of the system (usually the inverse spacing between the two lowest energy levels), and discretized into several hundred steps, so that the step duration is small with respect to this dynamical timescale, and the control is therefore quasi-continuous. The infinite dimensional Hilbert space is truncated so that $|n|\leq n_{\textrm{max}}$, where $n_{\textrm{max}}$ is chosen with respect to the initial and target states to avoid boundary effects. In the numerical simulations, the control usually involves $400$ steps and $n_{\textrm{max}}=10$. Thus the truncated space has a dimension of $2\times n_\textrm{max}+1=21$. Under these conditions, one iteration of the algorithm takes $0.17$ seconds, and it takes about $100$ iterations, i.e. $17$~seconds to obtain a cost function of order $10^{-4}$. The numerical simulations, written in Python, were conducted on a standard laptop computer.


\paragraph{State-to-state transfer.}


We first illustrate the optimal control of state-to-state transfer. The initial state of the BEC is represented by the state $\ket{\phi_{0+0}}$ and several target states are considered, which can be expressed in the canonical basis $\ket{\phi_{0+n}}$ with $q=0$. Attainable states range from individual momentum basis vectors to superposition of states~\cite{BEC2021}. Alternatively, they can also be Gaussian states, corresponding to a localized probability density in position and momentum within a lattice site~\cite{BEC2023}.
Here we define as Gaussian a state whose $x$ and $p$ probability density functions are normal distributions with standard deviations $\sigma_{x_0}$ and $\sigma_{p_0}$ equal to those of the ground state $\ket{\Psi_0(0)}$ of the lattice Hamiltonian. The Gaussian state tends to the exact ground state for $s \gg 1$, with $\sigma_{x_0}=s^{-1/4}$, and $\sigma_{p_0}=s^{1/4}/2$. A displaced Gaussian state $\ket{g(x_c,p_c)}$ has non-zero position and momentum averages within a lattice site, $\langle x\rangle =x_c$ and $\langle p\rangle=p_c$.
We can further define a squeezed Gaussian state $\ket{g(x_c,p_c,\xi)}$, for which the standard deviations are modified as $\sigma_x=\xi \sigma_{x_0}$ and $\sigma_p=\frac{\sigma_{p_0}}{\xi}$, where $\xi$ is the $x$- squeezing parameter ($\xi=1$ corresponding to a Gaussian state). The smaller $\xi$ becomes, the more the standard deviation $\sigma_x$ decreases and $\sigma_p$ increases.
Gaussian and squeezed states can be projected on the basis $\left( \ket{\phi_{0,+n}} \right)$ with the coefficients~\cite{BEC2023},
    \begin{align}
    c_{0,n}\left( x_c, p_c, \xi \right) = \left( \frac{2 \xi^2}{\pi \sqrt{s}} \right)^{1/4} e^{\imath x_c p_c /2} e^{-\imath n x_c} e^{-\xi^2 \left( n - p_c \right)^2 / \sqrt{s}}.
\end{align}

\begin{figure}
    \includegraphics[width=\textwidth]{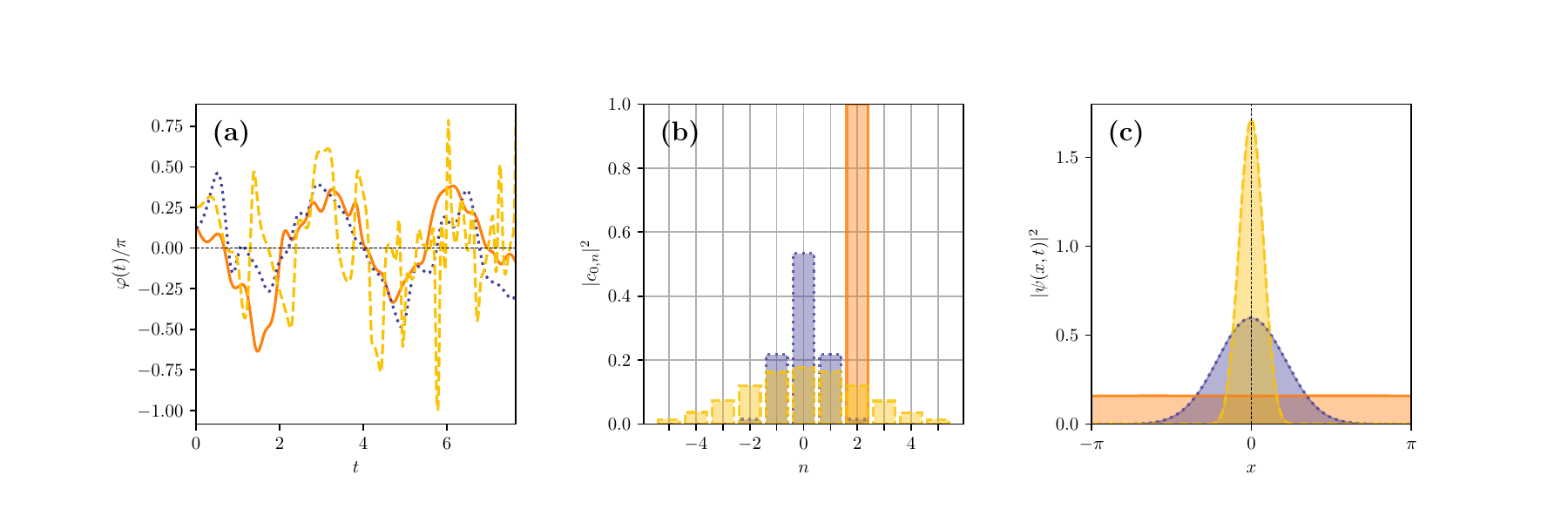}

    \caption{Examples of control for the state-to-state transfer of a BEC: In red the transfer from $\ket{\phi_{0+0}}$ to $\ket{\phi_{0+2}}$, in dotted-blue to the Gaussian state $\ket{g(0,0,1)}$ and in dashed-yellow to the squeezed state $\ket{g(0,0,1/3)}$. The transfers are computed for $q=0$ and $s=5$. (a) Control phase. (b) Bar diagram of the momentum distribution reached at time $t_f$. (c) Probability density in position within a lattice cell at final time.}
   \label{oct_BEC}
\end{figure}

Three numerical examples of optimal controls for state to state transfer are considered in Fig.~\ref{oct_BEC}. In each case, the initial state is $\ket{\phi_{0+0}}$, and we set $q=0$, $s=5$, $n_{max}=10$ and $t_f=7.6$, which corresponds to a duration of 150~$\mu$s. The target states are chosen to be $\ket{\phi_{0+2}}$, the centered Gaussian state $\ket{g(x_c=0,p_c=0,\xi=1)}$ and the centered squeezed Gaussian state $\ket{g(x_c=0,p_c=0,\xi=1/3)}$. The numerical results can be obtained from a code provided in the supplementary material.
%

\subsection{Experimental optimal control}
\label{sec:Experimental_optimal_control}

In this section, we present some illustrative optimal control results applied to a BEC manipulated in an optical lattice, following the previous formalism. Other examples can be found in~\cite{BEC2021,BEC2023,BEC2024}. We focus here on the experimental implementation of the control in a real system, namely the experimental setup at LCAR in Toulouse, the manipulation of the full quantum state of the BEC in the lattice, and a realization of an optimal control in an effective two-level quantum system.


\subsubsection{Experimental setup.}

In a typical implementation~\cite{BEC2021}, a BEC of $^{87}$Rb atoms is trapped in a one-dimensional optical lattice, created by two counter-propagating laser beams. The beams' wavelength, $\lambda=1063.9\,$nm, is chosen far from the main optical resonances of the atom, in order to minimize light scattering and heating of the condensate. This also sets the lattice spacing $d=\lambda/2\simeq 532\,$nm, and the characteristic energy scale $E_L=h^2/(2md^2)=h\cdot8.111\,$kHz, which also sets a characteristic timescale for the dynamics.

In the Schr\"odinger equation~\eqref{BEC_SE} governing the dynamics, both the dimensionless lattice depth $s(t)$ and its position $\varphi(t)$ can be varied arbitrarily in time, and therefore act as control parameters. This is achieved by using  Acousto-Optic-Modulators (AOMs) placed in the path of the lattice beams, which can adjust the amplitude and phase of the outgoing beam by varying the amplitude and phase of the RF signal applied to the AOM crystal. A common modulator varies the amplitude of a first laser beam which is then split to form the two arms of the lattice, after which one of the arms goes through an AOM which modulates its phase, thus varying the relative phase between the lattice beams and controlling $\varphi(t)$.

To ensure that optimal control solutions can be successfully applied in this system, it is important to consider the timescales involved. On the one hand, the dynamical timescale is determined by both the atoms' inertia and the depth of the sine potential: for a typical depth $s=5$, the energy spacing between the two lowest bands at $q=0$ is $1.975\,E_L$ (close to the level spacing in the harmonic approximation $\hbar \omega\simeq 2.236\,E_L$), corresponding to a characteristic duration of $T_0=62.4\,\mu$s.
    On the other hand, through a combination of bandwidth limitations from the driving electronics and the AOM itself, changes in the amplitude and phase of a beam exiting the AOM occur on a typical duration of $100\,$ns for sudden changes, corresponding to a spectral bandwidth of about 3 MHz. This sets the main limit on the speed with which the controls can be varied: the $3\,$MHz bandwidth nonetheless allows changes to be made almost instantaneously with respect to the dynamical timescale.

This leads to the typical choice for the control ramps applied experimentally: a duration of $100\,\mu\mathrm{s}\simeq 1.6T_0$, discretized in 400 intervals of $250\,$ns with a constant phase. Any change in the value of the phase occurs much faster than the inertial response of the atoms: the control is therefore quasi-continuous.

\medskip

In addition to these considerations on the lattice control, it is also crucial to consider other experimental effects that are not included in the modeling of the experiment:
\begin{itemize}
\item[$-$]~The laser wavelength $\lambda$ must be known precisely as it effectively enters in the dimensionless timescale $E_Lt/\hbar=\alpha t$. For commercially available fiber lasers, the wavelength is typically known to an accuracy of $10^{-4}$.
\item[$-$]~The $^{87}$Rb atoms used here experience repulsive interactions within the BEC, characterised by a scattering length $a=104 a_0$~\cite{DGObook}. These interactions can be described in the mean-field approximation by an additional, non-linear potential term in the Schrödinger equation \eqref{BEC_SE}, $V_\mathrm{int}=\beta|\psi(x,t)|^2$, yielding the Gross-Pitaevskii equation. The constant $\beta$ characterizes the non-linearity for the 1-D dynamics in the lattice. In the experiments presented here, it is typically small ($\beta<0.5$) due to the dilute nature of the BEC.
\item[$-$]~In a realistic experiment, the BEC loaded in the optical lattice has a finite size, which may also be affected by interactions. This corresponds to the occupation of a finite interval of quasimomenta around a central value (here $q=0$). In experiments shown here, about 100 lattice sites are populated, leading to an estimate of the quasi-momentum width $\Delta q\sim0.02$~\cite{Dubertrand2016}.
\item[$-$]~Last but not least, the lattice depth $s$ is a fixed parameter for the optimal control using $\varphi(t)$, but it must be known with a good precision to derive efficient controls. This means that it is crucial to have a precise calibration of the lattice depth~\cite{Cabrera2018} before optimizing the control, as well as excellent stability of the experimental setup.
\end{itemize}

\begin{figure}
         \centering
    \includegraphics[width=0.8\textwidth]{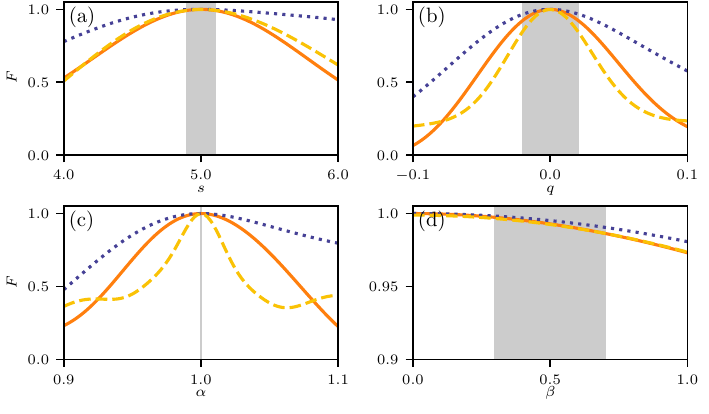}

    \caption{Examples of the influence of various parameters on the performance of the optimal control ramps for the state-to-state transfer of a BEC. (a) Fidelity of the preparation for a control ramp computed at an expected depth $s_0=5$, when the ramp is applied in a lattice of actual constant depth $s$, as the value of $s$ is varied. The full orange line (resp. blue dotted line and yellow dashed line) corresponds to the transfer from $\ket{\phi_{0+0}}$ to $\ket{\phi_{0+2}}$ (resp. to the Gaussian state $\ket{g(0,0,1)}$ and to the squeezed state $\ket{g(0,0,1/3)}$), here and throughout the figure. Likewise, throughout the figure, the control ramps are calculated for fixed values of the parameters: $s=5$, $q=0$, $\beta=0$ and $\alpha=1$, and are the ramps obtained in Fig.~\ref{oct_BEC}. (b) Similarly, fidelity of the preparation against a variation of the quasi-momentum: the control ramp is applied to the initial state $\ket{\phi_{q+0}}$, with $q$ varied. (c) Similarly, fidelity of the preparation against a variation of the timescale factor $\alpha$ (see text). (d) Fidelity of the preparation against a variation of the interaction parameter $\beta$ (see text). In each graph, the gray shaded area denotes typical experimental uncertainty intervals for the varied parameter.}
   \label{oct_BEC_parameters}
\end{figure}

To illustrate the role of these parameters, we show in Fig.~\ref{oct_BEC_parameters} how the fidelity of the state preparations studied in Sec.~\ref{sec:Numerical-optimal-control} are affected by variations around the value for which the optimal control is calculated, with relevant uncertainty ranges highlighted. For realistic values of the parameters, Fig.~\ref{oct_BEC_parameters} demonstrates that the timescale factor is very well characterized and that the effect of interaction is mostly negligible. It also highlights the importance of a well-calibrated lattice depth, and a small quasimomentum distribution width for the success of the state transfer. The quasimomentum effect is all the more important the more squeezed the target state is (\emph{i.e.} extended in momentum).

There are other important constraints on the controls available experimentally, namely on the maximum depth $s$ that can be applied to the atoms, and on the maximum available control time. The former is limited by the maximum laser intensity available, to $s\lesssim40$ for the experimental setup described here. The latter is constrained, when using $^{87}$Rb, by interaction-induced dynamical instabilities that may occur on a timescale of several milliseconds~\cite{DupontPNAS2023}.

Finally, we emphasize that in an experimental situation, it is not possible to directly measure the complex coefficients $c_{q,n}$ characterizing the quantum state. A single measurement of the BEC consists in imaging the absorption from a resonant infrared laser beam by the atoms, which is imaged on a CCD camera. This imaging is performed after a time-of-flight, during which the various momentum components of the BEC will separate spatially. Such a measurement, when good care is taken to remove any parasitic signal on the camera, will only provide a measurement of the probabilities $|c_{q,n}|^2$. A full characterization of the prepared state may therefore require multiple measurements, in order to extract relative phases~\cite{BEC2021}, or to perform a full state reconstruction~\cite{BEC2023}.

\subsubsection{Full quantum state control.}

A clear demonstration of optimal transfer between quantum states with control of probability amplitudes is provided by the preparation of energy eigenstates. The Bloch eigenstates corresponding to the energy spectrum for the lattice potential,  as shown in Fig.~\ref{levels_figure}, are defined by their coefficients $c_{q,n}^{(m)}$, stationary solutions of Eq.~\eqref{BEC_coef}, with specific amplitudes and signs. When such a state is prepared in the lattice, its stationary nature means that the momentum distribution (the measured probabilities $|c_{q,n}|^2$) do not evolve at subsequent times.

\begin{figure}
         \centering
    \includegraphics[width=0.8\textwidth]{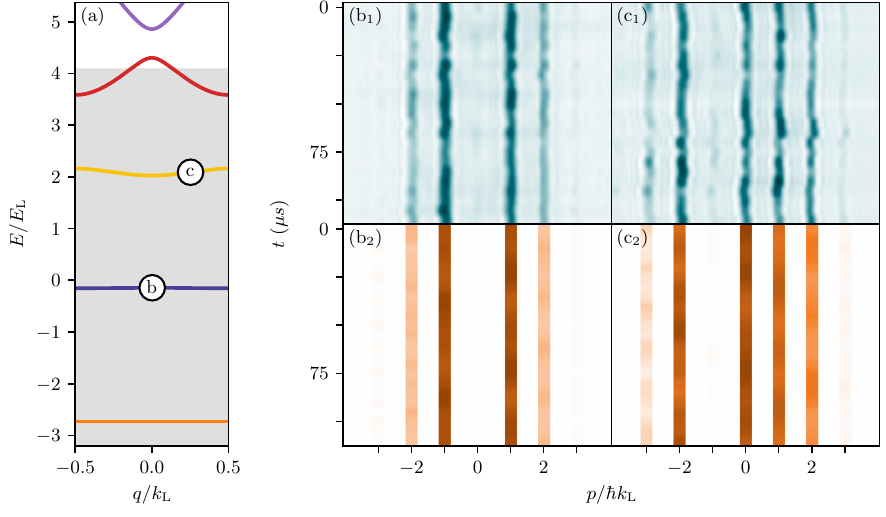}
    \caption{Preparation of lattice eigenstates. (a) Lattice band structure for $s=8.2$ (colored lines) with labels for the eigenstates prepared in (b) and (c). The shaded area denotes the sine potential depth. (b) Preparation of the $P$ band eigenstate at quasi-momentum $q = 0$ for a depth $s = 8.15 \pm 0.30$. $\mathrm{(b_1)}$ Experimental evolution of momentum distribution of the prepared state in the static lattice. $\mathrm{(b_2)}$ Corresponding theoretical evolution. (c) Same as (b) for the preparation of the $D$ band eigenstate  at quasi-momentum $q= 0.25 k_L$ for a depth  $s = 8.26 \pm 0.10$.}
   \label{oct_BEC_eigenstates}
\end{figure}

This control process is illustrated in Fig.~\ref{oct_BEC_eigenstates}. We first apply an optimal control ramp to transfer the lattice ground state $\ket{\Psi_0(0)}$ to the $P$ band Bloch state at $q=0$, as denoted in Fig~\ref{oct_BEC_eigenstates} \textbf{a}, in a lattice of depth $s=8.2$. After the preparation, we measure the momentum distribution obtained for increasing hold times, up to $110~\mu\mathrm{s}\simeq1.5T_0$. The result of this measurement is shown in Fig.~\ref{oct_BEC_eigenstates} \textbf{b$_1$}: the momentum distribution shows no significant evolution, as expected for an eigenstate. For comparison, Fig.~\ref{oct_BEC_eigenstates} \textbf{b$_2$} shows the numerical evolution of the momentum distribution, as expected from the theoretical final state of the optimal control ramp (which has a numerical fidelity of 99\% to the theoretical $P$ band eingenstate in good agreement with the experimental results).

It is also possible to prepare eigenstates in a subspace with non-zero quasi-momentum $q_0$, by taking advantage of a change of reference frame. We first prepare the plane wave superposition with coefficients $c_{q_0,n}^{(m)}$, starting from the lattice ground state. At the end of the preparation ramp $\varphi(t)$, instead of returning the phase to $\varphi=0$ (in which case we remain in the $q=0$ subspace where the prepared state is not an eigenstate), we set the lattice in linear motion with a ramp
$\varphi(t>t_f)=2q_0(t-t_f)$ (in reduced units). This effectively translates the state into the $q=q_0$ subspace in the reference frame of the lattice, and the prepared superposition is then an eigenstate. This is illustrated in Fig~\ref{oct_BEC_eigenstates} \textbf{c$_1$}, which shows the evolution of the momentum distribution after preparation of the $D$ band eigenstate at $q_0=0.25k_L$ (as denoted in Fig~\ref{oct_BEC_eigenstates} \textbf{a}) in the moving lattice. Again the distribution shows very little evolution, in good agreement with the numerically expected result shown in Fig~\ref{oct_BEC_eigenstates} \textbf{c$_2$}.
Further examples of state preparation and state characterization can be found in~\cite{BEC2021,BEC2023,BEC2024}.

\subsubsection{Two-level optimal control.}

Finally, the BEC system lends itself to the emulation of a two-level quantum system, as introduced in Sec.~\ref{BEC_model}.  The control protocol derived in Sec.~\ref{sec:OCT_spin_time_1_input_and_offset} can be used to perform a time-optimal transfer between the states,
\begin{align}
\ket{+}=\frac{1}{\sqrt{2}}(\ket{\phi_{0-1}}+\ket{\phi_{0-0}}),\nonumber \\
\ket{-}=\frac{1}{\sqrt{2}}(\ket{\phi_{0-1}}-\ket{\phi_{0-0}}).\nonumber
\end{align}
In this context, the lattice depth plays the role of the constant offset $\Delta=s/2$, while the phase variation $\dot{\varphi}=1+u(t)$ provides the variable control. To perform the bang-bang protocol of Sec.~\ref{sec:OCT_spin_time_1_input_and_offset}, we choose a lattice depth of $s\simeq0.5$, and a maximum control $u_0=0.5$. This yields typical control times for the experiment of $t_1\simeq 64\,\mu\mathrm{s}$ and $t_2\simeq 156\,\mu\mathrm{s}$.

In order to assess the result of the transfer, we need to characterize the initial and final states. However both $\ket{+}$ and $\ket{-}$ are approximate eigenstates of the Hamiltonian at a small depth, and they cannot be distinguished from a simple measurement of the (equally weighted) populations in $\ket{\phi_{0-1}}$ and $\ket{\phi_{0-0}}$. To circumvent this issue, we use a quench of these states into a deeper lattice $s_\mathrm{meas}\simeq6$, in which they are not eigenstates. The population dynamics in the deeper lattice then allow us to clearly distinguish $\ket{+}$ and $\ket{-}$. These state superpositions can be prepared using an optimal transfer ramp,  in the deeper lattice of depth $s_\mathrm{meas}$, both as initial states for the bang-bang protocol, and to characterize their evolution. We use a preparation ramp with numerical fidelity $F>99.5\%$, and a duration of $1.5T_0$ (about $84\,\mu\mathrm{s}$). The state is then characterized by recording the momentum distribution dynamics for $200\,\mu\mathrm{s}$ at $10\,\mu\mathrm{s}$ intervals.

\begin{figure}
         \centering
    \includegraphics[width=0.8\textwidth]{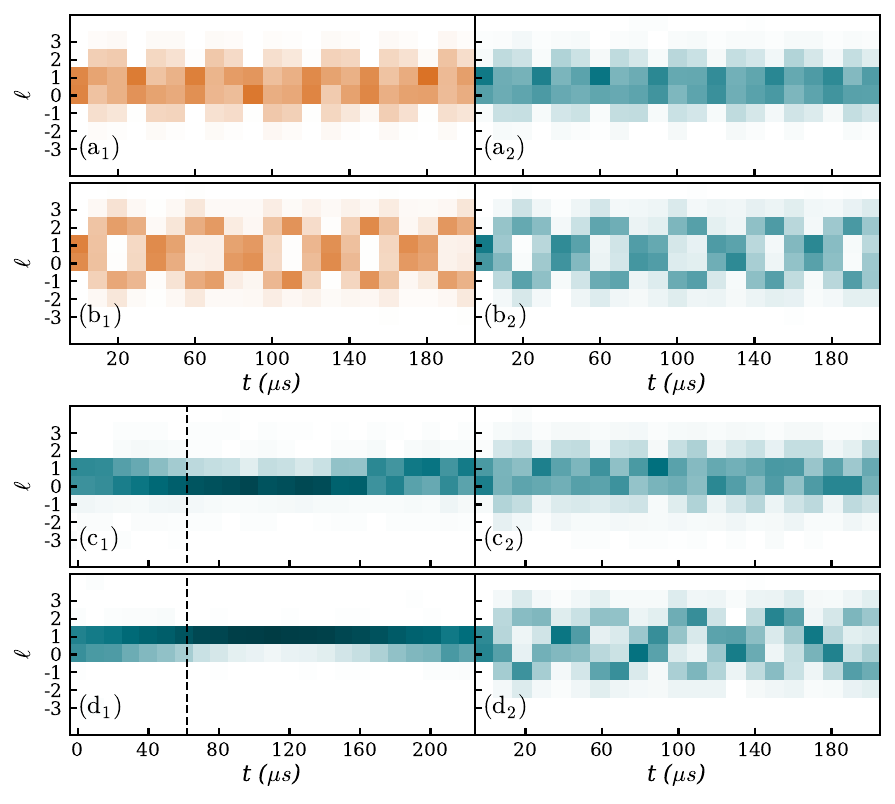}
    \caption{Time-optimal control of an effective two-level quantum system on the BEC platform. (a) Theoretical $\mathrm{(a_1)}$ and experimental $\mathrm{(a_2)}$ dynamics of the momentum distribution after preparation of the state $\ket{+}$ in the lattice of depth $s_\mathrm{meas}=6.07\pm0.05$. The numerical fidelity after preparation is $F>0.995$. (b) Same as (a) for the preparation and characterization of state $\ket{-}$. (c) Experimental evolution of the momentum distribution for a bang-bang transfer between $\ket{-}$ and $\ket{+}$ in a lattice of depth $s=0.57\pm0.03$ $\mathrm{(c_1)}$, followed by a quench to a deep lattice $s_\mathrm{meas}=5.90\pm0.09$ $\mathrm{(c_2)}$. The optimal control parameters are $u_0=0.5$, $t_1=63.66\,\mu\mathrm{s}$, and $t_2=159.5\,\mu\mathrm{s}$. (d) Same as (c) for a bang-bang transfer between $\ket{+}$ and $\ket{-}$. The color map for probabilities on all graphs extends from 0 to 1. 
    }
   \label{oct_BEC_LZexp}
\end{figure}

The realization of such a procedure is shown in Fig.~\ref{oct_BEC_LZexp}. Panels Fig.\ref{oct_BEC_LZexp} \textbf{a} and Fig.\ref{oct_BEC_LZexp} \textbf{b} present both the numerical and experimental evolution of the states $\ket{+}$ and $\ket{-}$ in the quenched lattice of depth $s_\mathrm{meas}$. This demonstrates that the change in relative phase can be clearly identified through the non-equilibrium dynamics in the deeper lattice, and shows at the same time that the two superposition states can be prepared efficiently as initial states for the bang-bang control.
Figure~\ref{oct_BEC_LZexp} \textbf{c} (resp. \textbf{d}) shows the experimental results from the time-optimal control for the transfer from $\ket{-}$ to $\ket{+}$ (resp. $\ket{+}$ to $\ket{-}$), followed by a quench to the depth $s_\mathrm{meas}$. The control is realized by applying two successive constant phase drifts $\dot{\varphi}=1\pm 0.5$ for times $t_1$ and $t_2$ and is identical for both transfers (only the initial state is changed). After the transfer, the quenched dynamics confirm that the states have been exchanged with good accuracy, the dynamics being almost identical to those of panels  Fig.~\ref{oct_BEC_LZexp} \textbf{a} and Fig.~\ref{oct_BEC_LZexp} \textbf{b}.

Note that a more thorough characterization of the initially prepared and final states can be achieved by state reconstruction techniques using the lattice dynamics data~\cite{BEC2023}. This goes beyond our purpose here, which has been to illustrate how the BEC platform can be used to emulate optimal control of a two-level quantum system.

\section{Conclusion}
\label{sec:conclusion}

In this introduction to the toolbox of quantum optimal control, we present both analytical and numerical methods based on the PMP. The key elements of this mathematical theory are described from an analogy with classical Lagrangian and Hamiltonian mechanics. It is then shown how these results can be used to design optimal control strategies for quantum systems. A comprehensive description of existing optimization algorithms is provided with a discussion of their advantages and areas of applicability. Particular attention is paid to shooting techniques and gradient-based algorithms that are directly derived from the PMP. Several problems of state-to-state transfer in a two-level quantum system have been analyzed in detail. The link between the optimal solution and the quantum speed limit is also explored in this case. The experimental implementation in the case of BEC in a one-dimensional optical lattice is described in a final section. Starting from the modeling of the quantum dynamics, we show step by step how the optimal control is computed and then implemented experimentally, to realize both two-level and many-level controls. Experimental constraints and limitations are also discussed.

The goal of this tutorial paper is to provide an overview of the toolbox of QOC that we hope is accessible to physicists with a background in quantum dynamics. Simple and concrete physical examples have been used throughout this paper to illustrate the different mathematical concepts. In particular, we have successively reused and adapted the same examples to illustrate different important points. Therefore, the examples presented in this introduction are not representative of the systems encountered in the literature. For instance, a key aspect which has not been discussed concerns the control of open quantum systems. Most experimental configurations must be modeled by taking into account the interaction of the system with its environment, and such features must be integrated into the optimization process~\cite{Koch2016}.
The degree to which OCT techniques have been applied to that end depends on the characteristics of the open quantum system considered.
In the Markovian regime in which the memory effect is neglected, optimal control procedures are now quite well understood. The main difficulty lies in the loss of complete controlability and therefore in the fact that certain target states are not reachable~\cite{dirr2009}. The situation is not at the same stage of maturity for non-Markovian dynamics. Although this aspect has been explored in a few examples, the usefulness of the memory effect as a resource for optimal control remains to be clarified.


A wide range of problems have been already solved in quantum optimal control, but with the advent of quantum computers and the progress of quantum technologies, new objectives are emerging. A first one concerns optimization performed on a quantum computer~\cite{hogg2000quantum,moll2018quantum,blekos2023review}. Such optimizations are based on quantum algorithms and the realization of quantum circuits in order to solve optimal control problems. The optimization can focus on the control design for the computer itself, but it can also be a totally independent control problem. Different quantum optimization algorithms can be distinguished extending from purely quantum algorithms, such as Grover's type algorithms~\cite{baritompa2005grover} and quantum annealing methods~\cite{das2005quantum,yarkoni2022quantum} or to hybrid algorithms based on a gradient descent~\cite{yang2017optimizing,cerezo2021variational,bonet2023performance}. In the latter case, the core of the algorithm is classical and it is the same as the one presented in Sec.~\ref{sec:GRAPE_algo}. The difference resides in the evaluation of the gradient, which is computed using quantum algorithms. This is particularly well adapted to quantum optimal control, but it does not offer any significant advantage, since the core of the algorithm remains classical. Breakthroughs are expected with pure quantum algorithms. Recently a few properties of the PMP have been combined with a Quantum Approximate Optimization Algorithm (QAOA)~\cite{PhysRevLett.126.070505,PhysRevApplied.16.054023}, but a general and versatile quantum algorithm based on the PMP remains to be found. Ideally, a PMP based quantum optimization algorithm should be designed for any type of cost function (not only restricted to bang-bang controls, like in~\cite{PhysRevLett.126.070505}), and it would solve the shooting problem using the advantages offered by quantum computations.

A second family of open questions concerns the inclusion of energy efficiency and sustainable development criteria in quantum technologies~\cite{PRXQuantum.3.020101}. The latter are not exempt from the challenge of global warming. Despite the fact that we are working with a very small number of quantum quantities, energy consumption is far from ideal. With the most powerful devices, we are beginning to achieve quantum supremacy, but we are far from having a quantum energy advantage. Part of the huge energy consumption comes from the cooling system needed to keep noise levels low, while a second source is due to controls. So far, we have mostly focused on time-optimal strategies, to avoid noise or dissipation effects. However, energy-optimal strategies can achieve a similar result with a drastic reduction  in energy consumption.  Optimizing energy consumption at all stages of quantum computing (and other quantum technologies) will be one of the major problems to be solved in the near future.

\noindent \textbf{Acknowledgments.} We gratefully acknowledge useful discussions with Pr. H. R. Jauslin. We thank the support from the Erasmus Mundus Master QuanTeem (Project number: 101050730), the project QuanTEdu-France (ANR-22-CMAS-0001) on quantum technologies  and the ANR project QuCoBEC (ANR-22-CE47-0008-02).

\appendix

\section{List of mathematical symbols and acronyms}\label{secappA}
\begin{itemize}
    \item[$ $] $X$ is the state of a dynamical system and $X_a$ is a vector component of $X$.
    \item[$ $] $\delta X$ is the infinitesimal difference between two states of a dynamical system.
    \item[$ $] $S$ is an action functional.
    \item[$ $] $\delta S$ is the functional derivative of the action $S$ for two trajectories close to each other.
    \item[$ $] $\Mc L$ is a Lagrangian.
    \item[$ $] $F$ is a vector function defining the system dynamics.
    \item[$ $] $u$ is a control parameter, $u_a$ is a vector component of $u$, and $u_n$ is the value at step $n$ of a piecewise constant control.
    \item[$ $] $U$ is the domain of definition of $u(t)$ (a subset of $\setR^m$).
    \item[$ $] $G$ is a terminal cost function.
    \item[$ $] $F_0$ is a running cost function.
    \item[$ $] $\mathcal{C}$ is the cost functional to minimize in an optimal control problem.
     \item[$ $] $\Lambda$ is the adjoint state, and $\Lambda_a$ is a vector component.
    \item[$ $] $\Lambda_0$ is the adnormal multiplier.
    \item[$ $] $H_p$ is the Pontryagin's Hamiltonian
    \item[$ $] $\Re$ and $\Im$ are respectively the real and imaginary parts of a complex number.
    \item[$ $] $\ket{\psi}$ is a quantum state.
    \item[$ $] $\hat \rho$ is a density matrix.
    \item[$ $] $\hat H$ is the Hamiltonian operator of a quantum system.
    \item[$ $] $\sigx, \sigy$ and $\sigz$ are the Pauli matrices.
    \item[$ $] $(x,y,z)$ are the coordinates of the Bloch vector for a two-level quantum system.
     \item[$ $] $\hat U(t_f,t_i)$ is the evolution operator from $t=t_i$ to $t=t_f$.
     \item[$ $] $\ket{\chi}$ is the adjoint state of $\ket{\psi}$.
\end{itemize}
The following acronyms are used in this paper:
\begin{itemize}
\item[$ $] OCT: Optimal Control Theory
\item[$ $] QOC: Quantum Optimal Control Theory
\item[$ $] PMP: Pontryagin Maximum Principle
\item[$ $] QSL: Quantum Speed Limit
\end{itemize}

\section{Lagrange Multiplier}\label{app_lagrange}
We recall in this section basic results on Lagrange multiplier in the finite-dimensional case. We are particularly interested in abnormal multipliers which are less described in the literature.

The method of Lagrange multiplier is a standard technique in finite-dimensional optimization problem that transforms a constrained optimization into an unconstrained one at the cost of an increase in dimension. Consider for instance the maximization (or minimization) of a smooth function $F_0$ on $\mathbb{R}^2$ with variables $(x,y)$. In absence of constraints, a necessary condition to fulfill for the extrema of the function is given by
$$
\nabla F_0 =\begin{pmatrix}
\frac{\partial F_0}{\partial x} \\
\frac{\partial F_0}{\partial y}
\end{pmatrix}
=
\begin{pmatrix}
0 \\
0
\end{pmatrix}.
$$
A slightly more difficult task is to find the maximum of $F_0$ under a constraint of the form $F(x,y)=0$ where $F$ is also a smooth function on $\mathbb{R}^2$. The extrema of $F_0$ are to be found on the level curve $F(x,y)=0$. In the generic case, the extremum is located at the point of $\mathbb{R}^2$ where the level curves  of $F$ and $F_0$ intersect at one point. At this point, the two curves have a common tangent and their gradient vectors are parallel. This gives the following condition
$$
\nabla F_0=-\lambda \nabla F,
$$
where $\lambda$ is a non-zero real parameter called a Lagrange multiplier. Note that the exact value of $\lambda$ is not important for finding the extrema. A systematic way to solve this problem is to introduce a new function $L$ on $\mathbb{R}^3$ as
$$
L(x,y,\lambda)=F_0(x,y)+\lambda F(x,y).
$$
The extrema are characterized by $\nabla L=0$, which leads to the same equations as those established previously. The extremum point is denoted $(x_0,y_0)$.

A singular behavior occurs when $\nabla F(x_0,y_0)=0$ and in this case it is necessary to adapt the procedure. A typical situation corresponds to the problem where the condition $F(x,y)=0$ is satisfied at only one isolated point. It is then clear that the value of $F_0$ at this point cannot be compared to its neighboring points. Consider for instance the function $F_0$ to be maximized $F_0(x,y)=x+y$ under the constraint $F(x,y)=x^2+y^2=0$. The point $(0,0)$ is the only point that satisfies the constraint. It therefore corresponds by construction to the point maximizing $F_0$. Note that we also have $\nabla F(0,0)=0$. This solution can be obtained by the general approach by modifying the definition of $L$ which is now a function on $\mathbb{R}^4$ such that
$$
L(x,y,\lambda,\lambda_0)=\lambda_0F_0(x,y)+\lambda F(x,y),
$$
where the real parameter $\lambda_0$ is called the abnormal multiplier. Note that $(\lambda,\lambda_0)$ is defined up to a real factor and that the two multipliers cannot be simultaneously equal to 0. The extrema are given by the conditions $\partial L/\partial x=\partial L/\partial y=\partial L/\partial \lambda=0$. When $\lambda_0\neq 0$, we recover the previous formulation of the optimization problem. New solutions appear when $\lambda_0=0$ and $\nabla F=0$. They have the peculiarity of not depending on $F_0$, i.e. the function to maximize. They are called abnormal extremal solutions. In the previous example, the point $(0,0)$ corresponds to such a solution for which $\lambda_0=0$.

\section{Pontryagin Maximum Principle}\label{appendixPMP}
We describe in a heuristic way the origin of the maximization condition of the PMP described in Thm.~\ref{thm:Pontryagin Maximum Principle}~\cite{bertsekasbook}. This also leads to a valuable geometric interpretation of the adjoint state and of the optimal control problem.

\begin{figure}[htbp]
\begin{center}
\includegraphics[width=10cm]{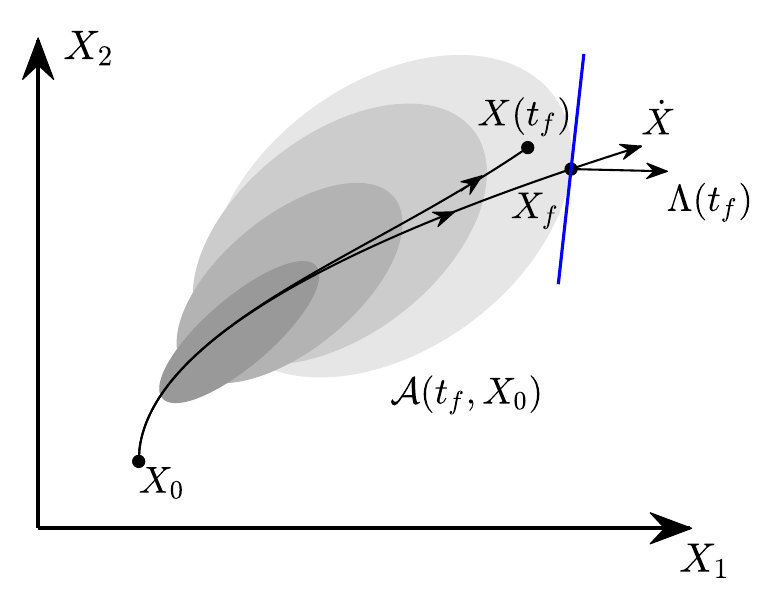}
\end{center}
\caption{Schematic description of a time-optimal control problem from $X_0$ to $X_f$. The vector space is $\mathbb{R}^2$ and the coordinates of $X$ are $(X_1,X_2)$. The set $\mathcal{A}(t_f,X_0)$ is the reachable set at time $t_f$ from $X_0$ (different reachable sets are plotted in shades of gray at different times). Two trajectories are plotted in black reaching respectively the points $X_f$ (optimal trajectory) and $X(t_f)$ (non-optimal trajectory). The blue line corresponds to the tangent to the reachable set in $X_f$. The adjoint state $\Lambda(t_f)$ is orthogonal to this line, while $\dot{X}$ is tangent to the trajectory.}
\label{fig10}
\end{figure}

We first consider a time-optimal control process where the goal is to steer the system from $X_0$ to $X_f$ in minimum time. We denote respectively by $u^\star$ and $X^\star$ a smooth optimal control and the corresponding trajectory such that $\dot{X}^\star=F(X^\star,u^\star)$. We consider another admissible control $u$ close to $u^\star$ which generates the dynamic $X(t)$ with $\dot{X}=F(X,u)$. The two trajectories are very close to each other and the small difference between the two is $\delta X=X-X^\star$. Starting from $\dot{X}^\star+\delta \dot{X}=F(X^\star+\delta X,u)$, a linearized equation of motion is used to calculate the dynamics around the reference trajectory  $X^\star$ (up to terms of order two in $\delta X$)
\begin{equation}\label{eqappdelta}
\delta \dot{X}=A\delta X+F(X^\star,u)-F(X^\star,u^\star),
\end{equation}
where the elements of the $n\times n$- matrix $A$ are given by $A_{ij}=\partial_{X_j} F_i|_{(X^\star,u^\star)}$. The rigorous derivation of Eq.~\eqref{eqappdelta} can be done under the assumption of regularity of $F$ and convexity of the set $F(X,u)$ for $u\in U$~\cite{bertsekasbook}. The solution of Eq.~\eqref{eqappdelta} can be expressed as
\begin{equation}\label{eqappdelta2}
\delta X(t)=\int_{0}^t V(t,t')[F(X^\star,u)-F(X^\star,u^\star)]dt',
\end{equation}
where $\delta X(0)=0$ and $V$ is the propagator from times 0 to $t$ associated to the differential equation $\dot{V}(t,0)=A(t) V(t,0)$, with $V(0,0)=I_n$~\cite{bryson1975applied}. The different elements of the optimal control problem are schematically represented in Fig.~\ref{fig10}.
We then introduce the reachable set $\mathcal{A}(t_f,X_0)$ at time $t_f$ from the state $X_0$ as the set of all the states $X(t_f)$ that can be reached by a trajectory starting from $X_0$ at time $t=0$, the trajectory being associated to an admissible control $u$. If the target state $X_f$ is attained exactly by the optimal solution at $t=t_f$ then $X_f$ is on the boundary of $\mathcal{A}(t_f,X_0)$. It is clear that if $X_f$ belongs to the interior of $\mathcal{A}(t_f,X_0)$ then a smaller time $t'<t_f$ can be found such that $X_f\in \mathcal{A}(t',X_0)$ and $t_f$ is not the minimum time to reach the target (see the reachable sets at different times in Fig.~\ref{fig10} to be convinced of this point). Assuming that the reachable set is convex in a neighborhood of $X_f$, we consider the plane tangent to $\mathcal{A}(t_f,X_0)$ in $X_f$ and we define the adjoint state $\Lambda(t_f)$ at time $t_f$ as a vector orthogonal to this plane and pointing outwards. For any final state $X(t_f)$ close to $X_f$ and associated to a non-optimal trajectory, we have $\delta X(t_f)\cdot \Lambda(t_f)\leq 0$ where $\delta X(t_f)=X(t_f)-X_f$.

We introduce a time-dependent vector $\Lambda(t)\in\mathbb{R}^n$ such that $\Lambda(t)^\intercal=\Lambda(t_f)^\intercal V(t_f,t)$. Using the relation $\dot{V}(t_f,t)=-V(t_f,t)A$, we arrive at $\dot{\Lambda}^\intercal=-\Lambda(t)^\intercal A$ or
\begin{equation}\label{eqappdelta3}
\dot{\Lambda}(t)=-A^\intercal \Lambda(t).
\end{equation}
As expected, $\Lambda(t)$ which is defined by a backward propagation of the dynamic can be identified with the adjoint state of the PMP. The Pontryagin Hamiltonian $H_P$ reads $H_P=\Lambda^\intercal F$. The corresponding Hamiltonian equation for the adjoint state can be written as
$$
\dot{\Lambda}_i(t)=-\frac{\partial H_P}{\partial X_i}=-\Lambda^\intercal \partial_{X_i}F(X,u)=-\sum_j\partial_{X_i}F_j(X,u)\Lambda_j,
$$
which is equivalent to Eq.~\eqref{eqappdelta3}. Starting from Eq.~\eqref{eqappdelta2} and the condition $\delta X(t_f)\cdot \Lambda(t_f)\leq 0$, we obtain
$$
\Lambda(t_f)^\intercal \int_0^{t_f}V(t_f,t')[F(X^\star,u)-F(X^\star,u^\star)]dt'\leq 0,
$$
which leads to
$$
\int_0^{t_f}\Lambda(t')^\intercal[F(X^\star,u)-F(X^\star,u^\star)]dt'\leq 0.
$$
This inequality can be transformed into
\begin{equation}\label{eqappmax}
\int_0^{t_f}[H_P(t')-H_P^\star(t')]dt'\leq 0,
\end{equation}
where $H_P=\Lambda^\intercal F(X^\star,u)$ and $H_P^\star=\Lambda^\intercal F(X^\star,u^\star)$.
Consider now for $u$ a control equal to $u^\star$ except on a very short time interval $[t,t+dt]$, we deduce that the condition \eqref{eqappmax} is satisfied if and only if
$$
H_P(t)\leq H_P^\star(t),
$$
for $t\in [0,t_f]$, i.e. $u^\star$ maximizes the function $H_P$ along the optimal trajectory.

We consider in a second step the relative position of $\Lambda(t_f)$ and $\dot{X}$ as represented in Fig.~\ref{fig10}. For the optimal solution, we deduce by construction that $H_P(t_f)=\Lambda(t_f)\cdot \dot{X}(t_f)\geq 0$. In this case, $X_f$ does not belong to $\mathcal{A}(X_0,t_f-dt)$ for any sufficiently small time step $dt>0$. Note that if $\dot{X}$ points inwards the reachable set, the trajectory will be time-maximal. When $H_P$ does not depend explicitly on time, the Pontryagin Hamiltonian is a constant of motion. It is straightforward to show this property if $U$ is an open set. In this case, we have $\frac{\partial H_P}{\partial u}=\Lambda\cdot \frac{\partial F}{\partial u}=0$ at any time $t$. We deduce that $\dot{H}_P=\dot{\Lambda}F+\Lambda\dot{F}$ can be expressed as
$$
\frac{d}{dt}H_P=-\Lambda \frac{\partial F}{\partial X}F+\Lambda\frac{\partial F}{\partial X}F=0.
$$
We denote by $-\Lambda_0$ the positive constant equal to $\Lambda\cdot \dot{X}$ at any time $t\in [0,t_f]$. A new Pontryagin Hamiltonian can then be defined as
$$
H_P=\Lambda\cdot \dot{X}+\Lambda_0,
$$
where $H_P=0$ along the optimal trajectory. The abnormal extremals for which $\Lambda_0=0$ can be identified here to the trajectory tangent to the boundary of the reachable set since in this case $\Lambda(t_f)\cdot \dot{X}(t_f)=0$.

\begin{figure}[htbp]
\begin{center}
\includegraphics[width=10cm]{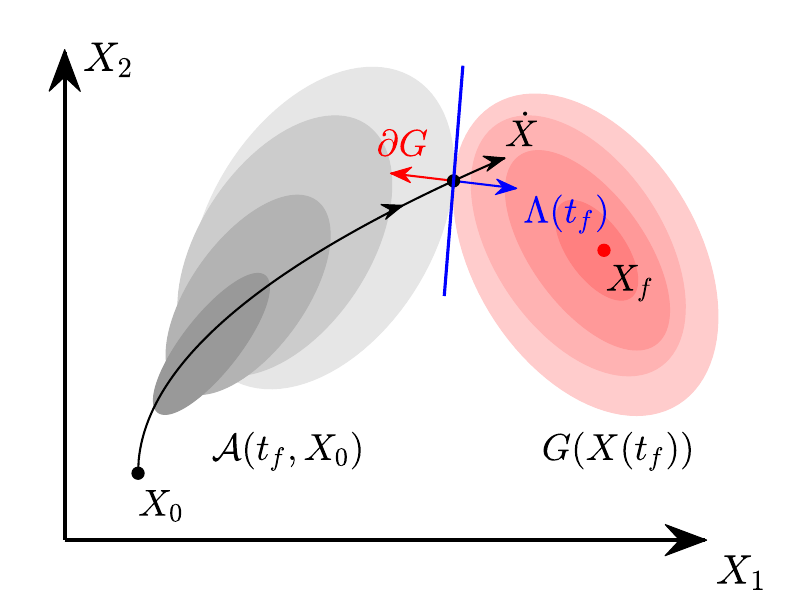}
\end{center}
\caption{Same as Fig.~\ref{fig10} but for the case of a fixed final time, the goal of the control problem being to minimize the distance to the target defined by the function $G$. The different reachable sets and the level surfaces of $G$ are plotted respectively in shades of gray and red. The blue line represents the common tangent plane to the boundary of the reachable set $\mathcal{A}(t_f,X_0)$ and the level curves of the distance $G(X(t_f))$. $\partial G$ denotes the gradient vector pointing outwards of the level surface at $X(t_f)$.}
\label{fig11}
\end{figure}

This argument can be extended to an optimal control problem with a fixed control time $t_f$ where the goal is to minimize the cost functional $\mathcal{C}=G(X(t_f))$. This cost can be for instance the distance from the target $X_f$ to $X(t_f)$. A geometric description of this case is given in Fig.~\ref{fig11}. We consider the level sets of the function $G$ as the set of points where $G(X)$ is a constant. The optimal situation corresponds to the case where the boundary of the reachable set at $t_f$ and the level set $G(X(t_f))$ are tangent in $X(t_f)$. We introduce the plane tangent to the two sets in $X(t_f)$. The adjoint state $\Lambda(t_f)$ is then defined up to a factor as the opposite of the gradient of the level set in $X(t_f)$, such that the two vectors point outwards of their respective sets. We denote by $\Lambda_0$ this negative constant and we finally have
$$
\Lambda(t_f)=\Lambda_0\frac{\partial G(X(t_f)}{\partial X(t_f)}.
$$
By definition of the optimal solution, we have
$$
G(X^\star(t_f))\leq G(X(t_f)).
$$
At first order in $\delta X$, we deduce that
\begin{equation}\label{eqappdelta4}
\partial_XG^\intercal \delta X(t_f)\geq 0.
\end{equation}
It is then straightforward to show that
$$
\Lambda(t_f)^\intercal \int_0^{t_f}V(t_f,t')[F(X^\star,u)-F(X^\star,u^\star)]dt'\leq 0,
$$
which gives
$$
\int_0^{t_f}\Lambda(t')^\intercal[F(X^\star,u)-F(X^\star,u^\star)]dt'\leq 0.
$$
We obtain the same inequality \eqref{eqappmax} by introducing the Pontryagin Hamiltonian.

\section{Description of the numerical optimization codes}\label{app_code}
This section aims to briefly present the different codes provided in the supplementary material. All codes are in Python and are executable independently of each other. The two GRAPE codes require only the use of the standard Python scientific libraries Numpy and SciPy. 

\paragraph{Shooting algorithm.} The code  \textit{shooting.py} implements the shooting method, an algorithm used in Sec.~\ref{sec:shooting_algo}. The integration of the equations of motion given by the PMP is performed using the Python package Nutopy~\cite{nutopy} and the function \textit{ivp.exp}. The optimization uses the function \textit{nle.solve}. The final part of the code returns two graphs, one with the Bloch coordinates $(x,y,z)$ of the two-level quantum system, and the other with the optimized control. We also provide a second code \textit{shooting2.py} without the package Nutopy, but with less precision.

\paragraph{GRAPE for a two-level quantum system.}
\textit{GRAPE.py} provides a minimal working code for the optimization of a control with GRAPE. It allows to derive the results of Sec.~\ref{sec:GRAPE_algo}, i.e. state-to-state transfer from the ground to the excited states. Several build-in functions are defined in the code, to perform the forward and backward propagation of the Schr\"odinger equation using the split-operator method. The optimization uses the Scipy function \textit{minimize} with the BFGS method. Here, the exact gradient is provided to the solver (with the function \textit{GradientGRAPE}), and the Hessian is estimated numerically from \textit{minimize}. An initial control must be provided to the algorithm. Many different choices would lead to the same result. Here, a cosine wave function is used. It is chosen different from zero, and smooth in order to obtain a good starting point for the algorithm. \textit{GRAPE2.py} considers the same control problem but with a polynomial parameterization of the control pulse. The computation of the gradient is modified accordingly.



\paragraph{GRAPE for a BEC system} The code \textit{GRAPE\_BEC.py} implements the algorithm GRAPE for state-to-state transfer in the case of BEC in a one-dimensional optical lattice. The results are those presented in Sec.~\ref{sec:Numerical-optimal-control}. The code is composed of several sections. The first is a class (``BEC'') which generates from the size of the system $N_k$, the value of the quasimomentum  $q$ and the depth of the lattice $s$, the Hamiltonian of the system (the matrices $H_0$, $H_1$ and $H_2$). The second section (``propagation''), also a class, defines the functions which return the fidelity and the correction to be made to the control given by the PMP. The third section groups together the functions of the code. The fourth section describes the system parameters, the constants, the control time, the values of $q$ and $s$, the initial state and the initial guess for the control. The fifth section aims to calculate the optimal control. The \textit{scipy.optimize.minimize} algorithm is used to iteratively find the solution to the state-to-state transfer problem, where the function to be minimized is ``Cost'' and the gradient ``dCost'' (based on class ``propagation''). The algorithm uses a L-BFGS-B method, it is therefore a second order algorithm since the Hessian is iteratively approximated at each step. The user can define a maximum number of iterations, a tolerance, and a bound for the control. The final sixth section displays the results for the three transfers, namely the controls, the population $\left\lvert c_{q,n} \right\rvert^2$, and the probability density.
At each iteration, we can also perform a line search approach to find the best value of the parameter $\epsilon$. This is done by computing a step that satisfies the Wolfe condition with \textit{scipy.optimize.line\_search}.

\section*{References}
\bibliographystyle{vancouver}

\end{document}